\documentclass[11pt, oneside]{article}   	
\usepackage{geometry}                		
\usepackage{graphicx}				
\usepackage{amssymb}

\usepackage[superscript,biblabel]{cite}

\title{{\bf Radiation by the superluminally moving current sheet in the magnetosphere of a neutron star}}
\author{Houshang Ardavan\footnote{Email address: ardavan@ast.cam.ac.uk}\\
Institute of Astronomy, University of Cambridge,\\
Madingley Road, Cambridge CB3 0HA, UK}
\date{}							

\begin{document}
\maketitle

\begin{abstract}

The mechanism by which the radiation received from obliquely rotating neutron stars is generated remains an open question half a century after the discovery of pulsars~\cite{Beskin2018, Melrose2021}.  In contrast, considerable progress has recently been made in determining the structure of the magnetosphere that surrounds these objects: numerical computations based on the force-free, magnetohydrodynamic and particle-in-cell formalisms have now firmly established that the magnetosphere of an oblique rotator entails a current sheet outside its light cylinder whose rotating distribution pattern moves with linear speeds exceeding the speed of light in vacuum~\cite{SpitkovskyA:Oblique, Contopoulos:2012, Tchekhovskoy2013, Philippov2018}.  However, the role played by the superluminal motion of this current sheet in generating the multi-wavelength focused pulses of radiation that we receive from neutron stars is unknown.  Here we insert the description of the current sheet provided by the numerical simulations~\cite{Tchekhovskoy:etal} in the classical expression for the retarded potential~\cite{JacksonJD:Classical} and thereby calculate the radiation field generated by this source in the time domain.  We find a radiation consisting of highly focused pulses whose (i) spectrum can extend from radio waves to gamma rays, (ii) brightness temperature can exceed $10^{40}$ ${}^\circ$K, (iii) linear polarization can be 100\%, (iv) two concurrent polarization position angles are approximately orthogonal often and swing through $180^\circ$ across the pulse profile in most cases, (v) circular polarization reverses sense across some components of the pulse profile, (vi) microstructure is determined by the thickness of the current sheet, and (vii) whose flux density diminishes with the distance $D$ from the star as $D^{-3/2}$ (rather than $D^{-2}$) in certain directions.  The intrinsically transient radiation process analysed here (and in ref.~\citeonline{Ardavan_JPP}) is thus capable of generating an emission whose features are strikingly similar to those of the emissions received from pulsars and magnetars and from the sources of fast radio bursts and gamma-ray bursts~\cite{Manchester1978,Kaspi2017,Petroff2019, Piron}.

\end{abstract}

\section{Main}

From the results obtained by numerical simulations of the magnetospheric structure of an obliquely rotating neutron star~\cite{SpitkovskyA:Oblique, Contopoulos:2012, Tchekhovskoy2013}, Tchekhovskoy et al.~\cite{Tchekhovskoy:etal} have derived a semi-analytic description of the distributions of the electric and magnetic fields that permeate the plasma surrounding these objects (see also ref.~\citeonline{Bogovalov1999}).  The described fields in conjunction with Maxwell's equations provide an explicit expression for the space-time distribution of the density of magnetospheric charges and currents including that of the current sheet (see \S 1 of the Supplementary Information).  The surface on which the current sheet is distributed spirals away from the light cylinder in the azimuthal direction at the same time as undulating in the latitudinal direction (see Fig.\ 1{\bf a}).  Its motion consists of a rotation with the angular frequency of rotation of the central neutron star, $\omega$, and a radial expansion with the speed of light in vacuum, $c$.  This is not incompatible with the requirements of special relativity because the superluminally moving distribution pattern of the current sheet is created by the coordinated motion of aggregates of subluminally moving charged particles~\cite{BolotovskiiBM:VaveaD, GinzburgVL:vaveaa, BolotovskiiBM:Radbcm}.

In this paper we treat the distribution of charges and currents that make up the current sheet at any given time as a prescribed volume source whose density can be inserted in the retarded solution of the inhomogeneous Maxwell's equations to find the radiation field it generates in unbounded free space.  The only role we assign to the rest of the magnetosphere, whose radiation field is negligibly weaker than that of the current sheet, is to maintain the propagation of this sheet.  The multi-wavelength focused pulses emitted by the current sheet escape the plasma surrounding the neutron star in the same way that the radiation generated by the accelerating charged particles invoked in most current attempts at modelling the emission mechanism of these objects does~\cite{Philippov2018,Philippov2019}.  

The current sheet is described by charge and current densities whose space-time distributions depend on the azimuthal coordinate $\varphi$ and time $t$ in the combination $\varphi-\omega t$ only.  The radiation field we are after can be built up, therefore, by the superposition of the fields of the uniformly rotating volume elements that constitute this source.  Superluminal counterpart of the field of synchrotron radiation, which plays the role of such a Green's function for the present problem, entails intersecting wave fronts that possess a two-sheeted cusped envelope (see Figs.\ 1{\bf b}, 1{\bf c} and 1{\bf d}).  Outside the envelope only one wave front passes through the observation point at any given observation time; but inside the envelope three distinct wave fronts, emitted at three different values of the retarded time, simultaneously pass through each observation point.  Coalescence of two of the contributing retarded times on the envelope of wave fronts results in the divergence of the Green's function on this surface.  At an observation point on the cusp locus of the envelope all three of the contributing retarded times coalesce and the Green's function has a higher-order singularity (see \S 2 of the Supplementary Information).

Constructive interference of the emitted waves and formation of caustics thus play a crucial role in determining the radiation field of the current sheet.  Not only the integral that defines the Green's function for the problem but every one of the repeated integrals in the classical expression for the retarded potential in the present case entails either singularities or nearby saddle points that coalesce and thereby result in further focusing of the radiation at certain observation points (see \S 3 of the Supplementary Information).

\section{Results}

The results reported below are consequences mainly of the shape and motion of the current sheet: two features of the magnetosphere which are the same not only both for a dipolar and a monopolar magnetic field at the surface of the star, but also both close to and far from the light cylinder~\cite{Tchekhovskoy:etal,Bogovalov1999}.

\subsection{Pulse profiles and polarization position angles}

Two examples of the longitudinal distributions of the Stokes parameters for the radiation generated by the current sheet are shown in Fig.\ 2.  The total intensity $I$ and the intensities of the linearly and circularly polarized components of the radiation, $L$ and $V$, are plotted for given values of the inclination angle $\alpha$ and colatitude $\theta_P$ of a far-field observer in units of
\begin{equation}
I_0=\left(\frac{\omega B_0 r_{s0}^2}{c R_P}\right)^2,
\label{E1}
\end{equation}
where $B_0$ is the magnitude of the star's dipolar field at its magnetic pole, $r_{s0}$ is the star's radius and $R_P$ is the distance of the observer from the star.  In these examples, and in Fig.\ 3, the origin of the azimuthal coordinate of the observation point is shifted to place the pulse window next to longitude zero.  The electric field of the radiation turns out to be the sum of two distinct parts with differing polarization position angles (see \S 4 of the Supplementary Information).  Figure 2 also shows the longitudinal distributions of the position angles of these two parts which we refer to as $P$ and $Q$ modes.  Here, those branches of the multi-valued function defining the position angle are adopted that yield continuous position-angle distributions across various components of a given pulse.

The variable $k_{\rm u}^{-1}$, whose value determines the wavelength of the small-amplitude modulations (microstructure) of the distributions shown in Fig.\ 2, represents a lower limit to the thickness of the current sheet in units of the light-cylinder radius.  The thickness assigned to the current sheet by the description in ref.~\citeonline{Tchekhovskoy:etal} is zero.  However, a superluminally moving source is necessarily volume-distributed~\cite{BolotovskiiBM:Radbcm}: it would give rise to a divergent field if it has no thickness.  We have circumvented this shortcoming of the description given in ref.~\citeonline{Tchekhovskoy:etal} by replacing the infinitely long integration range in the Fourier representation of the Dirac delta function describing the current sheet by the truncated wave-number interval $\vert k\vert\le k_{\rm u}$ (see Method).  

Figure 3 illustrates an example of a radically different type of pulse: one detectable near those observation points for which two nearby saddle points of the integral over the latitudinal distribution of the current sheet coalesce, thus giving rise to a much tighter focusing of the emitted waves.  Though their profiles over the entire pulse window look similar to those of other pulses (as in Fig.\ 3{\bf a}), such pulses display extraordinarily large amplitudes and short widths once their peaks are resolved (as in Fig.\ 3{\bf b}).  We shall see below that the extraordinary values of the amplitudes and widths of such pulses, illustrated by the example in Fig.~\ref{PF3}, are what underpin the high brightness temperatures and broad frequency spectra of the radiation generated by the current sheet.

Further examples of pulse profiles and position-angle distributions can be found in the Supplementary Information.  It should be added that at any given value of the inclination angle $\alpha$, the pulse observed at $\pi-\theta_P$ differs from that observed at $\theta_P$ only in that the intensity $V$ of its circularly polarized part is replaced by $-V$ and its longitude $\varphi_P$ is replaced by $\varphi_P+\pi$.  Moreover, the results for $\alpha > \pi/2$ follow from those for $\alpha < \pi/2$ by replacing $\theta_P$, $\varphi_P$ and $V$ by  $\pi-\theta_P$, $\varphi_P+\pi$ and $-V$, respectively.

\subsection{Brightness temperature}

By equating the magnitude of the Poynting flux of the radiation to the Rayleigh-Jeans law for the energy that a black body of the same temperature would emit per unit time per unit area into the frequency band $\Delta\nu$ centred on the frequency $\nu$, it can be shown that the brightness temperature of the radiation $T_{\rm b}$ is related to the Stokes parameter $I$ by
\begin{equation}
T_{\rm b}=1.23\times10^{10}I \,{\hat T}_{\rm b}\,\,{}^\circ{\rm K}\qquad{\rm with}\qquad{\hat T}_{\rm b}=\frac{{\hat B}_0^2d^4}{{\hat P}^2D^2{\hat\nu}^2\Delta{\hat\nu}}
\label{E2}
\end{equation}
in which $I$ is in units of $I_0$, ${\hat B}_0$ is the value of the magnetic field $B_0$ in units of $10^{12}$ Gauss, $d$ is the value of the star's radius $r_{s0}$ in units of $10^6$ cm, $D$ is the value of the observer's distance $R_P$ in kpc, ${\hat\nu}$ and $\Delta{\hat\nu}$ are the values of the radiation frequency and its bandwidth in units of $10^8$ Hz and $10^6$ Hz, respectively, and ${\hat P}=10^2/\omega$ is the value of the star's rotation period in seconds.  

Equation~(\ref{E2}) yields $T_{\rm b}=1.29\times10^{22}{\hat T}_{\rm b}$ ${}^\circ$K for the example depicted in Fig.\ 2{\bf a} and $T_{\rm b}=1.17\times10^{40}{\hat T}_{\rm b}$ ${}^\circ$K for the example depicted in Fig.\ 3.  (See \S 4 of the Supplementary Information for other examples including one for which $T_{\rm b}$ is as high as $10^{54}{\hat T}_{\rm b}$ ${}^\circ$K.)

\subsection{Frequency spectrum}

Given that the radiation field of the current sheet depends on the observation time $t_P$ and the azimuthal coordinate $\varphi_P$ of the observation point only in the combination $\varphi_P-\omega t_P$, the frequency spectrum of this radiation is equally well described by the Fourier decomposition of its longitudinal distribution.  In the present case, the content of this spectrum stems from two factors.  One factor is the thickness of the current sheet manifested in the sharp small-amplitude modulations of the pulse profile whose wavelengths are proportional to $k_{\rm u}^{-1}$ (see Fig.\ 1{\bf a}).  The other factor is the full width at half maximum ($\delta\varphi_P$) of the component of the pulse profile with the highest peak (see Fig.\ 3{\bf b}).  While the spectral distribution of the former lies in the radio band when $k_{\rm u}\gtrsim10^5$, that of the latter ranges from radio waves to gamma rays: the width $\delta\varphi_P=1.21\times10^{-26}$ radian of the pulse depicted in Fig.\ 3{\bf b}, for example, implies a frequency spectrum that extends as far as  $\nu\simeq\omega/(2\pi\delta\varphi_P)\simeq1.31\times10^{27}{\hat P}^{-1}$ Hz.

In the lower frequency range,  the Stokes parameter $I$ has the power-law dependence $\nu^{-\beta}$ on frequency with a spectral index $\beta$ that assumes the following values in various regimes: $2/3$, $1$, $4/3$, $5/3$, $2$, $7/3$ (see \S 4 of the Supplementary Information).

\subsection{Flux density and its rate of decay with distance}

In the present case where the length subtended by the longitudinal extent of the radiation beam embodying a high-frequency pulse can be shorter than $1$ cm at $D=1$ kpc, the flux density ${\cal S}$ of the radiation is related to the Stokes parameter $I$ by
\begin{equation}
{\cal S}=2.79\times10^{-3}\frac{I\,\delta\varphi_P}{D^2}\, {\cal S}_0\,\,\,\,\,\frac{{\rm erg}}{{\rm sec}\times{\rm cm}^2}\qquad{\rm with}\qquad{\cal S}_0=\left(\frac{{\hat B}_0 d^2}{{\hat P}}\right)^2,
\label{E3}
\end{equation}
in which $I$ is in units of $I_0$ and $\delta\varphi_P$ in radians.  In the case depicted in Fig.~\ref{PF3} where $\alpha=60^\circ$, $\theta_P=90^\circ$ and $k_{\rm u}=10^7$, for example, the flux density ${\cal S}$ has the value $32.1\, {\cal S}_0$ erg/(sec$\times$cm${}^2$) at $D=1$ kpc.

The separation between two nearby saddle points of the integral over the latitudinal distribution of the current sheet decreases as $R_P^{-1/2}$ with increasing distance $R_P$ of the observation point from the star.  The enhanced focusing of the radiation that is caused by this shortening of the separation between the saddle points gives rise to both a narrowing of the width and an augmenting of the peak intensity of the emitted pulse.  Because the peak intensity and the width of the pulse are modified with different rates, this effect results in a flux density that diminishes with increasing distance as $R_P^{-3/2}$ rather than $R_P^{-2}$ (or equivalently as $D^{-3/2}$ instead of $D^{-2}$).  The latitudinal width $\delta\theta_P$ of such non-spherically decaying pulses is of the order of $(R_P\omega/c)^{-1}$ radians in most cases.  But there are, in general, several latitudes near which such pulses can be observed; the number and locations of these latitudes are determined by the values of $\alpha$ and $R_P$ (see \S 4 of the Supplementary Information).

The violation of the inverse-square law encountered here is not incompatible with the requirements of the conservation of energy because the radiation process discussed in this paper is intrinsically transitive.  The difference in the fluxes of power across any two spheres centred on the star is in this case balanced by the change with time of the energy contained inside the shell bounded by those spheres (see Appendix C of ref.~\citeonline{Ardavan_JPP} for a demonstration of this feature).

Given their limited latitudinal extent, the non-spherically decaying pulses generated by the current sheet of a neutron star are more likely to be observed when, as a result of the precession of the star's rotation axis, the radiation beams embodying such pulses sweep past the Earth.  Using the decay rate $R_P^{-2}$, we would over-estimate the power emitted by the sources of the bursts of radiation we would receive in this way by as large a factor as $10^{14}$ if the neutron stars that generate the bursts lie at cosmological distances.  The enormously high energetic requirements normally attributed to the sources of fast radio bursts and gamma-ray bursts~\cite{Petroff2019,Piron} could therefore be artefacts of the invariably made assumption that the radiation fields of all sources necessarily decay as predicted by the inverse-square law.

\begin{figure}
\centerline{\includegraphics[width=15cm]{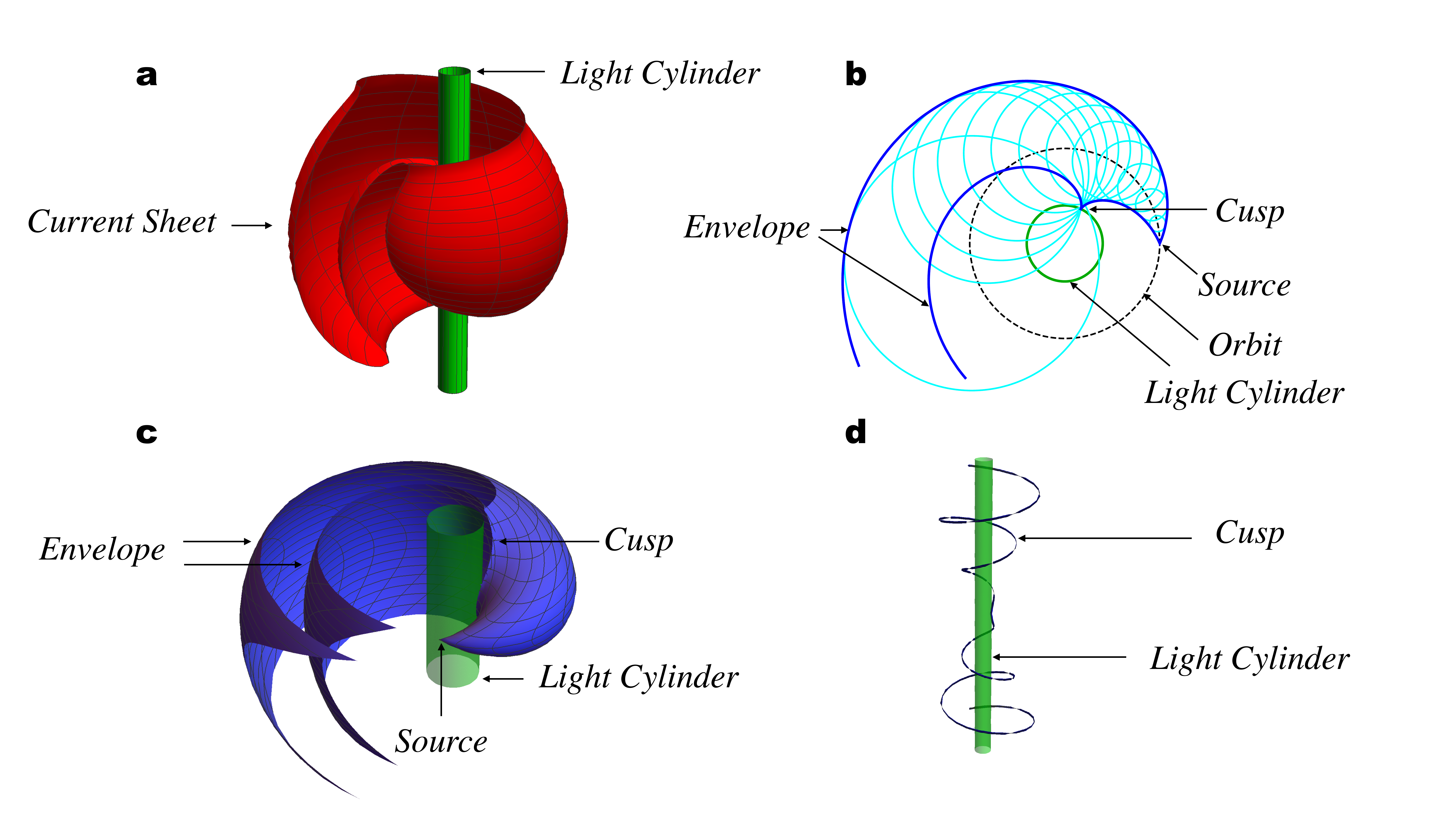}}
\caption{{\bf a}, Snapshot of a single turn of the current sheet about the light cylinder (the cylinder at which the speed of any entity co-rotating with the neutron star would equal the speed of light $c$). This surface undulates within the latitudinal interval $(\pi/2-\alpha,\pi/2+\alpha)$ each time it turns about the rotation axis, thus wrapping itself around the light cylinder as it extends to the outer edge of the magnetosphere ($\alpha$ which denotes the angle between the magnetic and rotation axes of the star has the value $\pi/3$ in this figure).  {\bf b}, Cross sections of the wave fronts (the circles in light blue) emanating from a volume element of the current sheet with fixed radial and latitudinal coordinates.  This figure is plotted for a source element the radius of whose orbit (the dotted circle) is $2.5$ times the radius $c/\omega$ of the light cylinder (the green circle).  Cross sections of the two sheets of the envelope of these wave fronts with the plane of the orbit (shown in dark blue) meet at a cusp and wind around the rotation axis, while moving away from it all the way to the far zone.  {\bf c}, Three-dimensional plot of the envelope of wave fronts emanating from a superluminally rotating source element and {\bf d}, the cusp of this envelope along which the two sheets of the envelope meet tangentially.  The cusp touches the light cylinder where it crosses the plane of the orbit and moves away from the axis of rotation as it and the envelope itself spiral into the far zone maintaining a symmetry with respect to this plane.} 
\label{PF1}
\end{figure}

\begin{figure}
\centerline{\includegraphics[width=15cm]{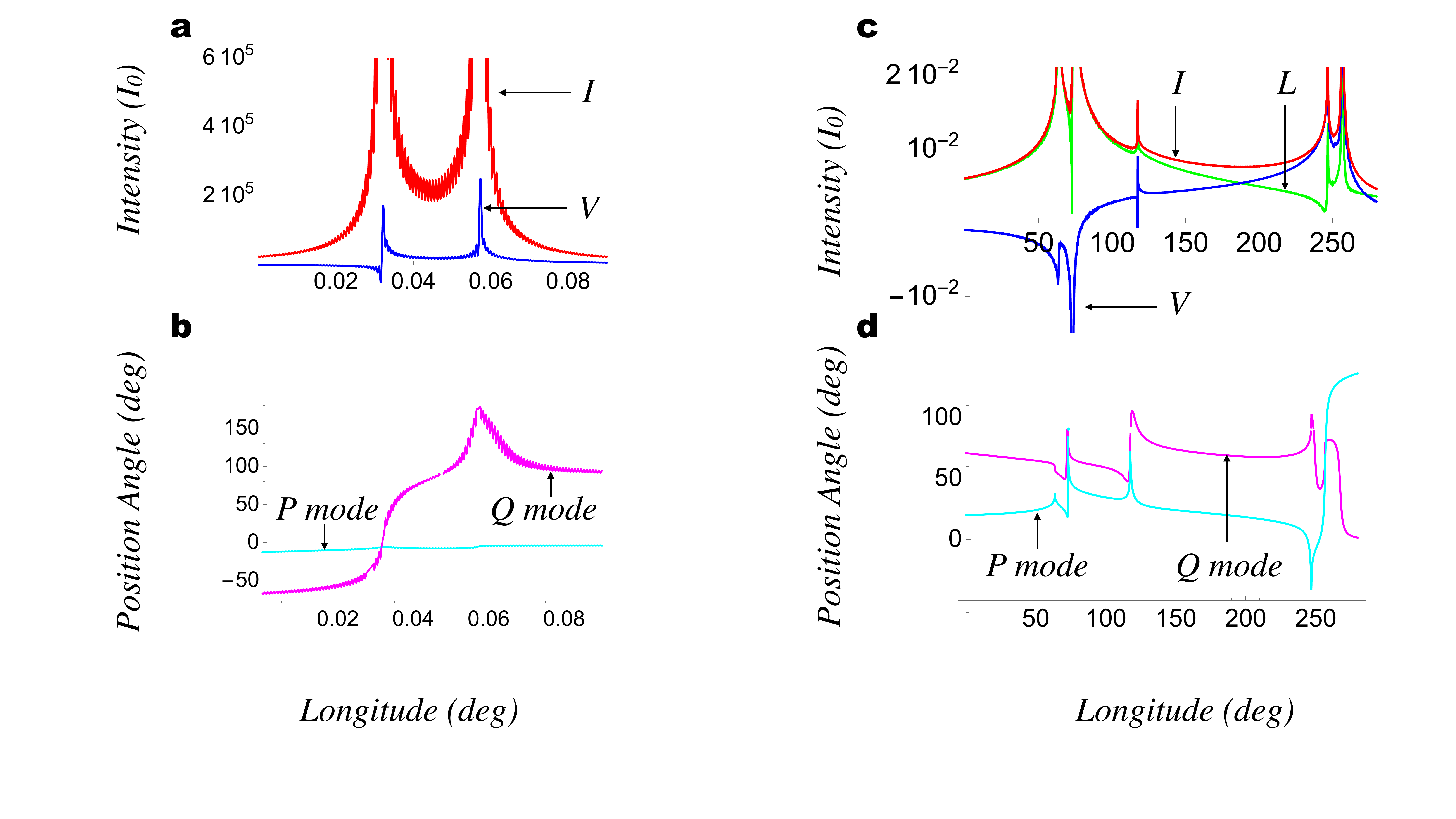}}
\caption{{\textbf a}, Longitudinal distributions of the intensities of the total ($I$) and circularly polarized ($V$) radiation generated by the current sheet in units of $I_0$ for the inclination angle $\alpha=5^\circ$ at a far-field observation point with the colatitude $\theta_P=5^\circ$.  Distribution of the intensity of the linearly polarized radiation ($L$) is essentially coincident with that of $I$ in this case.  The sharp modulations of the pulse profile (its microstructure) are discernible here because this figure is plotted for $k_{\rm u}=10^2$ which corresponds to a relatively thick current sheet.  The thinner the current sheet, the shorter are the wavelengths of these modulations.  At its peak, the right-hand component of this pulse has the intensity $I=1.05\times10^{12} I_0$ and the longitudinal width $6.76\times10^{-9}$ second when $k_{\rm u}$ is $10^7$, i.e., when the thickness of the current sheet is of the order of $10^{-7}c/\omega$.  {\bf b}, Position angles of the two polarization modes versus longitude for the pulse depicted in {\bf a}.  Note that not only does $V$ reverse sign across the left-hand component of this pulse but also the position angle of the $Q$ mode swings through $180^\circ$ across the profile of this component.  Note also the approximate orthogonality of the two modes over most of the pulse window. {\bf c}, The Stokes parameters $I$, $L$ and $V$ versus longitude for $\alpha=80^\circ$, $\theta_P=110^\circ$ and $k_{\rm u}=10^4$.  This is an example of a pulse with several widely separated peaks for which the Stokes parameters are comparable in magnitude over some longitudinal intervals.  Though hardly visible, small amplitude sharp modulations pervade also the distribution here.  {\bf d}, Position angles of the two polarization modes versus longitude for the pulse depicted in {\bf c} exemplifying position-angle distributions that undergo differing variations across different components of the pulse profile.  Part {\bf d} is plotted with $k_{\rm u}=10^7$ for clarity.} 
\label{PF2}
\end{figure}

\begin{figure}
\centerline{\includegraphics[width=15cm]{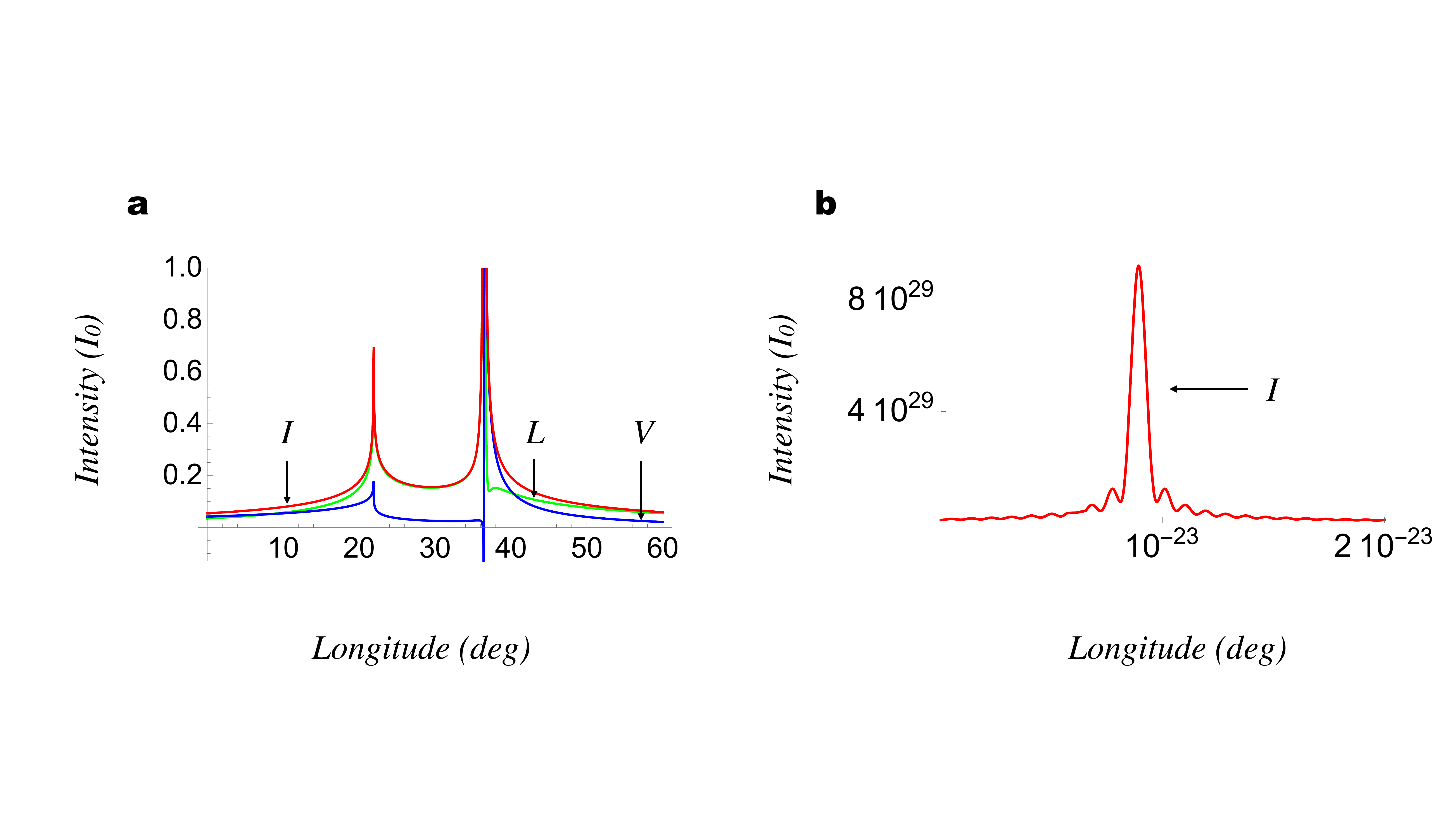}}
\caption{{\bf a}, Distributions of the Stokes parameters $I$, $L$ and $V$ in units of $I_0$ for the inclination angle $\alpha=60^\circ$, an observation point in the equatorial plane $\theta_P=90^\circ$, a distance of $10^{13}$ light-cylinder radii and $k_{\rm u}=10^7$.  In this example, the colatitude of the observation point lies close to one of the critical colatitudes for the given values of the inclination angle and distance (the colatitudes at which two nearby saddle points of the integral representing the retarded potential coalesce).  {\bf b}, The right-hand component of the pulse depicted in {\bf a} is here plotted over a sufficiently short longitudinal interval to resolve its peak and width.  The values of $\alpha$, $\theta_P$, $k_{\rm u}$ and distance in {\bf b} are the same as in {\bf a} but the origin of longitude is shifted in {\bf b} for clarity.  The shape of the pulse depicted in {\bf b} is the same in all cases of this type.} 
\label{PF3}
\end{figure}

\section{Method}

Both the outcomes of the numerical simulations of the magnetosphere of an oblique rotator~\cite{SpitkovskyA:Oblique, Contopoulos:2012, Tchekhovskoy2013, Philippov2018} and their semi-analytic description~\cite{Tchekhovskoy:etal} have appeared in the literature as space-time distributions of the electric and magnetic fields.  Once inserted in Maxwell's equations, the field distributions described in ref.~\citeonline{Tchekhovskoy:etal}  yield the following expressions for the charge and current densities, $\rho^{(cs)}$ and ${\bf j}^{(cs)}$, of the current sheet in terms of the spherical polar coordinates $(r_s, \theta,\varphi)$ and time $t$:   
\begin{equation}
\rho^{(cs)}=-\frac{j_0}{c {\hat r}_s^2}\sin\theta \, h\,\delta({\cal C}),
\label{E4}
\end{equation}
and
\begin{equation}
{\bf j}^{(cs)}=\rho^{(cs)}c\,{\hat{\bf e}}_{r_s}-\frac{j_0}{{\hat r}_s^3}\big[\sin\alpha\sin\left({\hat\varphi}+{\hat r}_s-{\hat r}_{s0}\right){\hat{\bf e}}_\theta+h\,{\hat{\bf e}}_\varphi\big]\delta({\cal C}),
\label{E5}
\end{equation}
where
\begin{equation}
{\cal C}=\sin\alpha\sin\theta\cos\left({\hat\varphi}+{\hat r}_s-{\hat r}_{s0}\right)+\cos\alpha\cos\theta,
\label{E6}
\end{equation}
\begin{equation}
{\hat\varphi}=\varphi-\omega t,
\label{E7}
\end{equation}
\begin{equation}
h=\sin\alpha\cos\theta\cos\left({\hat\varphi}+{\hat r}_s-{\hat r}_{s0}\right)-\cos\alpha\sin\theta,
\label{E8}
\end{equation}
\begin{equation}
j_0=0.365\omega B_0{\hat r}_{s0}^2w_1w_3/\pi,
\label{E9}
\end{equation}
\begin{equation}
w_1=\left\vert1-2\alpha/\pi\right\vert,\qquad w_3=1+0.2\sin^2\alpha,
\label{E10}
\end{equation}
$\delta$ is the Dirac delta function, ${\hat r}_s=r\omega/c$, ${\hat r}_{s0}=r_{s0}\omega/c$ and $({\hat{\bf e}}_{r_s}, {\hat{\bf e}}_\theta,{\hat{\bf e}}_\varphi)$ are the base vectors of the coordinate system (see \S 1 of the Supplementary Information).

For the purposes of the present analysis, it is essential that the finiteness of the duration of the source is taken into account (see Appendix B of ref.~\citeonline{Ardavan_JPP}).  If $\rho^{(cs)}$ and ${\bf j}^{(cs)}$ are turned on at $t=0$, then the coordinates $t$ and $\varphi$ in equations~(\ref{E4}) and (\ref{E5}) both range over $(0,\infty)$ but the values of the combination ${\hat\varphi}=\varphi-\omega t$ in which they occur has a limited range of length $2\pi$, e.g.,
\begin{equation}
0\le{\hat\varphi}<2\pi.
\label{E11}
\end{equation}
As can be seen from the alternative form $\varphi={\hat\varphi}+\omega t$ of equation~(\ref{E7}), ${\hat\varphi}$ is a Lagrangian coordinate that labels the rotating volume elements of the current distribution on each circle $r_s=$const, $\theta=$const, by their azimuthal positions at the time $t=0$.  This coordinate cannot range over a wider interval because the aggregate of volume elements that constitute a rotating source in its entirety can at most occupy an azimuthal interval of length $2\pi$ at any given time (e.g., at $t=0$). 

As pointed out in the text, we have replaced the range of integration in the following Fourier representation of the Dirac delta function that appears in equations (\ref{E4}) and (\ref{E5}), 
\begin{equation}
\delta({\cal C})=\frac{1}{2\pi}\int_{-\infty}^\infty{\rm d}k\,\exp({\rm i}\,k \,{\cal C}),
\label{E12}
\end{equation}
by the finite interval $(-k_{\rm u},k_{\rm u})$ to circumvent the divergence that arises from the vanishing of the thickness of the current sheet.  

In this paper, the retarded potential
\begin{equation}
\left[\matrix{{\mathbf A}({\bf x}_P,t_P)\cr\Phi({\bf x}_P,t_P)\cr}\right]=\frac{1}{c}\int{\rm d}^3 {\bf x}\,{\rm d}t\,\left[\matrix{{\mathbf j}^{(cs)}({\bf x},t)\cr{\rho^{(cs)} ({\bf x},t) c}\cr}\right]\frac{\delta(t-t_P+R/c)}{R},\quad t\ge0,\quad0\le{\hat\varphi}<2\pi,
\label{E13}
\end{equation}
that arises from the charge and current densities described by the modified versions of equations (\ref{E4}) and (\ref{E5}) is inserted in 
\begin{equation}
{\mathbf E}=-\nabla_P \Phi-\frac{1}{c}\frac{\partial{\mathbf A}}{\partial t_P},\quad\quad{\mathbf B}=\nabla_P{\mathbf{\times A}},
\label{E14}
\end{equation}
to obtain the corresponding expression for the generated radiation field $({\bf E},{\bf B})$, where $({\bf x},t)$ and $({\bf x}_P,t_P)$ are the space-time coordinates of the source points and the observation point $P$, respectively, and $R$ is the magnitude of the separation ${\bf R}={\bf x}_P-{\bf x}$ (see, e.g., ref.~\citeonline{JacksonJD:Classical}).  

To satisfy the required boundary conditions at infinity the free-space radiation field of an accelerated superluminal source has to be calculated (in the Lorenz gauge) by means of the retarded solution of the wave equation for the electromagnetic potential.  There is a fundamental difference between the classical expression for the retarded potential and the corresponding retarded solution of the wave equation that governs the electromagnetic field.  While the boundary contribution to the retarded solution of the wave equation for the potential that appears in Kirchhoff's surface-integral representation can always be rendered equal to zero by means of a gauge transformation that preserves the Lorenz condition, the corresponding boundary contribution to the retarded solution of the wave equation (or any other equation) for the field cannot be assumed to be zero {\it a priori}.  Not to exclude emissions whose intensity could decay more slowly than predicted by the inverse-square law, it is essential that the radiation field is derived from the retarded potential (see \S~3 of ref.~\citeonline{Ardavan_JPP} where this point is expounded).

Since the problem discussed in this paper entails the formation of caustics we cannot proceed to the far-field limit $\vert{\bf x}_P\vert\to\infty$  before evaluating the radiation field, as is customarily done in radiation theory.  The far-field approximation of the argument of the delta function in (\ref{E13}) would replace spherical wave fronts by planar wave fronts thereby relinquishing the possibility of their constructive interference.  Moreover, given the exceptionally short scales of the longitudinal (or equivalently temporal) variations of the present radiation, it would be intractably more difficult to obtain the time-domain results reported in this paper by means of a frequency-domain analysis (see, e.g., ref.~\citeonline{Achkasov2020}).  Further technical reasons why a conventional approach to the present problem does not work are discussed in  Appendix B of ref.~\citeonline{Ardavan_JPP}. 

We calculate the retarded potential (\ref{E13}) in \S3 of the Supplementary Information by first performing the integration with respect to $t$ at a fixed $(r_s,\theta,{\hat\varphi})$, i.e., by superposing the contributions of the uniformly rotating volume elements of the current sheet illustrated in Figs.~\ref{PF1}{\bf b}--{\bf d}.  A uniform asymptotic approximation to the value of the integral over $t$, which entails two nearby saddle points,~\cite{BleisteinN:Asei} is obtained by the time-domain version of the method of Chester, Friedman and Ursell.~\cite{BurridgeR:Asyeir,ChesterC:Extstd}  The divergences of the resulting expression (which acts as the Green's function for the problem) on the envelope of the emitted waves and its cusp stem from the relativistic restrictions inherent in Maxwell's equations and reflect the fact that no superluminal source can be point-like.~\cite{BolotovskiiBM:Radbcm}  Once the product of the charge--current density with this Green's function is integrated over the ${\hat\varphi}$ extent of the current sheet with the aid of Hadamard's regularisation technique,~\cite{HadamardJ:lecCau,HoskinsRF:GenFun} an expression is obtained that is singular only on the hyperboloid swept by the cusp loci of the envelopes of various source elements (see Fig.~\ref{PF1}{\bf d}).  This remaining singularity is also removed when the integration with respect to the radial extent of the current sheet is performed: an integration that receives its main contribution from the intersection of the hyperboloid in question with the current sheet.  The last integral (that with respect to $\theta$) is singularity-free and once again entails two nearby saddle points that coalesce for certain values of the colatitude $\theta_P$ of the observation point.

The five-dimensional integration (with respect to $t$, $r_s$, $\theta$, ${\hat\varphi}$ and $k$) required for evaluating the radiation field (\ref{E14}) has here been carried out analytically.  The only assumption made in the analysis presented in the Supplementary Information is that the radiation frequency appreciably exceeds the rotation frequency $\omega/2\pi$.

A final remark is in order: it is often presumed that the plasma equations used in the numerical simulations of the magnetospheric structure of an oblique rotator should, at the same time, predict any radiation that the resulting structure would be capable of emitting.~\cite{SpitkovskyA:Oblique, Contopoulos:2012}  Irrespective of the formalism on which they are based (whether force-free, MHD or particle-in-cell), the plasma equations used in these simulations are formulated in terms of the electric and magnetic fields (as opposed to potentials).  It has already been demonstrated in \S 3 of ref.~\citeonline{Ardavan_JPP}, however, that the gauge freedom offered by the solution of Maxwell's equations in terms of potentials plays an indispensable role in the prediction of the characteristics of the present radiation.  The absence of high-frequency radiation (and, specifically, the type of radiation described in this paper) is hardwired into the numerical simulations that have been performed to determine the magnetospheric structure of an oblique rotator by the imposition of the standard boundary conditions on the fields in the far zone (see \S 3 of ref.~\citeonline{Ardavan_JPP}).  

\section{Supplementary Information}

Supplementary Information for this article is included below in the form of a paper, a paper that presents the mathematical derivations of the results reported in the present article.

\end{document}


\maketitle

\tableofcontents

\section{Semi-analytic description of the magnetosphere}
\label{sec:magnetosphere}

In a reference frame marked by the spherical polar coordinates $(r_s,\theta,\varphi)$ for which $\theta=0$ coincides with the axis of rotation, the distributions of the electric and magnetic fields in the magnetosphere of an oblique rotator are, according to~\cite{Tchekhovskoy:etal}, given by
\begin{equation}
{\bf E}=-\frac{r_s\omega}{c}\sin\theta B_{r_s}{\hat{\bf e}}_\theta,
\label{2E1}
\end{equation}
\begin{equation}
B_{r_s}=w_3\left(w_1B_{r_s}^{(1)}+w_2B_{r_s}^{(2)}\right)
\label{2E2}
\end{equation}
\begin{equation}
B_\theta=0,\qquad B_\varphi=-\frac{r_s\omega}{c}\sin\theta B_{r_s},
\label{2E3}
\end{equation}
with
\begin{equation}
B_{r_s}^{(1)}=B_0\left(\frac{r_{s0}}{r_s}\right)^2\left[1+0.02(1-{\cal C}^2)^{1/2}+0.22(\vert{\cal C}\vert-1)-0.07(\vert{\cal C}\vert-1)^4\right]{\rm sgn}({\cal C}),
\label{2E4}
\end{equation}
\begin{equation}
B_{r_s}^{(2)}=B_0\left(\frac{r_{s0}}{r_s}\right)^2\sin\theta\cos\left(\varphi-\omega t+{\hat r}_s-{\hat r}_{s0}-\pi/6\right),
\label{2E5}
\end{equation}
\begin{equation}
{\cal C}=\sin\alpha\sin\theta\cos\left(\varphi-\omega t+{\hat r}_s-{\hat r}_{s0}\right)+\cos\alpha\cos\theta,
\label{2E6}
\end{equation}
\begin{equation}
w_1=\left\vert1-2\alpha/\pi\right\vert,
\label{2E7}
\end{equation}
\begin{equation}
w_2=1+0.17\vert\sin(2\alpha)\vert-w_1,
\label{2E8}
\end{equation}
\begin{equation}
w_3=1+0.2\sin^2\alpha,
\label{2E9}
\end{equation}
\refstepcounter{equation}
$$
{\hat r}_s=r_s\omega/c,\qquad{\hat r}_{s0}=r_{s0}\omega/c,
\eqno{(\theequation{\mathit{a},\mathit{b}})}
\label{2E10}
$$
where $t$ ($\ge 0$) is time, $\omega$ is the angular frequency of rotation of the star, $c$ is the speed of light in vacuum, $\alpha$ is the angle between the rotation and magnetic axes of the star (henceforth referred to as the inclination angle), $r_{s0}$ is the radius of the star and $B_0$ is the magnitude of the star's dipolar field at its magnetic pole.  The caret on $r_s$ and $r_{s0}$ (and any other variable with the dimension of a length) is used in this paper to designate a variable that is rendered dimensionless by being measured in units of the light-cylinder radius $c/\omega$.  

Inserting these fields in Maxwell's equations, we obtain the corresponding distributions of the charge density $\rho$ and the current density ${\bf j}$ in the magnetosphere:
\begin{equation}
\rho=\frac{1}{4\pi}{\bf \nabla\cdot E}=-\frac{\omega}{4\pi c}\left(2\cos\theta B_{r_s}+\sin\theta\frac{\partial B_{r_s}}{\partial\theta}\right),
\label{2E11}
\end{equation}
\begin{equation}
{\bf j}=\frac{c}{4\pi}\left({\bf\nabla\times B}-\frac{1}{c}\frac{\partial{\bf E}}{\partial t}\right)=\rho c\,{\hat{\bf e}}_{r_s}+\frac{c}{4\pi r_s}\left(\frac{1}{\sin\theta}\frac{\partial B_{r_s}}{\partial\varphi}{\hat{\bf e}}_\theta-\frac{\partial B_{r_s}}{\partial\theta}{\hat{\bf e}}_\varphi \right),
\label{2E12}
\end{equation}
in which
\begin{equation}
\frac{\partial B_{r_s}^{(1)}}{\partial\theta}=\left[\sin\alpha\cos\theta\cos\left(\varphi-\omega t+{\hat r}_s-{\hat r}_{s0}\right)-\cos\alpha\sin\theta\right]\frac{\partial B_{r_s}^{(1)}}{\partial{\cal C}},
\label{2E13}
\end{equation}
\begin{equation}
\frac{\partial B_{r_s}^{(1)}}{\partial\varphi}=-\sin\alpha\sin\theta\sin\left(\varphi-\omega t+{\hat r}_s-{\hat r}_{s0}\right)\frac{\partial B_{r_s}^{(1)}}{\partial{\cal C}},
\label{2E14}
\end{equation}
\begin{equation}
\frac{\partial B_{r_s}^{(2)}}{\partial\theta}=\cot\theta B_{r_s}^{(2)},
\label{2E15}
\end{equation}
\begin{equation}
\frac{\partial B_{r_s}^{(2)}}{\partial\varphi}=-\tan\left(\varphi-\omega t+{\hat r}_s-{\hat r}_{s0}-\pi/6\right)B_{r_s}^{(2)},
\label{2E16}
\end{equation}
\begin{equation}
\frac{\partial B_{r_s}^{(1)}}{\partial{\cal C}}=B_0\left(\frac{r_{s0}}{r_s}\right)^2\left[-\frac{0.02\,\vert{\cal C}\vert}{(1-{\cal C}^2)^{1/2}}+0.22-0.28(\vert{\cal C}\vert-1)^3+1.46\,\delta({\cal C})\right],
\label{2E17}
\end{equation}
the base vectors $({\hat{\bf e}}_{r_s},{\hat{\bf e}}_\theta,{\hat{\bf e}}_\varphi)$ are those of the spherical coordinate system $(r_s,\theta,\varphi)$ and $\delta$ is the Dirac delta function.  In deriving (\ref{2E12}), we have made use of the fact that here ${\bf E}$ depends on $\varphi$ and $t$ only in the combination $\varphi-\omega t$ and $r_s^2B_{r_s}$ depends on $\varphi$ and ${\hat r}_s$ only in the combination $\varphi+{\hat r}_s$, i.e., that $\partial{\bf E}/\partial t=-\omega\partial{\bf E}/\partial\varphi$ and $\partial(r_s^2B_{r_s})/\partial r_s=(\omega/c)\partial(r_s^2B_{r_s})/\partial\varphi$.

\begin{figure}
\centerline{\includegraphics[width=11cm]{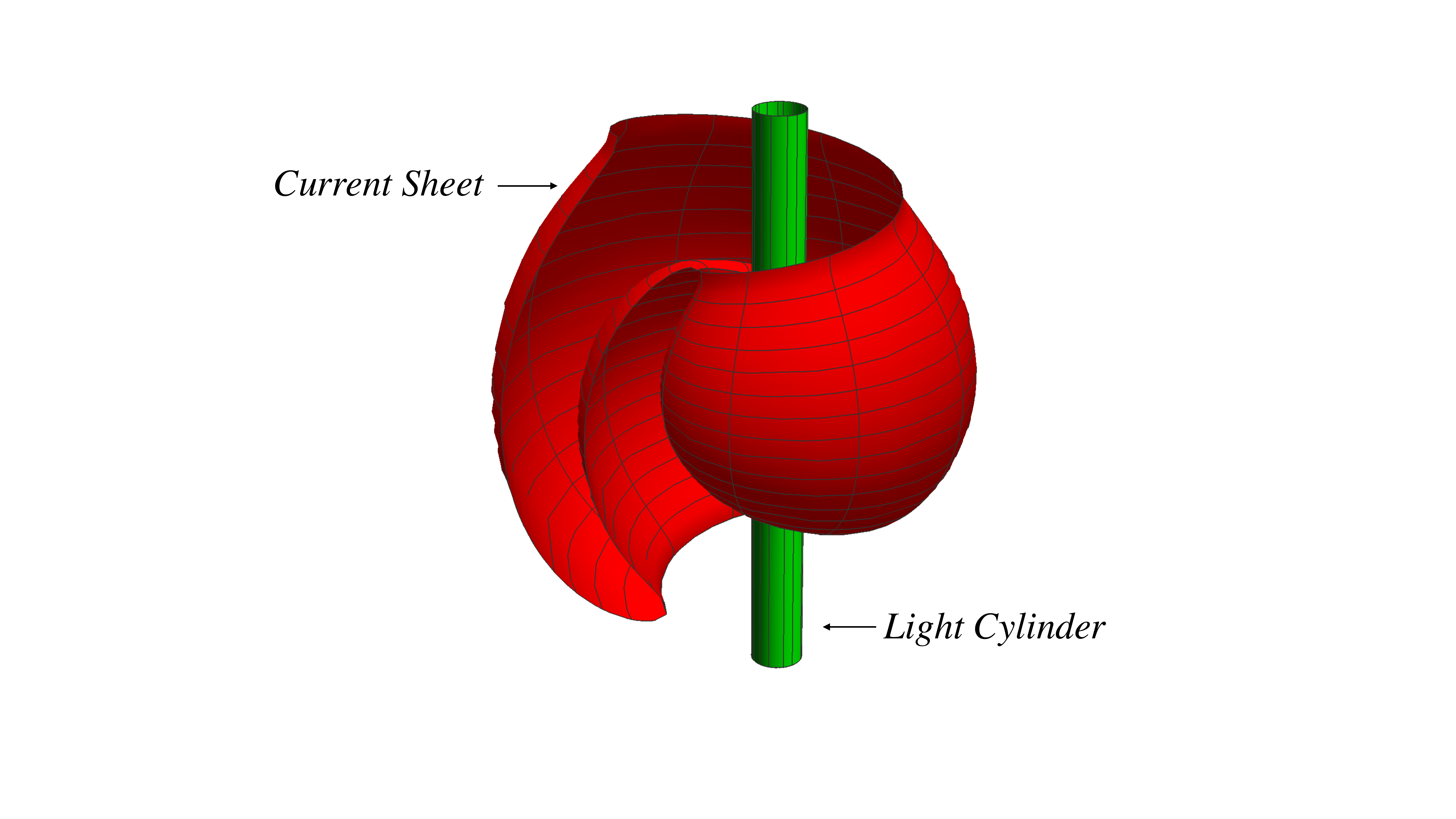}}
\caption{A single turn of the current sheet described by ${\cal C}=0$ for $\alpha=\pi/3$ (see~(\ref{2E6})).  At the same time as propagating radially outward with the speed of light in vacuum, $c$, the surface shown here rotates uniformly with the angular frequency of rotation of the central neutron star, i.e., with a linear speed $r\omega$ that exceeds $c$.  The superluminal motion of this current sheet is created by the coordinated movements of aggregates of subluminally moving charged particles.} 
\label{PF1}
\end{figure}

The magnetospheric current sheet shown in figure~\ref{PF1} occurs where the argument of the delta function in (\ref{2E12}), i.e., the function ${\cal C}$ defined by (\ref{2E6}), vanishes.  Since the radiation whose frequency appreciably exceeds the rotation frequency $\omega/2\pi$ is due entirely to this current sheet, we base our analysis of the radiation field in the following sections only on those terms in the above expressions for $\rho$ and ${\bf j}$ that involve the Dirac delta function.  We will see in \S~\ref{sec:characteristics} that the radiation generated by this sheet is in addition considerably more intense than that generated by the rest of the magnetosphere.  Disregarding the other terms in (\ref{2E11}) and (\ref{2E12}), we obtain
\begin{equation}
\rho^{(cs)}=-\frac{j_0}{c {\hat r}_s^2}\sin\theta \, h\,\delta({\cal C}),
\label{2E18}
\end{equation}
and
\begin{equation}
{\bf j}^{(cs)}=\rho^{(cs)}c\,{\hat{\bf e}}_{r_s}-\frac{j_0}{{\hat r}_s^3}\big[\sin\alpha\sin\left({\hat\varphi}+{\hat r}_s-{\hat r}_{s0}\right){\hat{\bf e}}_\theta+h\,{\hat{\bf e}}_\varphi\big]\delta({\cal C}),
\label{2E19}
\end{equation}
for this sheet's charge and current densities, where
\begin{equation}
{\hat\varphi}=\varphi-\omega t,
\label{2E20}
\end{equation}
\begin{equation}
h=\sin\alpha\cos\theta\cos\left({\hat\varphi}+{\hat r}_s-{\hat r}_{s0}\right)-\cos\alpha\sin\theta,
\label{2E21}
\end{equation}
and
\begin{equation}
j_0=0.365\omega B_0{\hat r}_{s0}^2w_1w_3/\pi.
\label{2E22}
\end{equation}
When the inclination angle $\alpha$ lies in the interval $[0,\pi/2]$, the above source densities are non-zero only in
\begin{equation}
\frac{\pi}{2}-\alpha\le\theta\le\frac{\pi}{2}+\alpha,
\label{2E23}
\end{equation}
as can be seen from the argument of the delta function that appears in (\ref{2E18}) and (\ref{2E19}).  It turns out that the symmetry of the current sheet with respect to $\theta$ and $\alpha$ enables us to infer the results for $\pi/2<\alpha<\pi$ from those for $0<\alpha<\pi/2$ (see \S~\ref{subsec:critical}).

For the purposes of quoting the results derived in~\cite{Ardavan_JPP}, it is convenient to express the base vectors of the spherical polar coordinates used above in terms of the base vectors $({\hat{\bf e}}_r,{\hat{\bf e}}_\varphi,{\hat{\bf e}}_z)$ of the cylindrical polar coordinates $(r,\varphi,z)$ defined by
\refstepcounter{equation}
$$
r=r_s\sin\theta,\qquad z=r_s\cos\theta,
\eqno{(\theequation{\mathit{a},\mathit{b}})}
\label{2E24}
$$
i.e., to replace $({\hat{\bf e}}_{r_s},{\hat{\bf e}}_\theta,{\hat{\bf e}}_\varphi)$ via
\refstepcounter{equation}
$$
{\hat{\bf e}}_{r_s}=\sin\theta\,{\hat{\bf e}}_r+\cos\theta\,{\hat{\bf e}}_z,\qquad {\hat{\bf e}}_\theta=\cos\theta\,{\hat{\bf e}}_r-\sin\theta\,{\hat{\bf e}}_z.
\eqno{(\theequation{\mathit{a},\mathit{b}})}
\label{2E25}
$$
Equations~(\ref{2E18}) and~(\ref{2E19}) can then be written as
\refstepcounter{equation}
$$
\rho^{(cs)}=-\frac{j_0}{{\hat r}_s^3}\,{\tilde\rho}\,\delta({\cal C}),\qquad{\bf j}^{(cs)}=-\frac{j_0}{{\hat r}_s^3}\,{\tilde{\bf j}}\,\delta({\cal C}),
\eqno{(\theequation{\mathit{a},\mathit{b}})}
\label{2E26}
$$
in which
\begin{equation}
{\tilde\rho}c={\hat r}_s\sin\theta\,h,
\label{2E27}
\end{equation}
\begin{equation}
{\tilde j}_r={\hat r}_s\sin^2\theta\,h+\sin\alpha\cos\theta\sin({\hat\varphi}+{\hat r}_s-{\hat r}_{s0}),
\label{2E28}
\end{equation}
\begin{equation}
{\tilde j}_\varphi=h,
\label{2E29}
\end{equation}
\begin{equation}
{\tilde j}_z={\hat r}_s\sin\theta\cos\theta\,h-\sin\alpha\sin\theta\sin({\hat\varphi}+{\hat r}_s-{\hat r}_{s0}),
\label{2E30}
\end{equation}
describe the dimensionless charge density ${\tilde\rho}$ and the cylindrical components of the dimensionless current density ${\tilde{\bf j}}$ of the magnetospheric current sheet as functions of $({\hat r}_s,\theta,{\hat\varphi})$ (see~(\ref{2E10}) and (\ref{2E20})).

\section{Formulation of the problem}
\label{sec:formulation}

\subsection{The free-space solution of Maxwell's equations that satisfies the required boundary conditions at infinity}
\label{subsec:boundaryConditions}
\setcounter{figure}{0}
\setcounter{equation}{0}

To satisfy the required boundary conditions at infinity the free-space radiation field of an accelerated superluminal source has to be calculated (in the Lorenz gauge) by means of the retarded solution of the wave equation for the electromagnetic potential.  There is a fundamental difference between the classical expression for the retarded potential and the corresponding retarded solution of the wave equation that governs the electromagnetic field.  While the boundary contribution to the retarded solution of the wave equation for the potential that appears in Kirchhoff's surface-integral representation can always be rendered equal to zero by means of a gauge transformation that preserves the Lorenz condition, the corresponding boundary contribution to the retarded solution of the wave equation (or any other equation) for the field cannot be assumed to be zero {\it a priori}.  Not to exclude emissions whose intensity could decay more slowly than predicted by the inverse-square law, it is essential that the radiation field is derived from the retarded potential (see \S~3 of~\cite{Ardavan_JPP} where this point is expounded).

Accordingly, we base the analysis in this paper on the classical expression
\begin{equation}
\left[\matrix{{\mathbf A}\cr{\Phi}\cr}\right]=\frac{1}{c}\int{\rm d}^3 {\bf x}\,{\rm d}t\,\left[\matrix{{\mathbf j^{(cs)}}\cr{\rho^{(cs)} c}\cr}\right]\frac{\delta(t-t_P+R/c)}{R}.
\label{3E1}
\end{equation}
for the retarded potentials that arise from the charge and current densities described by (\ref{2E26}) and insert this in 
\begin{equation}
{\mathbf E}=-\nabla_P \Phi-\frac{1}{c}\frac{\partial{\mathbf A}}{\partial t_P},\quad\quad{\mathbf B}=\nabla_P{\mathbf{\times A}},
\label{3E2}
\end{equation}
to obtain the corresponding expression for the generated fields:
\begin{equation}
\left[\matrix{{\mathbf E}\cr{\mathbf B}\cr}\right]=\frac{1}{c^2}\int{\rm d}^3 {\bf x}\,{\rm d}t\,\frac{\delta^\prime(t-t_P+R/c)}{R}\left[\matrix{{\mathbf j}^{(cs)}-\rho^{(cs)} c\,{\hat{\mathbf n}}\cr{\hat{\mathbf n}}{\mathbf\times}{\mathbf j}^{(cs)}\cr}\right],
\label{3E3}
\end{equation}
where $({\bf x},t)$ and $({\bf x}_P,t_P)$ are the space-time coordinates of the source points and the observation point $P$, respectively, $R$ is the magnitude of the separation ${\bf R}={\bf x}_P-{\bf x}$, ${\hat{\bf n}}={\bf R}/R$ is a unit vector along the radiation direction and $\delta^\prime$ denotes the derivative of the Dirac delta function with respect to its argument (see, e.g.,~\cite{JacksonJD:Classical}).  Since the problem we will be analysing entails the formation of caustics we cannot proceed to the far-field limit $\vert{\bf x}_P\vert\to\infty$ before evaluating the above integral, as is customarily done in radiation theory.  The far-field approximation of the argument of the delta function in (\ref{3E3}) would replace spherical wave fronts by planar wave fronts thereby relinquishing the possibility of their constructive interference.  As we will be applying our results to astronomical objects we can, however, approximate the unit vector ${\hat {\bf n}}$ by its far-field value
\begin{equation}
{\hat{\bf n}}_\infty=\sin\theta_P{\hat{\bf e}}_{r_P}+\cos\theta_P{\hat{\bf e}}_{z_P}
\label{3E4}
\end{equation}
at this stage, so that the magnetic field can be obtained from ${\bf B}={\hat{\bf n}}_\infty{\bf\times}{\bf E}$ once the electric field is known.  In (\ref{3E4}), $({\hat{\bf e}}_{r_P},{\hat{\bf e}}_{\varphi_P}, {\hat{\bf e}}_{z_P})$ are the cylindrical base vectors at the observation point $P$.

For the purposes of the present analysis, it is essential that the finiteness of the duration of the source is taken into account (see appendix B of~\cite{Ardavan_JPP}).   If $\rho^{(cs)}$ and ${\bf j}^{(cs)}$ are turned on at $t=0$, then the coordinates $t$ and $\varphi$ in~(\ref{2E26}) both range over $(0,\infty)$ but the values of the combination ${\hat\varphi}=\varphi-\omega t$ in which they occur has a limited range of length $2\pi$, e.g.,
\begin{equation}
0\le{\hat\varphi}<2\pi.
\label{3E5}
\end{equation}
As can be seen from the alternative form $\varphi={\hat\varphi}+\omega t$ of~(\ref{2E20}), ${\hat\varphi}$ is a Lagrangian coordinate that labels the rotating volume elements of the current distribution on each circle $r=$const, $z=$const, by their azimuthal positions at the time $t=0$.  This coordinate cannot range over a wider interval because the aggregate of volume elements that constitute a rotating source in its entirety can at most occupy an azimuthal interval of length $2\pi$ at any given time (e.g., at $t=0$).  In this section, we mark the spatial coordinates of the source points by cylindrical polar coordinates and eliminate $t$ in favour of ${\hat\varphi}$.

Thus changing the variables of integration in (\ref{3E3}) to $({\bf x},t)=(r,\varphi, z, {\hat\varphi})$ 
and introducing the dimensionless coordinates ${\hat r}=r\omega/c$ and ${\hat z}=z\omega/c$, we obtain
\begin{equation}
{\bf E}=\frac{1}{\omega} \sum_{m=1}^\infty\int{\hat r}{\textrm d}{\hat r}\,{\textrm d}{\hat\varphi}\,{\textrm d}{\hat z}\,\int_{{\hat\varphi}+2(m-1)\pi}^{{\hat\varphi}+2m\pi}{\textrm d}{\varphi}\,\frac{\delta^\prime(g-\phi)}{{\hat R}}({\mathbf j}^{(cs)}-\rho^{(cs)} c\,{\hat{\bf n}}_\infty),
\label{3E6}
\end{equation}
where
\begin{equation}
{\hat R}=[({\hat z}-{\hat z}_P)^2+{{\hat r}_P}^2+{\hat r}^2-2{\hat r}_P{\hat r}\cos(\varphi-\varphi_P)]^{1/2},
\label{3E7}
\end{equation}
the function $g({\hat r},\varphi,{\hat z};{\hat r}_P,\varphi_P,{\hat z}_P)$ is defined by
\begin{equation}
g\equiv\varphi-\varphi_P+{\hat R},
\label{3E8}
\end{equation}
and the variable $\phi$ in the argument of the delta function stands for
\begin{equation}
\phi\equiv{\hat\varphi}-{\hat\varphi}_P\qquad{\rm with}\qquad{\hat\varphi}_P\equiv\varphi_P-\omega t_P.
\label{3E9}
\end{equation}
We have expressed the range of $\varphi$ integration as a sum of the intervals of length $2\pi$ that the element initially located at ${\hat\varphi}$ traverses during each of its individual rotations: $m$ is a positive integer enumerating successive rotation periods (the first rotation period being designated by $m=1$) and the summation extends over the set of rotations executed by the source over its lifetime. 

\subsection{The Green's function for the problem and its loci of singularities}
\label{subsec:Green's}

To put the current density ${\mathbf j}^{(cs)}=j^{(cs)}_r{\hat{\mathbf e}}_r+j^{(cs)}_\varphi{\hat{\mathbf e}}_\varphi+j^{(cs)}_z{\hat{\mathbf e}}_z$ into a form suitable for performing the integration with respect to $\varphi$, we need to express the $\varphi$-dependent base vectors $({\hat{\mathbf e}}_r,{\hat{\bf e}}_\varphi,{\hat{\mathbf e}}_z)$ associated with the source point $(r,\varphi,z)$ in terms of the constant base vectors $({\hat{\mathbf e}}_{r_P},{\hat{\mathbf e}}_{\varphi_P},{\hat{\mathbf e}}_{z_P})$ at the observation point $(r_P,\varphi_P,z_P)$:
\begin{equation}
\left[\matrix{{\hat{\mathbf e}}_r\cr {\hat{\mathbf e}}_\varphi\cr {\hat{\mathbf e}}_z\cr}\right]=\left[\matrix{\cos(\varphi-\varphi_P)&\sin(\varphi-\varphi_P)&0\cr
-\sin(\varphi-\varphi_P)&\cos(\varphi-\varphi_P)&0\cr
0&0&1\cr}\right]\left[\matrix{{\hat{\bf e}}_{r_P}\cr {\hat{\bf e}}_{\varphi_P}\cr {\hat{\bf e}}_{z_P}\cr}\right].
\label{3E10}
\end{equation}
Once the resulting expression,
\begin{eqnarray}
{\bf j}^{(cs)}&=&[j^{(cs)}_r\cos(\varphi-\varphi_P)-j^{(cs)}_\varphi\sin(\varphi-\varphi_P)]{\hat{\mathbf e}}_{r_P}\nonumber\\
&&+[j^{(cs)}_r\sin(\varphi-\varphi_P)+j^{(cs)}_\varphi\cos(\varphi-\varphi_P)]{\hat{\mathbf e}}_{\varphi_P}+j^{(cs)}_z{\hat{\mathbf e}}_{z_P},
\label{3E11}
\end{eqnarray}
is inserted in (\ref{3E6}) and $\delta^\prime(g-\phi)$ is written as $- \partial\delta(g-\phi)/\partial{\hat\varphi}$ (see (\ref{3E9})), we arrive at
\begin{equation}
{\mathbf E}=-\frac{1}{\omega}\sum_{j=1}^3\int_{\mathcal S}{\hat r}{\textrm d}{\hat r}\,{\textrm d}{\hat\varphi}\,{\textrm d}{\hat z}\,\frac{\partial G_j}{\partial{\hat\varphi}}{\bf u}_j,
\label{3E12}
\end{equation}
in which
\begin{equation}
\left[\matrix{G_1\cr G_2\cr G_3\cr}\right]=\sum_{m=1}^\infty\int_{{\hat\varphi}+2(m-1)\pi}^{{\hat\varphi}+2m\pi} {\rm d}\varphi\,{\delta(g-\phi)\over {\hat R}}\left[\matrix{\cos(\varphi-\varphi_P)\cr \sin(\varphi-\varphi_P)\cr 1\cr}\right]
\label{3E13}
\end{equation}
denotes the outcome of the remaining integration with respect to $\varphi$ and
\begin{equation}
{\bf u}_j=-\frac{j_0}{{\hat r}_s^3}\,{\tilde{\bf u}}_j\,\delta({\cal C})
\label{3E14}
\end{equation}
with
\begin{equation}
{\tilde{\bf u}}_1={\tilde j}_r{\hat{\bf e}}_{r_P}+{\tilde j}_\varphi{\hat{\bf e}}_{\varphi_P},
\label{3E15}
\end{equation}
\begin{equation}
{\tilde{\bf u}}_2={\tilde j}_r{\hat{\bf e}}_{\varphi_P}-{\tilde j}_\varphi{\hat{\bf e}}_{r_P},
\label{3E16}
\end{equation}
\begin{equation}
{\tilde{\bf u}}_3=-{\tilde\rho}c\sin\theta_P{\hat{\bf e}}_{r_P}+({\tilde j}_z-{\tilde\rho}c\cos\theta_P){\hat{\bf e}}_{z_P}
\label{3E17}
\end{equation}
(see (\ref{2E27})--(\ref{2E30})).  Note that since ${\hat\varphi}+2(m-1)\pi$ designates the same source element as ${\hat\varphi}+2m\pi$ the dependence on ${\hat\varphi}$ of the limits of integration in~(\ref{3E13}) does not contribute toward the values of the derivatives of $G_j$ with respect to ${\hat\varphi}$.

The function $G_j({\hat r},{\hat\varphi},{\hat z};{\hat r}_P,{\hat\varphi}_P,{\hat z}_P)$ here acts as the Green's function for the present problem.  It describes the Li\'enard-Wiechert field that arises from an individual volume element of the rotating distribution pattern of the source.  If we specialize the current distribution to a rotating point charge $q$, i.e., let $j_r=j_z=0$ and $j_\varphi=r^\prime\omega q\delta(r-r^\prime)\delta({\hat\varphi})\delta(z)$ with a constant $r^\prime$, then (\ref{3E13}) at an observation point in the far zone would describe the familiar field of synchrotron radiation when $r^\prime<c/\omega$ and a synergic field combining attributes of both synchrotron and \v Cerenkov emissions when $r^\prime>c/\omega$.  

\begin{figure}
\centerline{\includegraphics[width=11cm]{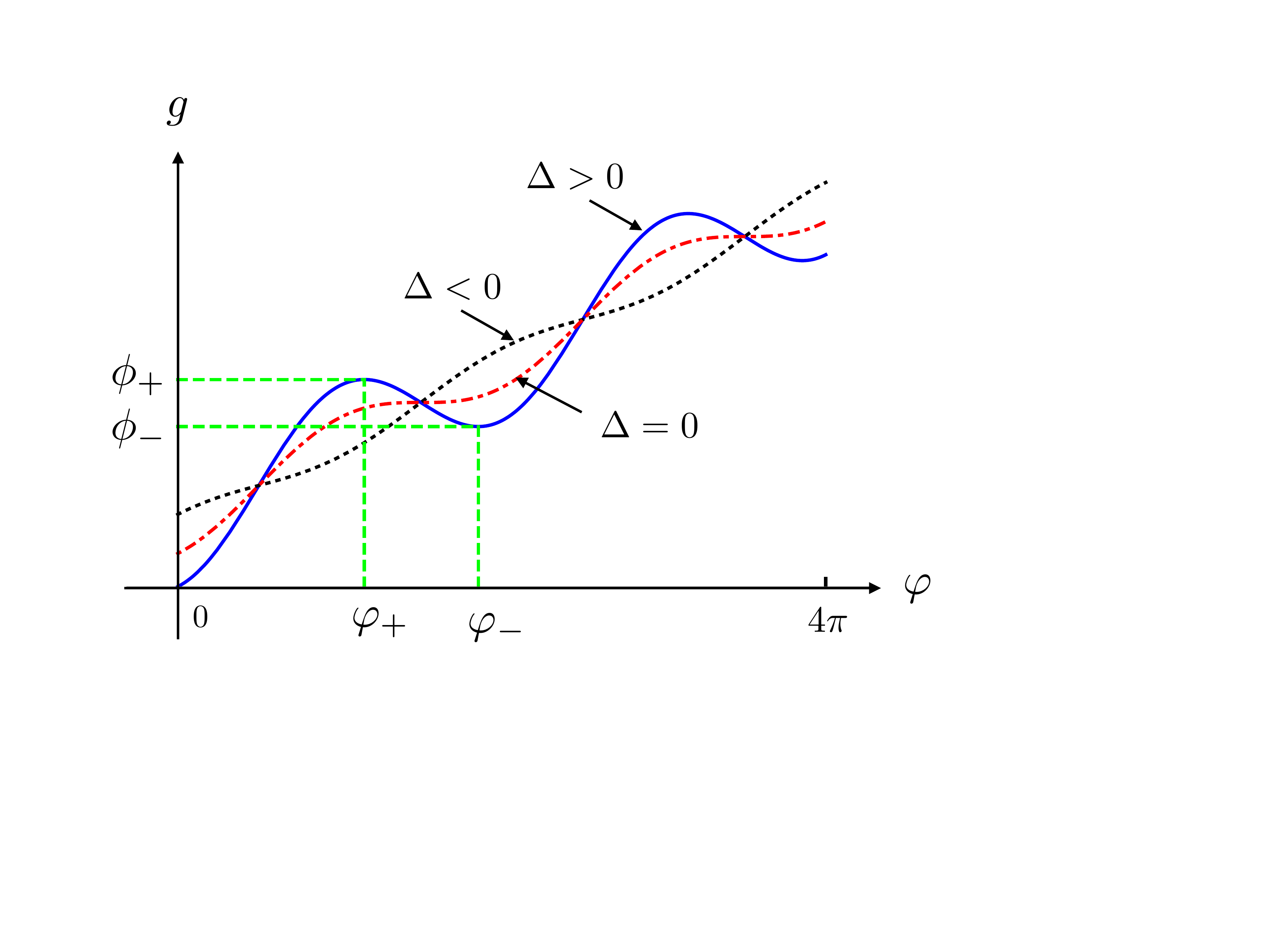}}
\caption{Generic forms of the function $g(\varphi)$ for source points whose $({\hat r},{\hat z})$ coordinates lie across the boundary $\Delta=0$ delineating the projection of the cusp curve of the bifurcation surface onto the $({\hat r},{\hat z})$ plane.  Depending on whether $\phi$ lies outside or inside the interval $(\phi_-,\phi_+)$, contributions are made toward the observed field (i.e., the argument $g(\varphi)-\phi$ of the Dirac delta function in (\ref{3E13}) vanishes) at either one or three retarded positions of the source.  For a horizontal line $g=\phi$ that either approaches an extremum of $g(\varphi)$ from inside the interval $(\phi_-,\phi_+)$ or passes through an inflection point of $g(\varphi)$, two or all three of the retarded positions in question coalesce and so their contributions interfere constructively to form caustics.  This figure is for ${\hat r}=3$ and only shows two rotation periods.  At higher speeds, the difference between the values of $\phi_+$ and $\phi_-$ can be large enough for a horizontal line $g=\phi$ to intersect $g(\varphi)$ over more than one rotation period (see figure~36 in~\cite{Ardavan_JPP}).  Contributions toward the observed field can thus arise, not only from one or three, but from any odd number of retarded positions of the source.  There are contributions from more than three retarded times whenever the rotation period of the source is shorter than the time taken by the collapsing sphere $\vert{\bf x}-{\bf x}_P\vert=c(t-t_P)$, centred on the observation point $P$, to cross the orbit of the source.}
\label{PF2}
\end{figure}

Depending on the value of 
\begin{equation}
\Delta=({{\hat r}_P}^2-1)({\hat r}^2-1)-({\hat z}-{\hat z}_P)^2
\label{3E18}
\end{equation}
for a given source point $(r,{\hat\varphi},z)$ with $r\omega>c$, the $\varphi$-dependence of the function $g$ that appears in the definition of the Green's function $G_j$ in~(\ref{3E13}) has one of the generic forms shown in figure~\ref{PF2}.  As can be seen from the curve labelled $\Delta>0$ in this figure, there are values, 
\begin{equation}
\varphi_\pm=\varphi_P+2m\pi-\arccos\left(\frac{1\mp\Delta^{1/2}}{{\hat r}{\hat r}_P}\right),
\label{3E19}
\end{equation}
of the retarded position of the source point at which
\begin{equation}
\frac{\partial g}{\partial\varphi}=1+\frac{{\hat r}{\hat r}_P\sin(\varphi-\varphi_P)}{{\hat R}}
\label{3E20}
\end{equation}
vanishes and so $G_j$ diverges.  These turning points of $g$ occur at source points for which $\partial(R\vert_{\varphi={\hat\varphi}+\omega t})/\partial t=-c$, i.e., the source points that approach the observer, along the radiation direction ${\hat{\bf n}}$, with the speed of light at the retarded time.  The inflection point of $g$ (see the curve labelled $\Delta=0$ in figure~\ref{PF2}), at which
\begin{equation}
\frac{\partial^2g}{\partial\varphi^2}\Big\vert_{\varphi=\varphi_\pm}=\mp\frac{\Delta^{1/2}}{{\hat R}_\pm}
\label{3E21}
\end{equation}
in addition vanishes, occurs at source points that approach the observer not only with the wave speed but also with zero acceleration at the retarded time, i.e., for which both $\partial(R\vert_{\varphi={\hat\varphi}+\omega t})/\partial t=-c$ and $\partial^2(R\vert_{\varphi={\hat\varphi}+\omega t})/\partial t^2=0$ at the time when $g\vert_{\varphi={\hat\varphi}+\omega t}=\phi$ and $\partial g/\partial\varphi=\partial^2 g/\partial\varphi^2=0$.  In (\ref{3E21}),
\begin{equation}
{\hat R}_\pm= [({\hat z}-{\hat z}_P)^2+{\hat r}^2+{{\hat r}_P}^2-2(1\mp\Delta^{1/2})]^{1/2}
\label{3E22}
\end{equation}
is the value of ${\hat R}$ at the extrema $\varphi_\pm$ of $g$. 

\begin{figure}
\centerline{\includegraphics[width=11cm]{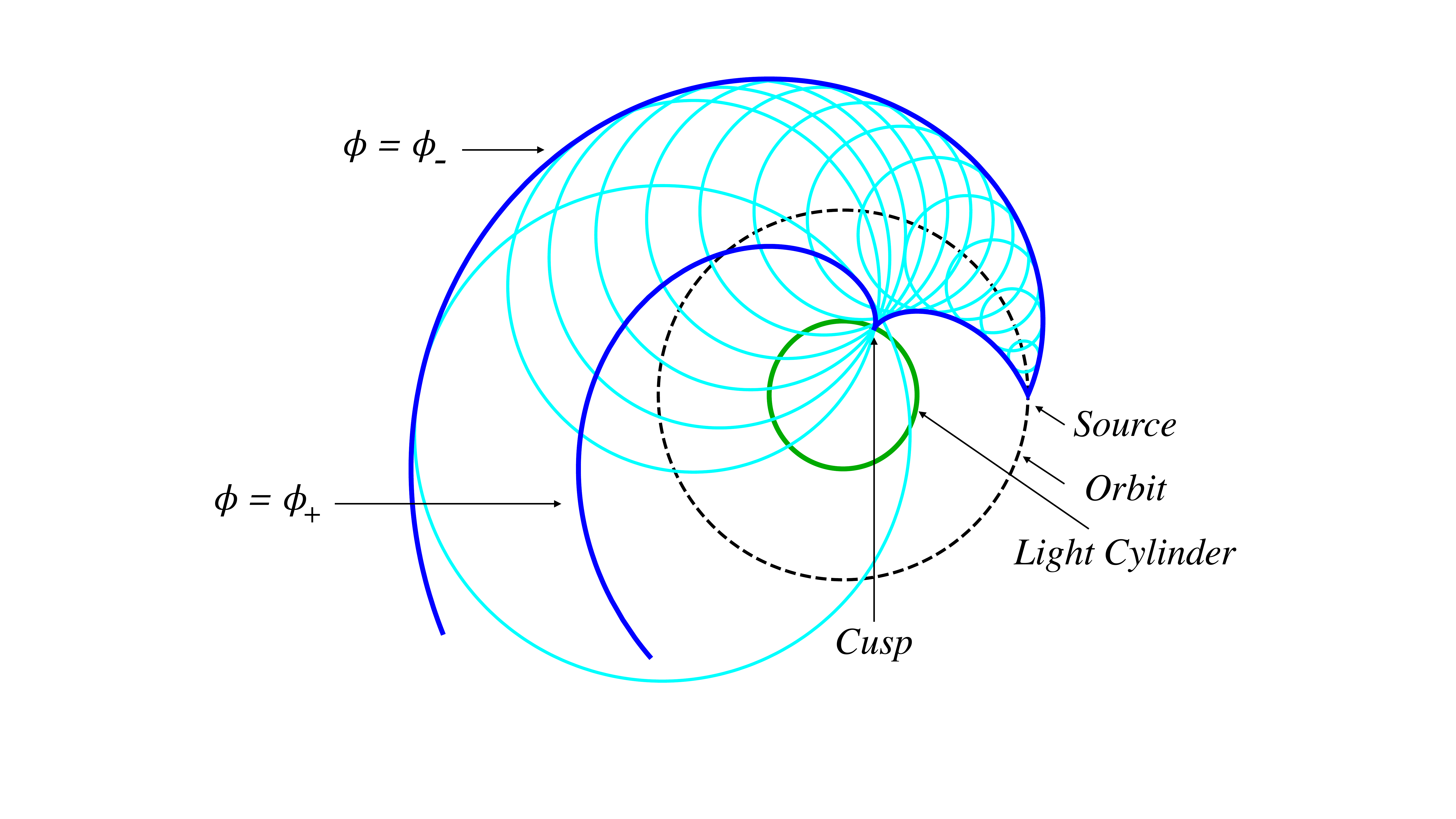}}
\caption{Cross sections with the plane ${\hat z}_P={\hat z}$ of the spherical wave fronts emanating from a rotating source point.  This source has an angular frequency of rotation, $\omega$, that is constant and a speed, $r\omega$, that exceeds the speed of light $c$ in vacuum.  The larger circle depicts the orbit of the source and the smaller circle the light cylinder $r=c/\omega$.  The heavier curves show the intersection of the envelope of these wave fronts (see figure~\ref{PF4}) with the plane of rotation.}
\label{PF3}
\end{figure}

\begin{figure}
\centerline{\includegraphics[width=12cm]{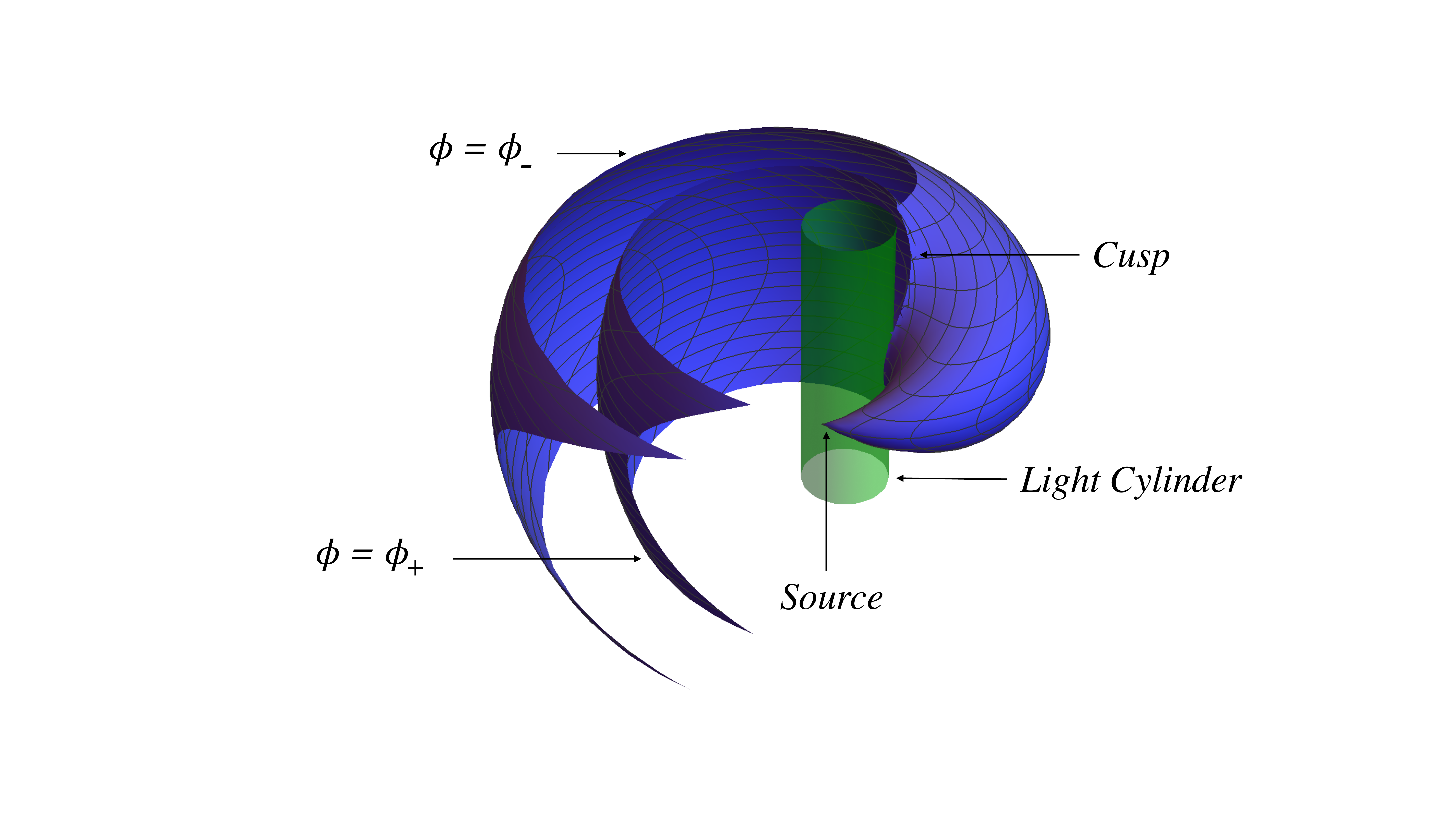}}
\caption{Three-dimensional view (in the space $({\hat r}_P,{\hat\varphi}_P,{\hat z}_P)$ of observation points) of the envelope of wave fronts emanating from the rotating source point $({\hat r},{\hat\varphi},{\hat z}$).  This envelope consists of two sheets that tangentially meet along a cusp (see figure~\ref{PF5}).  The singular sheet, i.e., the sheet that issues from the source point with an initial conical shape, is that described by ${\hat\varphi}_P={\hat\varphi}-\phi_-({\hat r}_P,{\hat z}_P;{\hat r}, {\hat z})$.}
\label{PF4}
\end{figure}

\begin{figure}
\centerline{\includegraphics[width=12cm]{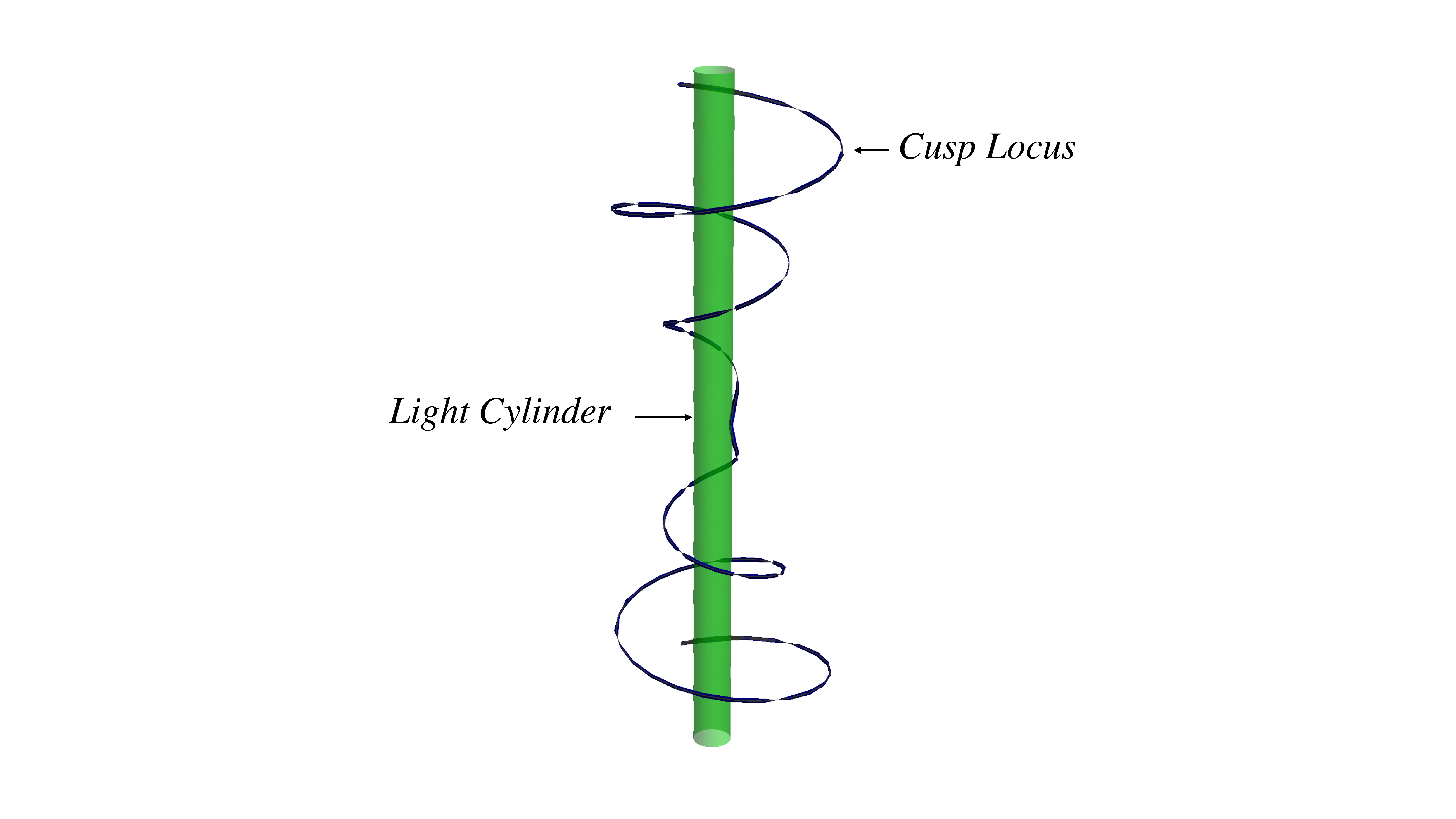}}
\caption{The cusp along which the two sheets of the envelope of wave fronts meet and are tangent to one another.  This cusp touches and is tangent to the light cylinder ${\hat r}_P=1$ on the plane ${\hat z}_P={\hat z}$ and spirals outward into the far field on the hyperboloid generated by the revolution of  the curve $\Delta({\hat r}_P,{\hat z}_P;{\hat r}, {\hat z})=0$.}
\label{PF5}
\end{figure}

The envelope of the wave fronts emanating from a given rotating source element $({\hat r},{\hat\varphi},{\hat z})$, on which $\partial g/\partial\varphi$ vanishes, consists of the rigidly rotating two-sheeted surface ${\hat\varphi}-{\hat\varphi}_P=g(\varphi_\pm)$ in the space $({\hat r}_P,{\hat\varphi}_P,{\hat z}_P)$ of observation points. This surface, which is shown in figures.~\ref{PF3} and \ref{PF4}, is described by
\begin{equation}
\phi_\pm\equiv{\hat\varphi}_\pm-{\hat\varphi}_P=\varphi_\pm-\varphi_P+{\hat R}_\pm
\label{3E23}
\end{equation}
(see~(\ref{3E8}), (\ref{3E9}), (\ref{3E19}) and (\ref{3E22})).  The two sheets of this surface tangentially meet along a cusp on which $\partial^2g/\partial\varphi^2$ as well as $\partial g/\partial\varphi$ vanishes (see figures~\ref{PF4} and \ref{PF5}).  Three distinct wave fronts, emitted at three differing values of the retarded time, pass through any given observation point inside the envelope.  At an observation point located on the envelope or its cusp, respectively two or all three of these waves coalesce and interfere constructively (see figure~\ref{PF3}).

\subsection{Bifurcation surface of an observation point}
\label{subsec:bifurcation}

\begin{figure}
\centerline{\includegraphics[width=12cm]{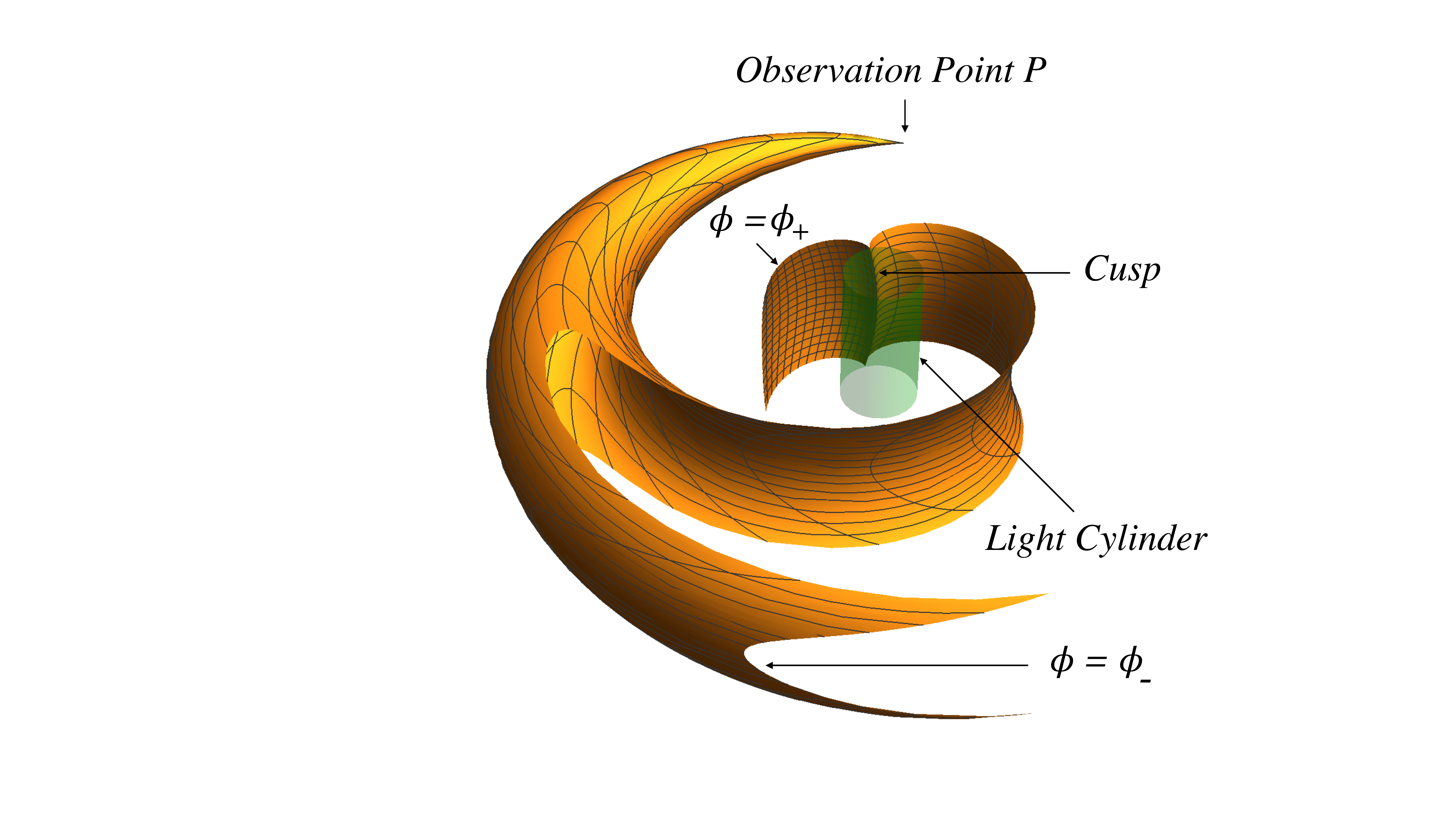}}
\caption{The two sheets $\phi=\phi_\pm$ of the bifurcation surface issuing from the observation point $P$, the cusp $C$ of this surface and the light cylinder ${\hat r}=1$.  In contrast to the envelope of wave fronts which resides in the space of observation points, the surface shown here resides in the space $(r,{\hat\varphi},z)$ of source points: it is the locus of source points that approach $P$, along the radiation direction, with the speed of light at the retarded time.  The two sheets of this surface meet along a cusp that tangentially touches the light cylinder at ${\hat z}={\hat z}_P$ and moves outward spiralling around the rotation axis on the hyperboloid generated by the revolution of the curve $\Delta({\hat r},{\hat z};{\hat r}_P,{\hat z}_P)=0$ (see figure~\ref{PF5}).  The source points on this cusp approach the observer along the radiation direction not only with the speed of light but also with zero acceleration at the retarded time. If the position of the observation point is such that the cusp shown here intersects the current sheet, there will be wave fronts with differing emission times that are received simultaneously: while the source points outside the bifurcation surface make their contributions toward the value of the observed field at a single instant of retarded time, the source points inside this surface make their contributions at $3$ (or $5,7,\cdots$) distinct instants of retarded time.}
\label{PF6}
\end{figure}

Reciprocally, the locus in the space of source points $({\hat r},{\hat\varphi},{\hat z})$ on which $\partial g/\partial\varphi$ vanishes is a two-sheeted cusped surface issuing from the fixed observation point $P$ (see figure~\ref{PF6}).  We refer to this locus, which is described by (\ref{3E23}) for fixed values of $({\hat r}_P,{\hat\varphi}_P,{\hat z}_P)$ rather than fixed values of $({\hat r},{\hat\varphi},{\hat z})$, as the bifurcation surface of the observation point $P$.  The two sheets $\phi=\phi_+$ and $\phi=\phi_-$ of this surface meet along the following cusp:
\begin{eqnarray}
    &C: \left\{
      \begin{array}{c}
     {\hat r}={\hat r}_C({\hat z})=[1+({\hat z}-{\hat z}_P)^2/({{\hat r}_P}^2-1)]^{1/2},\\
     \varphi=\varphi_C({\hat z})=\varphi_P+2m\pi-\arccos[1/({\hat r}{\hat r}_P)],
\end{array} \right.
\label{3E24}
\end{eqnarray}
where $m$ is the same integer as that appearing in (\ref{3E13}).   In this paper we refer to both $C$ and its projection onto the $(r_s,\theta)$ plane as the \emph{cusp locus of the bifurcation surface}; whether it is $C$ itself or its projection that is referred to will be clear from the context.
 
The source points inside the bifurcation surface, close to its cusp, make their contributions toward the observed value of the field at three distinct retarded positions in their trajectory (where a horizontal line $g=\phi$ in figure~\ref{PF2} intersects the curve $\Delta>0$ between its extrema), while those outside the bifurcation surface make their contributions at a single retarded position (where the curve $\Delta<0$ is intersected by $g=\phi$ in figure~\ref{PF2}).  For the source points on the bifurcation surface (i.e., those for which $g=\phi_\pm$ in figure~\ref{PF2}), two of the contributing retarded positions coalesce at the extrema of the curve $\Delta>0$ in figure~\ref{PF2} giving rise to a divergent value of the Green's function at $P$ (see figures 9 and 10 of~\cite{Ardavan_JPP}).  For the source points located on the cusp locus $C$ of the bifurcation surface (i.e., those for which $\Delta=0$ in figure~\ref{PF2}), all three of the contributing retarded positions coalesce at the inflection point of the curve $\Delta=0$ in figure~\ref{PF2} giving rise to a higher-order singularity in $G_j$.  

\subsection{A uniform asymptotic approximation to the value of the Green's function near the cusp locus of the bifurcation surface}
\label{subsec:Expansion}

The time-domain version~\cite{BurridgeR:Asyeir} of the method of Chester et al.~\cite{ChesterC:Extstd} can be employed to derive a uniform asymptotic approximation to the value of $G_j$ for the source points close to the cusp $C$ of the bifurcation surface.  The result, which corresponds to that derived in \S 4.5 of~\cite{Ardavan_JPP} for the case $n=1$, is
\begin{equation}
G_j=\left\{\begin{array}{lll}
G_j^{\rm{in}}  &        \Delta>0, \,\,\vert\chi\vert<1\\
G_j^{\rm{out}}  &      \Delta\geq0, \,\,\vert\chi\vert\geq1\\
G_j^{\rm{sub}}  &       \Delta<0, \,\,\vert\chi^\prime\vert>1
\end{array}
\right.
\label{3E25}
\end{equation}
with
\begin{equation}
G_j^{\rm{in}}\simeq\sum_{m=1}^\infty 2{\mathcal H}_\infty c_1^{-2}(1-\chi^2)^{-1/2}[p_j\cos(\textstyle{\frac{1}{3}}\arcsin\chi)-c_1q_j\sin(\textstyle{\frac{2}{3}}\arcsin\chi)],\quad\vert\chi\vert<1,\quad
\label{3E26}
\end{equation}
\begin{eqnarray}
G_j^{\rm{out}}\simeq\sum_{m=1}^\infty {\mathcal H}_\infty c_1^{-2}(\chi^2-1)^{-1/2}[p_j\sinh(\textstyle{\frac{1}{3}}{\textrm{arccosh}}\vert\chi\vert)+c_1q_j{\textrm{sgn}}(\chi)\sinh(\textstyle{\frac{2}{3}}{\textrm{arccosh}}\vert\chi\vert)],\nonumber\\*
\vert\chi\vert>1,\qquad\qquad
\label{3E27}
\end{eqnarray}
\begin{equation}
G_j^{\rm{sub}}\simeq\sum_{m=1}^\infty {\mathcal H}_\infty c_1^{-2}({\chi^\prime}^2+1)^{-1/2}[p_j\cosh(\textstyle{\frac{1}{3}}{\textrm{arcsinh}}\chi^\prime)+\vert c_1\vert q_j\sinh(\textstyle{\frac{2}{3}}{\textrm{arcsinh}}\chi^\prime)],\quad\vert\chi^\prime\vert>1,
\label{3E28}
\end{equation}
where
\refstepcounter{equation}
$$
\chi=\frac{3(\phi-c_2)}{2{c_1}^3},\qquad
\chi^\prime=\frac{3(\phi-c_2)}{2\vert c_1\vert^3},
\eqno{(\theequation{\mathit{a},\mathit{b}})}
\label{3E29}
$$
\refstepcounter{equation}
$$
c_1=[{\textstyle\frac{3}{4}}(\phi_+-\phi_-)]^{1/3},\qquad c_2={\textstyle\frac{1}{2}}(\phi_++\phi_-),
\eqno{(\theequation{\mathit{a},\mathit{b}})}
\label{3E30}
$$
\refstepcounter{equation}
$$
\phi_\pm={\hat\varphi}_\pm-{\hat\varphi}_P\quad{\rm with}\quad{\hat\varphi}_P=\varphi_P-\omega t_P,
\eqno{(\theequation{\mathit{a},\mathit{b}})}
\label{3E31}
$$
\begin{eqnarray}
\left[\matrix{p_1\cr p_2\cr p_3\cr}\right]&&=\frac{1}{{\hat r}{\hat r}_P}\left(\frac{c_1}{2\Delta^{1/2}}\right)^{1/2}
\left[\matrix{{\hat R}_+^{-{\textstyle\frac{1}{2}}}+{\hat R}_-^{-{\textstyle\frac{1}{2}}}+\Delta^{1/2}({\hat R}_-^{-{\textstyle\frac{1}{2}}}-{\hat R}_+^{-{\textstyle\frac{1}{2}}})\cr -({\hat R}_-^{{\textstyle\frac{1}{2}}}+{\hat R}_+^{{\textstyle\frac{1}{2}}})\cr {\hat r}{\hat r}_P({\hat R}_-^{-{\textstyle\frac{1}{2}}}+{\hat R}_+^{-{\textstyle\frac{1}{2}}})\cr}
\right],\quad
\label{3E32}
\end{eqnarray}
\begin{eqnarray}
\left[\matrix{q_1\cr q_2\cr q_3\cr}\right]=\frac{1}{{\hat r}{\hat r}_P(2c_1\Delta^{1/2})^{1/2}}
\left[\matrix{{\hat R}_-^{-{\textstyle\frac{1}{2}}}-{\hat R}_+^{-{\textstyle\frac{1}{2}}}+\Delta^{1/2}({\hat R}_-^{-{\textstyle\frac{1}{2}}}+{\hat R}_+^{-{\textstyle\frac{1}{2}}})\cr {\hat R}_+^{{\textstyle\frac{1}{2}}}-{\hat R}_-^{{\textstyle\frac{1}{2}}}\cr {\hat r}{\hat r}_P({\hat R}_-^{-{\textstyle\frac{1}{2}}}-{\hat R}_+^{-{\textstyle\frac{1}{2}}})\cr}
\right],
\label{3E33}
\end{eqnarray}
\begin{equation}
{\mathcal H}_\infty={\rm H}[{\hat R}_P-\omega t_P+2m\pi]-{\rm H}[{\hat R}_P-\omega t_P+2(m-1)\pi],
\label{3E34}
\end{equation}
and ${\rm H}(x)$ denotes the Heaviside step function.   

The two-dimensional loci $\chi=\pm1$ across which the above expression for the Green's function $G_j$ changes form correspond to the two sheets $\phi_\pm$ of the bifurcation surface, respectively.  As a source point $(r,{\hat\varphi},z)$ in the vicinity of the cusp $C$ approaches the bifurcation surface from inside, i.e., as $\chi\to1-$ or $\chi\to-1+$, $G_j^{\rm in}$ diverges.  However, as a source point approaches either one of the sheets of the bifurcation surface from outside, the numerator and the denominator in (\ref{3E27}) vanish simultaneously and $G_j^{\rm out}$ tends to a finite limit:
\begin{equation}
G_j^{\rm out}\big\vert_{\phi=\phi_\pm}= G_j^{\rm out}\big\vert_{\chi=\pm1}=\textstyle{\frac{1}{3}}{\cal H}_\infty c_1^{-2}\left(p_j\pm2c_1 q_j\right).
\label{3E35}
\end{equation}
Note that $c_1$, and hence $p_j$ and $q_j$, are independent of $m$ (see (\ref{3E19}), (\ref{3E23}) and (\ref{3E30})).  Thus the Green's function $G_j$ is singular only on the inner side of the bifurcation surface (see figures 9 and 10 of~\cite{Ardavan_JPP}).

\subsection{Hadamard's finite part of the divergent integral representing the field}
\label{subsec:Hadamard}

It follows from (\ref{3E25}) and (\ref{3E26}) that the factor $\partial G_j/\partial{\hat\varphi}$ in the integrand of the integral (\ref{3E12}) diverges as $(1-\chi^2)^{-3/2}$ and so has a non-integrable singularity on the bifurcation surface where $\chi^2$ equals 1.  This singularity has arisen because we differentiated the retarded potential (\ref{3E1}) under the integral sign when calculating the field.  Had we evaluated the integral in (\ref{3E1}) prior to differentiating it we would have found a singularity-free expression.  Interchanging the orders of integration and differentiation is mathematically permissible when the integrand is discontinuous only if one treats the resulting integral as a generalized function and so one handles any non-integrable singularities that consequently arise by means of Hadamard's regularization technique (see~\cite{HadamardJ:lecCau},~\cite{HoskinsRF:GenFun} and the illustrative example in appendix A of~\cite{Ardavan_JPP}). 

Hadamard's procedure consists of performing an integration by parts and discarding the divergent (integrated) term in the resulting expression.  The remaining finite part is the value that Hadamard's regularization assigns to the integral; in the present case, it is the value we would have obtained if we had first evaluated the finite integral representing the retarded potential and had differentiated the result $[\Phi({\textbf x}_P,t_P),{\bf A}({\textbf x}_P,t_P)]$ of that evaluation subsequently.  (The more direct approach, in which the potential is first evaluated and then differentiated, cannot of course be carried out for any realistic source distribution analytically.)

The ${\hat\varphi}$ coordinates ${\hat\varphi}_\pm$ of the two sheets of the bifurcation surface depend on the observation time $t_P$ [see (\ref{3E31}) and (\ref{3E9})], so that these two sheets move across the ${\hat\varphi}$ extent of the source distribution as $t_P$ elapses.  If the position of the observation point is such that the cusp locus of the bifurcation surface intersects the source distribution, the two sheets of this surface (which tangentially meet at the cusp) will divide the volume of the source into a part that lies inside and a part that lies outside the bifurcation surface.  The Lagrangian coordinates ${\hat\varphi}$ designating the initial azimuthal positions of the constituent volume elements of a source that fully occupies an annular region range over the interval $0\le{\hat\varphi}<2\pi$.  The $({\hat r},{\hat z})$ coordinates of these source elements either fall in $\Delta\ge0$ or in $\Delta<0$.  The elements in $\Delta\ge0$ are always divided into two sets: a set inside the bifurcation surface for which ${\hat\varphi}_-\le{\hat\varphi}\le{\hat\varphi}_+$  and so the Green's function $G_j$ has the form $G_j^{\rm in}$ and a set outside for which ${\hat\varphi}$ lies either in $(0,{\hat\varphi}_-)$ or in $({\hat\varphi}_+,2\pi)$ and so $G_j$ has the form $G_j^{\rm out}$ [see (\ref{3E25})].  On the other hand, if the position of the observation point is such that $\Delta<0$ for all values of $({\hat r},{\hat z})$ within the magnetosphere, then the source lies entirely outside the bifurcation surface and $G_j$ has the form $G_j^{\rm sub}$.  Note that, for certain space-time coordinates of the observation point $P$, the values of ${\hat\varphi}_-$ and ${\hat\varphi}_+$ that lie in the interval $(0,2\pi)$ could correspond to different rotation periods, i.e., to different values of $m$ [see (\ref{3E19}), (\ref{3E22}) and (\ref{3E23})].  To simplify the notation, here we adopt an observation time $t_P$ at which the values of ${\hat\varphi}_-$ and ${\hat\varphi}_+$ that lie in the interval $(0,2\pi)$ correspond to the same rotation period $m$. 

Breaking up the volume of integration in the expression for the radiation field ${\mathbf E}$ into the domains of validity of $G_j^{\rm in}$, $G_j^{\rm out}$ and $G_j^{\rm sub}$, we can therefore write the ${\hat\varphi}$-integral over ${\bf u}_j$ in (\ref{3E12}) as
\begin{eqnarray}
{\mathbf I}_{\hat\varphi}&\equiv&\int_0^{2\pi}{\textrm d}{\hat\varphi}\,{\mathbf u}_j\frac{\partial G_j}{\partial{\hat\varphi}}\nonumber\\*
&=&{\rm H}(\Delta)\left[\left(\int_0^{{\hat\varphi}_-}+\int_{{\hat\varphi}_+}^{2\pi}\right){\textrm d}{\hat\varphi}\,{\mathbf u}_j\frac{\partial G_j^{\rm out}}{\partial{\hat\varphi}}+\int_{{\hat\varphi}_-}^{{\hat\varphi}_+}{\textrm d}{\hat\varphi}\,{\mathbf u}_j\frac{\partial G_j^{\rm in}}{\partial{\hat\varphi}}\right]\nonumber\\*
&&+{\rm H}(-\Delta)\int_0^{2\pi}{\textrm d}{\hat\varphi}\,{\mathbf u}_j\frac{\partial G_j^{\rm sub}}{\partial{\hat\varphi}}.
\label{3E36}
\end{eqnarray}
If we now integrate every term of the above expression by parts, recall that ${\hat\varphi}=0$ labels the same source point as does ${\hat\varphi}=2\pi$, and use the fact that the exact version of $G_j$ given in (\ref{3E13}) is periodic in ${\hat\varphi}$ as well as in $\varphi$ (with the same period $2\pi$), we arrive at
\begin{eqnarray}
{\mathbf I}_{\hat\varphi}&=&{\rm H}(\Delta)\Bigg\{\left[{\mathbf u}_j\left(G_j^{\rm in}-G_j^{\rm out}\right)\right]_{{\hat\varphi}={\hat\varphi}_-}^{{\hat\varphi}={\hat\varphi_+}}-\left(\int_0^{{\hat\varphi}_-}+\int_{{\hat\varphi}_+}^{2\pi}\right){\textrm d}{\hat\varphi}\,\frac{\partial{\mathbf u}_j}{\partial{\hat\varphi}}G_j^{\rm out}\nonumber\\*
&&-\int_{{\hat\varphi}_-}^{{\hat\varphi}_+}{\textrm d}{\hat\varphi}\,\frac{\partial{\mathbf u}_j}{\partial{\hat\varphi}}G_j^{\rm in}\Bigg\}-{\rm H}(-\Delta)\int_0^{2\pi}{\textrm d}{\hat\varphi}\,\frac{\partial{\mathbf u}_j}{\partial{\hat\varphi}}G_j^{\rm sub},
\label{3E37}
\end{eqnarray}
an expression that reduces to 
\begin{eqnarray}
{\mathbf I}_{\hat\varphi}&=&{\rm H}(\Delta)\left[{\mathbf u}_{nj}\left(G_j^{\rm in}-G_j^{\rm out}\right)\right]_{{\hat\varphi}={\hat\varphi}_-}^{{\hat\varphi}={\hat\varphi_+}}
-\int_0^{2\pi}{\textrm d}{\hat\varphi}\,\frac{\partial{\mathbf u}_j}{\partial{\hat\varphi}}G_j,\nonumber\\
\label{3E38}
\end{eqnarray}
once the integrals over $G_j^{\rm in}$, $G_j^{\rm out}$ and $G_j^{\rm sub}$ are combined in the light of (\ref{3E25}).  

We have seen in the last paragraph of \S~\ref{subsec:Expansion} that $G_{nj}^{\rm in}$ diverges at ${\hat\varphi}={\hat\varphi}_\pm$.  The Hadamard finite part of ${\mathbf I}_{\hat\varphi}$ is therefore given by the right-hand side of (\ref{3E38}) without the divergent terms involving $G_j^{\rm in}\vert_{{\hat\varphi}={\hat\varphi}_-}$ and $G_j^{\rm in}\vert_{{\hat\varphi}={\hat\varphi}_+}$:  
\begin{equation}
{\rm  Fp}\{{\mathbf I}_{\hat\varphi}\}=-{\rm H}(\Delta){\mathbf u}_jG_j^{\rm out}\big\vert_{{\hat\varphi}={\hat\varphi}_-}^{{\hat\varphi}={\hat\varphi_+}}-\int_0^{2\pi}{\textrm d}{\hat\varphi}\,\frac{\partial{\mathbf u}_j}{\partial{\hat\varphi}}G_j,
\label{3E39}
\end{equation}
where ${\rm Fp}\{{\mathbf I}_{\hat\varphi}\}$ denotes the Hadamard finite part of the divergent integral ${\mathbf I}_{\hat\varphi}$ (see~\cite{HadamardJ:lecCau,HoskinsRF:GenFun}).  

Once the integral with respect to ${\hat\varphi}$ in (\ref{3E12}) is equated to the expression on the right-hand side of (\ref{3E39}), we find that
\begin{equation}
{\mathbf E}={\mathbf E}^{\rm v}+{\mathbf E}_+^{\rm b}-{\mathbf E}_-^{\rm b}
\label{3E40}
\end{equation}
with
\begin{equation}
{\mathbf E}^{\rm v}=\frac{1}{\omega}\sum_{j=1}^3\int\,{\hat r}{\textrm d}{\hat r}\,{\textrm d}{\hat\varphi}\,{\textrm d}{\hat z}\,G_j\frac{\partial{\mathbf u}_j}{\partial{\hat\varphi}},
\label{3E41}
\end{equation}
and
\begin{equation}
{\mathbf E}_\pm^{\rm b}=\frac{1}{\omega}\sum_{j=1}^3\int\,{\hat r}{\textrm d}{\hat r}\,{\textrm d}{\hat z}\,{\rm H}(\Delta)\,G_j^{\rm out}{\mathbf u}_j\Big\vert_{{\hat\varphi}={\hat\varphi}_\pm} .
\label{3E42}
\end{equation}
The term ${\bf E}^{\rm v}$ constitutes the contribution from the entire volume of the source while the terms ${\bf E}_\pm^{\rm b}$ denote the contributions from the discontinuities of the Green's function on the two sheets $\phi=\phi_\pm$ of the bifurcation surface, respectively.  We will see that the terms ${\bf E}_\pm^{\rm b}$ describe an unconventional radiation field with characteristics that turn out to differ from any previously known radiation fields.

\section{Radiation field of the current sheet}
\label{sec:radiation}
\setcounter{equation}{0}
\setcounter{figure}{0}

\subsection{The contribution from discontinuities of the Green's function}
\label{subsec:contribution}

Since the description of the current sheet given in \S~\ref{sec:magnetosphere} is in terms of spherical polar coordinates, it is more convenient, for the purposes of evaluating the contributions ${\bf E}^b_\pm$ to the radiation field ${\bf E}$, to change the integration variables in (\ref{3E42}) from $({\hat r},{\hat z})$ to $({\hat r_s},\theta)$ while continuing to designate the orientations of the current density and the field by means of the cylindrical base vectors $({\hat{\bf e}}_{r_P},{\hat{\bf e}}_{\varphi_P},{\hat{\bf e}}_{z_P})$.  Equations (\ref{2E27})--(\ref{2E30}), (\ref{3E14}), (\ref{3E35}) and (\ref{3E42}) then jointly yield   
\begin{equation}
{\bf E}_\pm^{\rm b}=-\frac{j_0}{3\omega}\sum_{j=1}^3\int_{\frac{\pi}{2}-\alpha}^{\frac{\pi}{2}+\alpha}{\rm d}\theta\int{\rm d}{\hat r}_s\,{\rm H}(\Delta)\frac{\sin\theta(p_j\pm2c_1q_j)}{{\hat r}_s c_1^2}{\tilde{\bf u}}_j\delta({\cal C})\Big\vert_{{\hat\varphi}={\hat\varphi}_\pm},
\label{4E1}
\end{equation}
in which $\Delta$, $c_1$, $p_j$, $q_j$ and ${\hat\varphi}_\pm$ are expressed as functions of $({\hat r}_s,\theta)$ by the insertion of (\ref{2E24}) in (\ref{3E18}), ({\ref{3E19}) ,(\ref{3E22}) and (\ref{3E29})--(\ref{3E34}).

It can be seen from (\ref{2E6}) that the argument of the delta function in (\ref{4E1}) vanishes when
\begin{equation}
\cos({\hat\varphi}+{\hat r}_s-{\hat r}_{s0})=-\cot\alpha\cot\theta,
\label{4E2}
\end{equation}
i.e., when ${\hat\varphi}$ assumes one of the following values
\begin{equation}
{\hat\varphi}_l=(-1)^l\arccos(\cot\alpha\cot\theta)-{\hat r}_s+{\hat r}_{s0}+(2n+1)\pi,
\label{4E3}
\end{equation}
where $l$ is either $1$ or $2$ and $n$ is the integer for which the requirement $0\le{\hat\varphi}_l\le2\pi$ set by (\ref{3E5}) is met.  Hence, an alternative form of this delta function, suitable for first performing the integration with respect to ${\hat r}_s$ in (\ref{4E1}), is
\begin{equation}
\delta({\cal C})=\frac{1}{\sin\alpha\sin\theta(1-\cot^2\alpha\cot^2\theta)^{1/2}}\sum_{l=1}^2\delta({\hat\varphi}-{\hat\varphi}_l),
\label{4E4}
\end{equation}
an expression that holds true for any given values of ${\hat r}_s$ and $\theta$.

Inserting (\ref{3E15})--(\ref{3E17}) and  (\ref{4E4}) in (\ref{4E1}) and using~(\ref{3E15})--(\ref{3E17}) and  (\ref{4E3}) to evaluate ${\tilde{\bf u}}_j$ at ${\hat\varphi}={\hat\varphi}_l$, we obtain
\begin{equation}
{\bf E}_\pm^{\rm b}=\frac{j_0}{3\omega\sin\alpha}\sum_{l=1}^2\sum_{j=1}^3\int_{\frac{\pi}{2}-\alpha}^{\frac{\pi}{2}+\alpha}{\rm d}\theta\int{\rm d}{\hat r}_s\,{\rm H}(\Delta)\frac{{\cal H}_\infty (p_j\pm2c_1q_j){\bf v}_{lj}}{{\hat r}_s c_1^2(1-\cot^2\alpha\cot^2\theta)^{1/2}}\delta({\hat\varphi}_\pm-{\hat\varphi}_l),
\label{4E5}
\end{equation}
where
\begin{equation}
{\bf v}_{l1}=[{\hat r}_s\cos\alpha\sin\theta+(-1)^l\sin\alpha\cos\theta(1-\cot^2\alpha\cot^2\theta)^{1/2}\,]{\hat{\bf e}}_{r_P}+\cos\alpha\csc\theta\,{\hat{\bf e}}_{\varphi_P},
\label{4E6}
\end{equation}
\begin{equation}
{\bf v}_{l2}=-\cos\alpha\csc\theta\,{\hat{\bf e}}_{r_P}+[{\hat r}_s\cos\alpha\sin\theta+(-1)^l\sin\alpha\cos\theta(1-\cot^2\alpha\cot^2\theta)^{1/2}\,]{\hat{\bf e}}_{\varphi_P},
\label{4E7}
\end{equation}
and
\begin{equation}
{\bf v}_{l3}=[{\hat r}_s\cos\alpha\cos\theta-(-1)^l\sin\alpha\sin\theta(1-\cot^2\alpha\cot^2\theta)^{1/2}\,]{\hat{\bf e}}_{z_P}-{\hat r}_s\cos\alpha\,{\hat{\bf n}}_\infty
\label{4E8}
\end{equation}
(see also~(\ref{2E21}), (\ref{2E27})--(\ref{2E30}) and (\ref{3E4})).  The step function ${\cal H}_\infty$ in (\ref{4E5}) ensures that the contribution from the $m$th rotation cycle reaches a far-field observer at $({\hat R}_P, \theta_P,\varphi_P)$ during the interval ${\hat R}_P+2(m-1)\pi\le\omega t_P\le{\hat R}_P+2m\pi$ of observation time (see~(\ref{3E34})).  If the observation time is set at midpoint of this interval, i.e.,
\begin{equation}
\omega t_P={\hat R}_P+(2m-1)\pi,
\label{4E9}
\end{equation}
the argument of the delta function in (\ref{4E5}) assumes the form
\begin{eqnarray}
f_l^\pm&=&{\hat\varphi}_\pm-{\hat\varphi}_l\big\vert_{\omega t_P={\hat R}_P+(2m-1)\pi}\nonumber\\*
&=&{\hat R}_\pm-{\hat R}_P-\arccos\left(\frac{1\mp\Delta^{1/2}}{{\hat r}{\hat r}_P}\right)+{\hat r}_s-{\hat r}_{s0}-(-1)^l\arccos(\cot\alpha\cot\theta)+\varphi_P-2n\pi\nonumber\\*
\label{4E10}
\end{eqnarray}
(see~(\ref{3E9}), (\ref{3E19}), (\ref{3E23}) and (\ref{4E3})).   Note that there is no loss of generality in fixing the observation time: because $t_P$ only appears in the combination $\varphi_P-\omega t_P$, the temporal variation of the radiation is equally well described by its dependence on $\varphi_P$.

Replacing the delta function in (\ref{4E5}) by its Fourier representation
\begin{equation}
\delta\left({\hat\varphi}_\pm-{\hat\varphi}_l\big\vert_{\omega t_P={\hat R}_P+(2m-1)\pi}\right)=\delta(f_l^\pm)
=\frac{1}{2\pi}\int_{-\infty}^\infty {\rm d}k\exp({\rm i}k f_l^\pm),
\label{4E11}
\end{equation}
we therefore obtain 
\begin{eqnarray}
{\bf E}_\pm^{\rm b}&=&\frac{j_0}{6\pi\omega\sin\alpha}\sum_{l=1}^2\sum_{j=1}^3\int_{\frac{\pi}{2}-\alpha}^{\frac{\pi}{2}+\alpha}{\rm d}\theta\int_{-\infty}^\infty{\rm d}k\nonumber\\*
&&\times\int{\rm d}{\hat r}_s\,{\rm H}(\Delta)\frac{(p_j\pm2c_1q_j){\bf v}_{lj}}{{\hat r}_s c_1^2(1-\cot^2\alpha\cot^2\theta)^{1/2}}\exp({\rm i}k f_l^\pm),
\label{4E12}
\end{eqnarray}
in which we have interchanged the orders of integration with respect to ${\hat r}_s$ and $k$.  Note that the range of integration over ${\hat r}_s$ is set by $\Delta\ge0$. 

A pair of integration variables more suitable than $({\hat r}_s,\theta)$ for evaluating (\ref{4E12}) are $\eta$ and $\tau$ defined by
\begin{equation}
\eta=\Delta^{1/2}=\left[{\hat r}_s^2({\hat r}_P^2\sin^2\theta-1)+2{\hat r}_s{\hat z}_P\cos\theta-{\hat R}_P^2+1\right]^{1/2}
\label{4E13}
\end{equation}
(see~(\ref{3E18})) and
\begin{equation}
\tau=\arccos(\csc\alpha\cos\theta)
\label{4E14}
\end{equation}
since the Jacobian 
\begin{equation}
\frac{\partial({\hat r}_s,\theta)}{\partial(\eta,\tau)}=\frac{\eta\sin\alpha(1-\cot^2\alpha\cot^2\theta)^{1/2}}{[({\hat r}_P^2-1)({\hat R}_P^2\sin^2\theta-1)+({\hat r}_P^2\sin^2\theta-1)\eta^2]^{1/2}}
\label{4E15}
\end{equation}
of this transformation removes the singularities of the integrand that occur on the boundaries ($\theta=\pi/2\pm\alpha$ and $\Delta=0$ where $c_1=0$) of the integration domain.  If we now express ${\hat r}_s$ and $\theta$ everywhere in terms of $\eta$ and $\tau$ by inverting (\ref{4E13}) and (\ref{4E14}) we obtain
\begin{equation}
\theta=\arccos(\sin\alpha\cos\tau),
\label{4E16}
\end{equation}
\begin{equation}
{\hat r}_s=\frac{[({\hat r}_P^2-1)({\hat R}_P^2\sin^2\theta-1)+({\hat r}_P^2\sin^2\theta-1)\eta^2]^{1/2}-{\hat z}_P\cos\theta}{{\hat r}_P^2\sin^2\theta-1},
\label{4E17}
\end{equation}
and so can rewrite (\ref{4E12}) as
\begin{eqnarray}
{\bf E}_\pm^{\rm b}&=&\frac{j_0}{6\pi\omega}\sum_{l=1}^2\sum_{j=1}^3\int_0^{\pi}{\rm d}\tau\int_{-\infty}^\infty{\rm d}k\nonumber\\*
&&\times\int_0^\infty{\rm d}\eta\,\frac{\eta(p_j\pm2c_1q_j){\bf v}_{lj}}{{\hat r}_s c_1^2[({\hat r}_P^2-1)({\hat R}_P^2\sin^2\theta-1)+({\hat r}_P^2\sin^2\theta-1)\eta^2]^{1/2}}\exp({\rm i}k f_l^\pm),\qquad
\label{4E18}
\end{eqnarray}
where ${\hat r}_s$ and $\theta$ henceforth stand for the functions of $(\eta,\tau)$ given by (\ref{4E16}) and (\ref{4E17}).  We shall see below that the limiting value of the ratio $\eta/c_1$ at $\eta=0$, where $c_1$ vanishes, is finite.

The task of the rest of this section is to evaluate the right-hand side of (\ref{4E18}) by treating it as a repeated integral.

\subsection{Evaluation of the asymptotic values of the integrals over $\eta$ by the method of stationary phase}
\label{subsec:eta}

It follows from (\ref{4E10}) in conjunction with (\ref{4E16}) and (\ref{4E17}) that $\partial f_l^\pm/\partial\eta$ vanishes at $\eta=0$, so that the main contribution towards the asymptotic value for large $\vert k\vert$ of the integral over $\eta$ in (\ref{4E18}) comes from the vicinity of the cusp locus $C$ of the bifurcation surface at which $\eta$ is zero (see, e.g.,~\cite{BleisteinN:Asei}).  We can therefore approximate the functions $f_l^\pm$ in the phase of the integrand of (\ref{4E18}) by the following leading terms in their Taylor expansions in powers of $\eta$:
\begin{equation}
f_l^\pm\simeq f_{lC}+\textstyle{\frac{1}{2}}a\,\xi^2\pm\textstyle{\frac{1}{3}}\xi^3,
\label{4E19}
\end{equation}   
where
\begin{equation}
\xi=\frac{\eta}{({\hat r}_P^2{\hat r}_{sC}^2\sin^2\theta-1)^{1/2}},
\label{4E20}
\end{equation}
\begin{eqnarray}
f_{lC}&=&({\hat r}_P^2{\hat r}_{sC}^2\sin^2\theta-1)^{1/2}-{\hat R}_P+{\hat r}_{sC}-\arccos\left(\frac{1}{{\hat r}_P{\hat r}_{sC}\sin\theta}\right)\nonumber\\*
&&-(-1)^l\arccos(\cot\alpha\cot\theta)+\varphi_P-{\hat r}_{s0}-2n\pi,
\label{4E21}
\end{eqnarray}
\begin{equation}
a=({\hat r}_P^2{\hat r}_{sC}^2\sin^2\theta-1)\left[\frac{({\hat r}_P^2{\hat r}_{sC}^2\sin^2\theta-1)^{1/2}+{\hat r}_{sC}}{{\hat r}_{sC}({\hat r}_P^2-1)^{1/2}({\hat R}_P^2\sin^2\theta-1)^{1/2}}-\frac{1}{({\hat r}_P^2{\hat r}_{sC}^2\sin^2\theta-1)^{1/2}}\right],
\label{4E22}
\end{equation}
and
\begin{equation}
{\hat r}_{sC}=\frac{({\hat r}_P^2-1)^{1/2}({\hat R}_P^2\sin^2\theta-1)^{1/2}-{\hat z}_P\cos\theta}{{\hat r}_P^2\sin^2\theta-1}
\label{4E23}
\end{equation}
denotes the value of ${\hat r}_s$ at the cusp locus $C$ (see~(\ref{4E17})).  Note that the third-order term has to be included in the above expansion to take account of the difference between $f_l^+$ and $f_l^-$.  Note, moreover, that according to (\ref{3E30}$a$) and (\ref{4E10}), $c_1$ can be written as $[\textstyle{\frac{3}{4}}(f_l^+-f_l^-)]^{1/3}$ which implies, in conjunction with (\ref{4E19}), that $c_1\simeq2^{-1/3}\xi$ near $\eta=0$.

If we now apply the method of stationary phase to the $\eta$-integral, i.e., insert (\ref{4E19}) in (\ref{4E18}) and replace the amplitude of the exponential in the integrand of (\ref{4E18}) by its value at the stationary point $\eta=0$ (see, e.g.,~\cite{BleisteinN:Asei}), we obtain the following expression for the leading term in the asymptotic expansion of this integral for large $\vert k\vert$: 
\begin{equation}
{\bf E}_\pm^{\rm b}\simeq\frac{j_0}{3\pi\omega}\sum_{l=1}^2\int_0^{\pi}{\rm d}\tau\int_{-\infty}^\infty{\rm d}k\,b\exp({\rm i}k f_{lC})
\int_0^\infty{\rm d}\xi\,(\xi^{-1}{\bf P}_l\pm{\bf Q}_l)\exp[{\rm i}k(\textstyle{\frac{1}{2}}a\,\xi^2\pm\textstyle{\frac{1}{3}}\xi^3)],
\label{4E24}
\end{equation}
where
\begin{equation}
b=\frac{{\hat r}_P^2{\hat r}_{sC}^2\sin^2\theta-1}{{\hat r}_{sC}({\hat r}_P^2-1)^{1/2}({\hat R}_P^2\sin^2\theta-1)^{1/2}},
\label{4E25}
\end{equation}
\begin{equation}
{\bf P}_l=2^{-1/3}\sum_{j=1}^3p_j{\bf v}_{lj}\Big\vert_{\eta=0},
\label{4E26}
\end{equation}
\begin{equation}
{\bf Q}_l=2^{1/3}\sum_{j=1}^3q_j{\bf v}_{lj}\Big\vert_{\eta=0},
\label{4E27}
\end{equation}
and use has been made of the fact that $\eta/c_1=2^{1/3}({\hat r}_P^2{\hat r}_{sC}^2\sin^2\theta-1)^{1/2}$in the limit $\eta\to0$.  (Note that $b$ tends to $1$ as ${\hat R}_P$ tends to infinity.)  The above asymptotic approximation for large $\vert k\vert$ is justified since the frequency of the radiation we are interested in has a much higher value than the rotation frequency $\omega/2\pi$ of the central neutron star. 

Hence, the unconventional contribution ${\bf E}_+^b-{\bf E}_-^b$ to the radiation field ${\bf E}$ that was encountered in (\ref{3E40}) can be written as
\begin{equation}
{\bf E}^{\rm uc}={\bf E}_+^b-{\bf E}_-^b\simeq \frac{j_0}{3\pi\omega}\sum_{l=1}^2\int_0^{\pi}{\rm d}\tau\int_{-\infty}^\infty{\rm d}k\,b\exp({\rm i}k f_{lC})(I_P{\bf P}_l+I_Q{\bf Q}_l),
\label{4E28}
\end{equation}
once the integration with respect to $\xi$ has been carried out analytically to obtain 
\begin{eqnarray}
I_Q&=&2\int_0^\infty{\rm d}\xi\,\exp(\textstyle{\frac{1}{2}}{\rm i}k a\xi^2)\cos\left(\textstyle{\frac{1}{3}}k\xi^3\right)\nonumber\\*
&=&\frac{2\pi}{\vert k\vert^{1/3}}\exp\left(\textstyle{\frac{1}{12}}{\rm i}k a^3\right){\rm Ai}\left(-\textstyle{\frac{1}{4}}k^{2/3}a^2\right),
\label{4E29}
\end{eqnarray}
in terms of the Airy function ${\rm Ai}$ and
\begin{eqnarray}
I_P&=&2{\rm i}\int_0^\infty{\rm d}\xi\,\xi^{-1}\exp(\textstyle{\frac{1}{2}}{\rm i}k a\xi^2)\sin\left(\textstyle{\frac{1}{3}}k\xi^3\right)\nonumber\\*
&=&{\rm sgn}(k)\bigg[\frac{{\rm i}\pi}{3}-\frac{3^{1/6}}{2}\Gamma\left(\textstyle{\frac{2}{3}}\right)a\,k^{1/3}\, {}_3F_4
\left(\matrix{1/6&5/12&11/12&{}\cr
1/3&1/2&5/6&7/6\cr}
;-\frac{k^2a^6}{144}\right)\nonumber\\*
&&+\frac{{\rm i}}{2^2\times3^{7/6}}\Gamma\left(\textstyle{-\frac{2}{3}}\right)a^2k^{2/3}\, {}_3F_4
\left(\matrix{1/3&7/12&13/12&{}\cr
1/2&2/3&7/6,&4/3\cr}
;-\frac{k^2a^6}{144}\right)\nonumber\\*
&&+\frac{5\pi{\rm i}}{2^6\times3^{1/3}\Gamma\left(-2/3\right)}a^4k^{4/3}\, {}_3F_4
\left(\matrix{2/3&11/12&17/12&{}\cr
5/6&4/3&3/2&5/3\cr}
;-\frac{k^2a^6}{144}\right)\nonumber\\*
&&-\frac{7\Gamma\left(-2/3\right)}{2^5\times3^{13/6}\times5}a^5k^{5/3}\, {}_3F_4
\left(\matrix{5/6&13/12&19/12&{}\cr
7/6&3/2&5/3&11/6\cr}
;-\frac{k^2a^6}{144}\right)\bigg]
\label{4E30}
\end{eqnarray}
in terms of the generalized hypergeometric function ${}_3F_4$ (see, e.g.,~\cite{Olver}).  Mathematica has been used to perform the above integrations by first employing a change of integration variable to cast $I_Q$ and $I_P$ in the forms of sine and cosine Fourier transforms and using the relations in \S5.5 of~\cite{Olver} to simplify the gamma functions in the resulting expressions. 

\subsection{Dominance of the contribution from large values of $\vert k\vert$}
\label{subsec:k}

In this section we assess the expectation that the main contribution toward the value of the radiation field should come from large values of $\vert k\vert$.  We will do this by evaluating the $k$-integral in (\ref{4E28}) once with the exact value of its integrand and another time with the asymptotic value of its integrand for large $\vert k\vert$ and comparing the outcomes of these two evaluations.

The functions ${\bf P}_l$, ${\bf Q}_l$ and $b$ in the integrand of (\ref{4E28}) are independent of $k$.  Once the Airy function that appears in the expression for $I_Q$ is expressed in terms of Bessel functions, the $k$-integral multiplying ${\bf Q}_l$ assumes the form of a tabulated Fourier transform (cf.,~\cite{Erdelyi}) and so can be evaluated exactly to obtain
\begin{equation}
\int_{-\infty}^\infty{\rm d}k\,\exp\left({\rm i}k f_{lC}\right)I_Q=\frac{8\pi}{a^2}{\cal F}_1^{\rm e}\left(\frac{12}{a^3}f_{lC}+1\right),
\label{4E31}
\end{equation}
where 
\begin{eqnarray}
{\cal F}_1^{\rm e}(x)&=&\frac{{\rm H}(\vert x\vert-1)}{2(x^2-1)^{1/2}}\left[\left(\vert x\vert+\sqrt{x^2-1}\right)^{1/3}-\left(\vert x\vert+\sqrt{x^2-1}\right)^{-1/3}\right]\nonumber\\*
&&+\frac{2{\rm H}(1-\vert x\vert)}{(1-x^2)^{1/2}}\cos\left(\frac{1}{3}\arcsin\vert x\vert\right).
\label{4E32}
\end{eqnarray}
The $k$-integral multiplying ${\bf P}_l$ in (\ref{4E28}) also consists of the sum of five Fourier transforms each of which can be evaluated explicitly by means of Mathematica.  The result is
\begin{equation}
\int_{-\infty}^\infty{\rm d}k\,\exp\left({\rm i}k f_{lC}\right)I_P=\frac{12}{a^3}{\cal F}_2^{\rm e}\left(\frac{12}{a^3}f_{lC}\right),
\label{4E33}
\end{equation}
where   
\begin{eqnarray}
{\cal F}_2^{\rm e}(x)&=&-\frac{2\pi}{3x}+\frac{2^{2/3}\pi}{3x^{4/3}}{\rm H}(\vert x\vert-2)\bigg[{}_4F_3\left(\matrix{1/6&5/12&2/3&11/12\cr
1/3&1/2&5/6&{}\cr}
;\frac{4}{x^2}\right)\nonumber\\*
&&+\frac{1}{(2x)^{1/3}}{}_4F_3\left(\matrix{1/3&7/12&5/6&13/12\cr
1/2&2/3&7/6&{}\cr}
;\frac{4}{x^2}\right)\nonumber\\*
&&-\frac{5}{6x}{}_4F_3\left(\matrix{2/3&11/12&7/6&17/12\cr
5/6&4/3&3/2&{}\cr}
;\frac{4}{x^2}\right)\nonumber\\*
&&-\frac{7}{3(2x)^{4/3}}{}_4F_3\left(\matrix{5/6&13/12&4/3&19/12\cr
7/6&3/2&5/3&{}\cr}
;\frac{4}{x^2}\right)\bigg]\nonumber\\*
&&-\frac{\pi}{3}{\rm H}(2-\vert x\vert)\bigg[\frac{1-{\rm sgn}(x)}{\sqrt{6\vert x\vert}}{}_4F_3\left(\matrix{5/12&7/12&11/12&13/12\cr
1/2&3/4&5/4&{}\cr}
;\frac{x^2}{4}\right)\nonumber\\*
&&+\frac{35}{108}\sqrt{\frac{\vert x\vert}{6}}[1-{\rm sgn}(x)]{}_4F_3\left(\matrix{11/12&13/12&17/12&19/12\cr
5/4&3/2&7/4&{}\cr}
;\frac{x^2}{4}\right)\nonumber\\*
&&-\frac{2}{x}\,{}_4F_3\left(\matrix{1/6&1/3&2/3&5/6\cr
1/4&1/2&3/4&{}\cr}
;\frac{x^2}{4}\right)+\frac{4}{9}{}_4F_3\left(\matrix{2/3&5/6&7/6&4/3\cr
3/4&5/4&3/2&{}\cr}
;\frac{x^2}{4}\right)\bigg].\nonumber\\*
\label{4E34}
\end{eqnarray}
Note that repeated use has to be made of the relations in \S 5.5 of~\cite{Olver} to cast the result in the above simplified form.

The function ${\cal F}_1^{\rm e}$ diverges as $(1-x^2)^{-1/2}$ at $\vert x\vert=1-$.  Moreover, the generalized hypergeometric functions appearing in (\ref{4E34}) are singular in the limit where their arguments approach unity.  While the singularities of the individual terms of (\ref{4E34}) at $x=2$, $x=-2-$ and $x=0$ cancel out when added together, the function ${\cal F}_2^{\rm e}$ diverges as $(x+2)^{-1/2}$ at $x=-2+$ (see figure~\ref{PF7}). 

\begin{figure}
\centerline{\includegraphics[width=13cm]{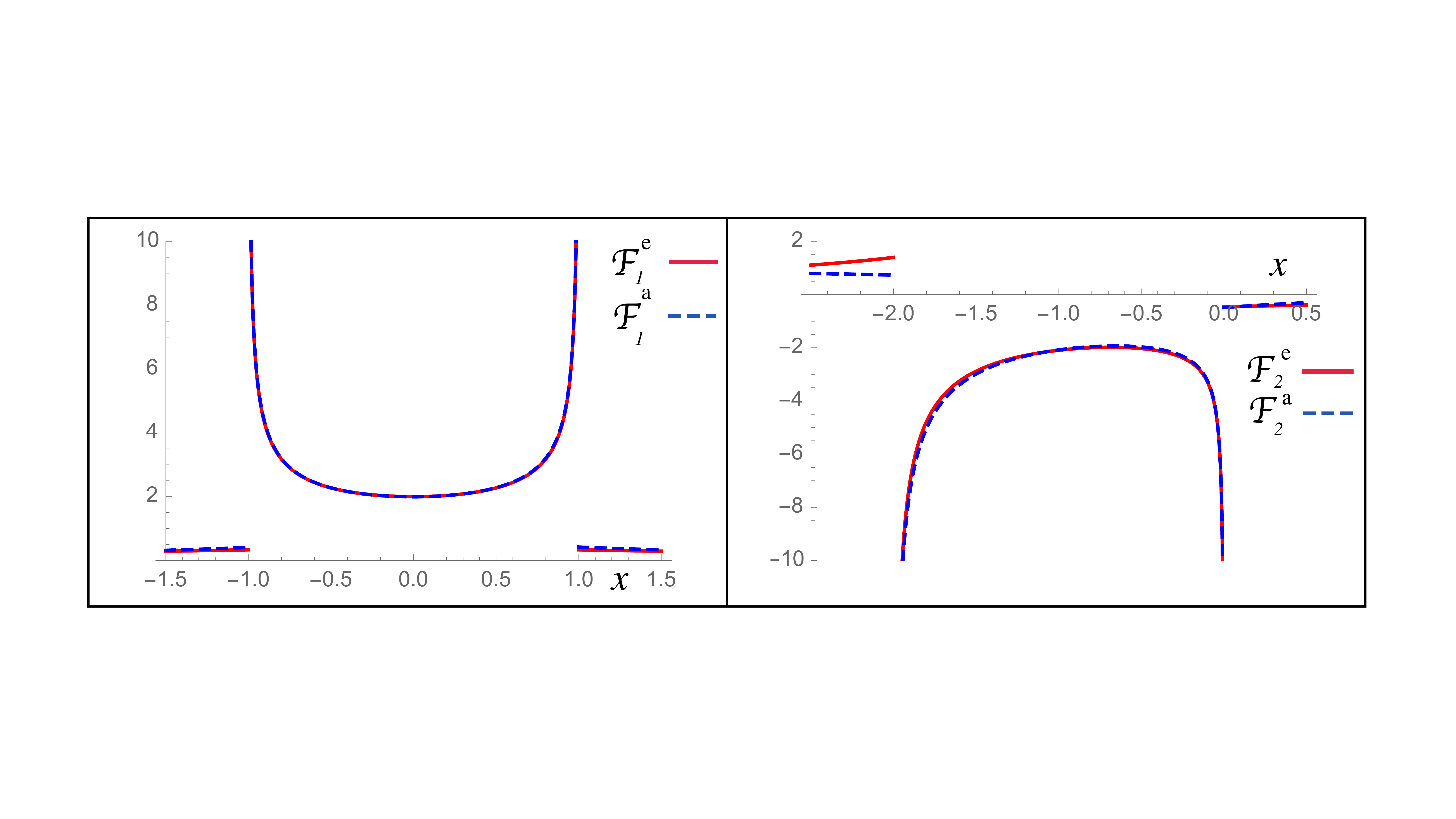}}
\caption{The solid (red) lines in this figure depict the exact values (${\cal F}_1^{\rm e}$ and ${\cal F}_2^{\rm e}$) of the functions derived in (\ref{4E32}) and (\ref{4E34}).  The dashed (blue) lines depict the approximate versions (${\cal F}_1^{\rm a}$ and ${\cal F}_2^{\rm a}$) of these functions given by (\ref{4E41}) and (\ref{4E42}) for $k_1=3/(5a^3)$ and $k_2=12/(5a^3)$.}
\label{PF7}
\end{figure}

The factor $a$ that multiplies $k$ in the arguments of the Airy function in (\ref{4E29}) and the generalized hypergeometric functions in (\ref{4E30}) is non-zero and positive everywhere and approaches the finite value
\begin{equation}
a\simeq\csc\theta_P\csc\theta(1-\cos\theta_P\cos\theta)
\label{4E35}
\end{equation}
as ${\hat R}_P$ tends to infinity (see(\ref{4E22})).  To calculate the contribution of large $\vert k\vert$ toward the values of the integrals in (\ref{4E31}) and (\ref{4E33}) we can therefore replace the functions $I_P$ and $I_Q$ that appear in the integrands of these integrals by the following leading terms 
\begin{equation}
I_P^\infty=\lim_{k\to\pm\infty}I_P=\left(\frac{2\pi}{a^3\vert k\vert}\right)^{1/2}\exp\left(\pm{\textstyle\frac{3}{4}}{\rm i}\pi\right)\left[\exp\left({\textstyle\frac{1}{6}}{\rm i}a^3k\right)\pm{\textstyle\frac{1}{3}}{\rm i}\right],
\label{4E36}
\end{equation} 
\begin{equation}
I_Q^\infty=\lim_{k\to\pm\infty}I_Q=\left(\frac{2\pi}{a \vert k\vert}\right)^{1/2}\exp\left(\mp{\textstyle\frac{1}{4}}{\rm i}\pi\right)\left[\exp\left({\textstyle\frac{1}{6}}{\rm i}a^3k\right)\pm{\rm i}\right],
\label{4E37}
\end{equation}
in their asymptotic expansions for large $\vert k\vert$ before performing the integrations.  (Note that (\ref{4E37}) is found more easily by a direct asymptotic evaluation of the integral over $\xi$ in the first member of (\ref{4E30}) than by the calculation of the limiting value of the second member of this equation.)  To calculate the corresponding contribution from small values of $\vert k\vert$ we should replace $I_P$ and $I_Q$ by the following leading terms 
\begin{equation}
I_P^0=\lim_{\vert k\vert\to0}I_P=\frac{{\rm i}\pi}{3}{\rm sgn}(k),\qquad I_Q^0=\lim_{\vert k\vert\to0}I_Q=\frac{2\pi}{3^{2/3}\Gamma(2/3)\vert k\vert^{1/3}},
\label{4E38}
\end{equation}
in their expansions about $k=0$ (see~\cite{Olver}). 

It turns out that the functions ${\cal F}_1^{\rm e}$ and ${\cal F}_2^{\rm e}$, defined in (\ref{4E31})--(\ref{4E34}), can be accurately approximated by
\begin{equation}
{\cal F}_1^{\rm a}\left(\frac{12}{a^3}f_{lC}+1\right)=\frac{a^2}{4\pi}\Re\int_{k_1}^\infty{\rm d}k\,\exp\left({\rm i}k f_{lC}\right)I^\infty_Q
\label{4E39}
\end{equation}  
and
\begin{equation}
{\cal F}_2^{\rm a}\left(\frac{12}{a^3}f_{lC}\right)=\frac{a^3}{6\pi}\Re\left\{\int_0^{k_2}{\rm d}k\,\exp\left({\rm i}k f_{lC}\right)I^0_P+\int_{k_2}^\infty{\rm d}k\,\exp\left({\rm i}k f_{lC}\right)I^\infty_P\right\},
\label{4E40}
\end{equation}  
respectively, with appropriate values of the free parameters $k_1$ and $k_2$.  We have here replaced the integrals over $0\le \vert k\vert\le k_{1,2}$ and $k_{1,2}\le \vert k\vert<\infty$ by twice the real parts of the integrals over $0\le k\le k_{1,2}$ and $k_{1,2}\le k<\infty$ because the contributions from $k<0$ toward the values of the $k$-integrals in (\ref{4E31}) and (\ref{4E33}) equal the complex conjugates of the contributions from $k>0$.  Evaluation of the integrals in (\ref{4E39}) and (\ref{4E40}) results in
\begin{eqnarray}
{\cal F}_1^{\rm a}(x)&=&\sqrt{\frac{3}{2}}\left\{\frac{1}{\vert x+1\vert^{1/2}}\left[\frac{1}{2}-C\left(\sqrt{\frac{a^3k_1}{6\pi}\vert x+1\vert}\right)\right]
\right.\nonumber\\*
&&\left.+\frac{{\rm sgn}(x+1)}{\vert x+1\vert^{1/2}}\left[\frac{1}{2}-S\left(\sqrt{\frac{a^3k_1}{6\pi}\vert x+1\vert}\right)\right]\right.\nonumber\\*
&&\times\left.\frac{1}{\vert x-1\vert^{1/2}}\left[\frac{1}{2}-C\left(\sqrt{\frac{a^3k_1}{6\pi}\vert x-1\vert}\right)\right]
\right.\nonumber\\*
&&\left.-\frac{{\rm sgn}(x-1)}{\vert x+1\vert^{1/2}}\left[\frac{1}{2}-S\left(\sqrt{\frac{a^3k_1}{6\pi}\vert x-1\vert}\right)\right]\right\},
\label{4E41}
\end{eqnarray}
and
\begin{eqnarray}
{\cal F}_2^{\rm a}(x)&=&-\frac{4\pi}{3x}\sin^2\left(\frac{a^3k_2}{24}x\right)-\sqrt{\frac{2}{3}}\pi\left\{\frac{1}{\vert x+2\vert^{1/2}}\left[\frac{1}{2}-C\left(\sqrt{\frac{a^3k_2}{6\pi}\vert x+2\vert}\right)\right]
\right.\nonumber\\*
&&\left.+\frac{{\rm sgn}(x+2)}{\vert x+2\vert^{1/2}}\left[\frac{1}{2}-S\left(\sqrt{\frac{a^3k_2}{6\pi}\vert x+2\vert}\right)\right]+\frac{1}{3\vert x\vert^{1/2}}\left[\frac{1}{2}-C\left(\sqrt{\frac{a^3k_2}{6\pi}\vert x\vert}\right)\right]
\right.\nonumber\\*
&&\left.-\frac{{\rm sgn}(x)}{3\vert x\vert^{1/2}}\left[\frac{1}{2}-S\left(\sqrt{\frac{a^3k_2}{6\pi}\vert x\vert}\right)\right]\right\}
\label{4E42}
\end{eqnarray}
where $C$ and $S$ are the Fresnel cosine and sine integrals, respectively.

The exact (${\cal F}_1^{\rm e},{\cal F}_2^{\rm e}$) and approximate (${\cal F}_1^{\rm a},{\cal F}_2^{\rm a}$) versions of the functions ${\cal F}_1$ and ${\cal F}_2$ are compared in figure~\ref{PF7} for $k_1={\textstyle\frac{3}{5}}a^{-3}$ and $k_2={\textstyle\frac{12}{5}}a^{-3}$.  The goodness of the fit in this figure and the shortness of the intervals $0\le k\le k_{1,2}$ relative to $k_{1,2}\le k<\infty$ show that (i) the difference in values of the exact and approximate versions of these two functions is negligibly small in the intervals over which ${\cal F}_1$ and ${\cal F}_2$ make their main contributions toward the value of the integrand of the $\tau$-integral in (\ref{4E28}) and (ii) the contributions of large $\vert k\vert$ toward the values of ${\cal F}_1^{\rm e}$ and ${\cal F}_2^{\rm e}$ is by far greater than those of small $\vert k\vert$.  In order that we can derive an analytic expression for the radiation field ${\bf E}^{\rm uc}$, we will accordingly base the analysis in the rest of this section on the following approximate values of $I_P$ and $I_Q$
\begin{equation}
I^{\rm a}_P= {\rm H}(k_2-\vert k\vert)I_P^0+{\rm H}(\vert k\vert-k_2)I_Q^\infty
\label{4E43}
\end{equation}
and
\begin{equation}
I^{\rm a}_Q={\rm H}(\vert k\vert-k_1) I_Q^\infty
\label{4E44}
\end{equation}
that were derived in (\ref{4E36})--(\ref{4E38}) and let $k_1$ and $k_2$ have the values that were used in plotting figure~\ref{PF7}. 

Replacing $I_P$ and $I_Q$ in (\ref{4E28}) with their approximate values $I^{\rm a}_P$ and $I^{\rm a}_Q$ and interchanging the orders of integration with respect to $k$ and $\tau$, we write (\ref{4E28}) as
\begin{eqnarray}
{\bf E}^{\rm uc}&\simeq&\frac{2j_0}{3\omega}\sum_{l=1}^2\int_0^\infty{\rm d}k\int_0^{\pi}{\rm d}\tau\,b\,\bigg\{\bigg[\frac{{\rm i}}{3}{\rm H}(k_2-k){\bf P}_l+\sqrt{\frac{2}{\pi a k}}\exp\left(\frac{{\rm i}\pi}{4}\right)\nonumber\\*
&&\times\left({\rm H}(k-k_1){\bf Q}_l-\frac{1}{3a}{\rm H}(k-k_2){\bf P}_l\right)\bigg]\exp\left({\rm i}k f_{lC}\right)\nonumber\\*
&&+\sqrt{\frac{2}{\pi a k}}\exp\left(-\frac{{\rm i}\pi}{4}\right)\left({\rm H}(k-k_1){\bf Q}_l-\frac{1}{a}{\rm H}(k-k_2){\bf P}_l\right)\exp\left({\rm i}k{\bar f}_{lC}\right)\bigg\},
\label{4E45}
\end{eqnarray}
where
\begin{equation}
{\bar f}_{lC}=f_{lC}+{\textstyle\frac{1}{6}}a^3,
\label{4E46}
\end{equation}
and proceed to evaluate the integral over $\tau$ first.  That the unconventional radiation field ${\bf E}^{\rm uc}$ is given by the real part of the above expression (in which the integrals over $-\infty<k<\infty$ are written as twice the integrals over $0\le k<\infty$) is understood.  

\subsection{Critical points of the phase functions $f_{lC}$ and ${\bar f}_{lC}$}
\label{subsec:critical}

The first two derivatives with respect to $\tau$ of the functions $f_{lC}$ and ${\bar f}_{lC}$--defined by (\ref{4E21}) and (\ref{4E43})--which appear in (\ref{4E42}) are given in appendix~\ref{AppA}. 

It follows from (\ref{A1})--(\ref{A5}) that the nature of the critical points of $f_{lC}$ is determined by the value of the coordinate $\theta_P$ of the observation point: this function can have two turning points (a maximum and a minimum), can have a single inflection point, or can be monotonic.  As indicated by (\ref{4E21}), the changes $\theta\to\pi-\theta$, $\theta_P\to\pi-\theta_P$ and $\varphi_P\to\varphi_P+\pi$ transform $f_{2C}(\tau)$ into $f_{1C}(\tau)$.  This means that $f_{2C}$ has the same kind of critical points as $f_{1C}(\tau)$ but in a different hemisphere (in $\theta>\pi/2$ and $\theta_P>\pi/2$ instead of $\theta<\pi/2$ and $\theta_P<\pi/2$ and vice versa).  Moreover, the function $f_{lC}$ for $\alpha>\pi/2$ follows from that for $\alpha<\pi/2$ if we replace $\theta$ by $\pi-\theta$ at the same time as replacing $\alpha$ by $\pi-\alpha$ (see~(\ref{4E21})).  It is sufficient, therefore, to consider only the cases in which $\alpha<\pi/2$.  Note that, owing to the presence of the factor $w_1$ in the expression for the density of the current sheet, the field ${\bf E}^{\rm uc}$ is zero for $\alpha=\pi/2$ (see (\ref{2E7})). 

\begin{figure}
\centerline{\includegraphics[width=13cm]{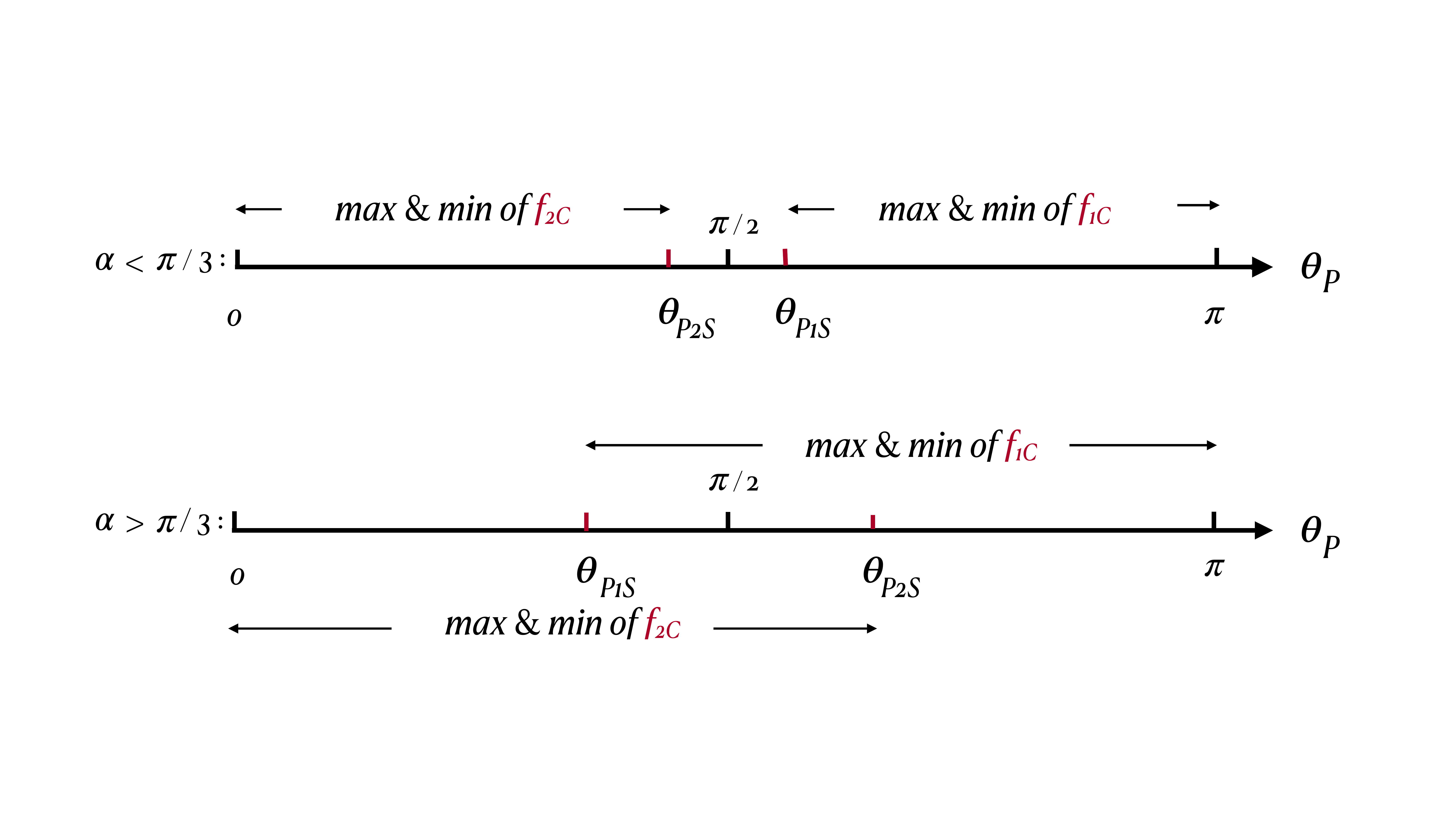}}
\caption{The relative positions of the critical angles $\theta_{P1S}$ and $\theta_{P2S}$ along the $\theta_P$-axis for $\alpha<\pi/3$ and $\alpha>\pi/3$.  The ranges of values of $\theta_P$ for which $f_{1C}$ and $f_{2C}$ have two turning points as functions of $\tau$ are also shown.  Outside the shown intervals, both $f_{1C}$ and $f_{2C}$ vary monotonically with $\tau$.}
\label{PF8}
\end{figure}

\begin{figure}
\centerline{\includegraphics[width=16cm]{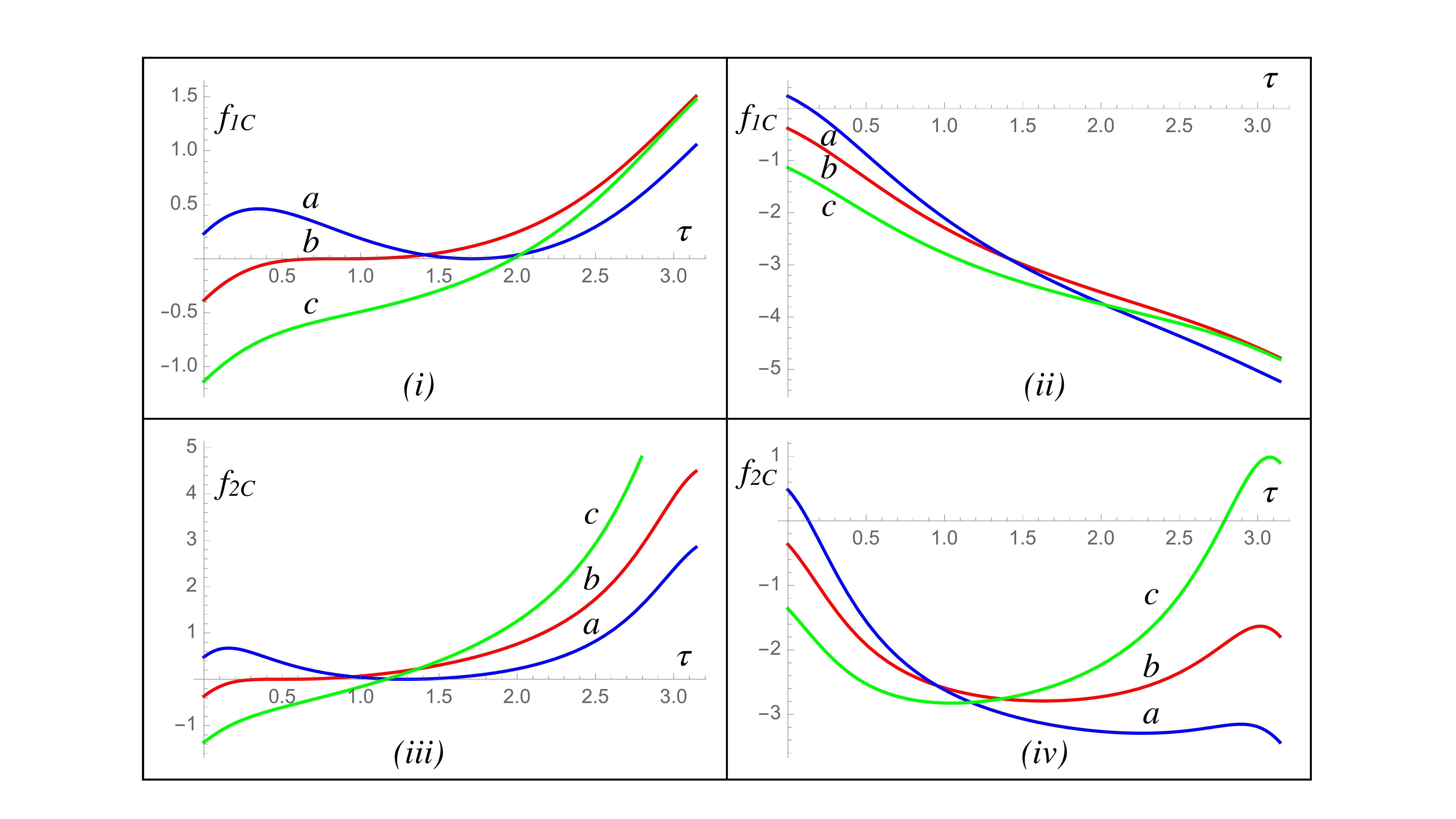}}
\caption{Dependence of the phase functions $f_{1C}$ and $f_{2C}$ on the integration variable $\tau$ for $\alpha=\pi/4$ in frames (i) and (ii) and for $\alpha=3\pi/8$ in frames (iii) and (iv).  In frames (i) and (ii), $\theta_P>\theta_{P1S}>\theta_{P2S}$ for the blue curves $a$, $\theta_P=\theta_{P1S}>\theta_{P2S}$ for the red curves $b$ and $\theta_{P2S}<\theta_P<\theta_{P1S}$ for the green curves $c$.  In frames (iii) and (iv), $\theta_{P1S}<\theta_P<\theta_{P2S}$ for the blue curves $a$, $\theta_P=\theta_{P1S}<\theta_{P2S}$ for the red curves $b$ and $\theta_P<\theta_{P1S}<\theta_{P2S}$ for the green curves $c$.  Note that a change in the coordinate $\varphi_P$ of the observation point shifts the above curves (which have here been plotted for ${\hat R}_P=10^6$) vertically without changing their shapes. These curves illustrate that while in the case of $\pi/3<\alpha<\pi/2$ both $f_{1C}$ and $f_{2C}$ have maxima, minima or inflection points, in the case of $0<\alpha<\pi/3$ either $f_{1C}$ or $f_{2C}$ is a monotonic function of $\tau$.}
\label{PF9}
\end{figure}

In this paper, we denote the values of $\tau$, $\theta_P$ and $\varphi_P$ at which
\begin{equation}
f_{lC}=\frac{\partial f_{lC}}{\partial\tau}=\frac{\partial^2 f_{lC}}{\partial\tau^2} =0
\label{4E47}
\end{equation}
by $\tau_{lS}$, $\theta_{PlS}$, and $\varphi_{PlS}$, respectively.  It turns out that $\theta_{P1S}$ and $\theta_{P2S}$ always lie on opposite sides of $\theta_P=\pi/2$ (see figure~\ref{PF8}): they satisfy $\theta_{P2S}<\pi/2<\theta_{P1S}$ for $0<\alpha<\pi/3$ and $\theta_{P1S}<\pi/2<\theta_{P2S}$ for $\pi/3<\alpha<\pi/2$ when the observation point lies at infinity.  For $\alpha=\pi/3$ they have the values $\theta_{P1S}=\pi/2-1/{\hat R}_P$ and $\theta_{P2S}=\pi/2+1/{\hat R}_P$ when ${\hat R}_P\gg1$. 

Plots of $f_{lC}$ versus $\tau$ for various values of the free parameters $\theta_P$ and $\alpha$ display the following features.  When $\alpha<\pi/3$, (i) $f_{1C}$ is monotonic for $0<\theta_P<\theta_{P1S}$ and has a maximum and a minimum for $\theta_{P1S}<\theta_P<\pi$, (ii) $f_{2C}$ has a maximum and a minimum for $0<\theta_P<\theta_{P2S}$ and is monotonic for $\theta_{P2S}<\theta_P<\pi$ and (iii) the extrema of $f_{1C}$ and $f_{2C}$ coincide at an inflection point for $\theta_P=\theta_{P1S}$ and $\theta_P=\theta_{PS2}$, respectively.  Hence, both $f_{1C}$ and $f_{2C}$ are monotonic for $\theta_{P2S}<\theta_P<\theta_{P1S}$ when $0<\alpha<\pi/3$.  On the other hand, when $\pi/3<\alpha<\pi/2$, (i) $f_{1C}$ is monotonic in $0<\theta_P<\theta_{P1S}$ and has two turning points in $\theta_{P1S}<\theta_P<\pi$, (ii) $f_{2C}$ has a maximum and a minimum in $0<\theta_P<\theta_{P2S}$ and is monotonic in $\theta_{P2S}<\theta_P<\pi$ and (iii) once again the extrema of $f_{1C}$ and $f_{2C}$ coincide at an inflection point for $\theta_P=\theta_{P1S}$ and $\theta_P=\theta_{PS2}$, respectively.  Hence $f_{1C}$ and $f_{2C}$ each have a maximum and a minimum for $\theta_{P2S}<\theta_P<\theta_{P1S}$ when $\pi/3<\alpha<\pi/2$.  Figure~\ref{PF9} illustrates some of the generic forms assumed by $f_{1C}$ and $f_{2C}$ as functions of $\tau$ inside the integration domain $0<\tau<\pi$. 

Changes in values of the free parameter $\varphi_P$ simply shift the curve representing $f_{lC}$ versus $\tau$ up or down without altering its shape (see~(\ref{4E21})).  But changes in values of the remaining free parameter ${\hat R}_P$ does alter the relative positions of the two turning points of $f_{lC}$ when $\theta_P$ is close to $\theta_{PlS}$.  The length of the interval separating the $\tau$ coordinates of the maximum and minimum of $f_{lC}$ decreases with increasing ${\hat R}_P$ in a case where $\theta_P$ is close to $\theta_{PlS}$ and so this interval is small.  If we denote the $\tau$ coordinates of the maximum and minimum of $f_{lC}$ by $\tau_{l{\rm max}}$ and $\tau_{l{\rm min}}$, respectively, then it turns out that $\vert\tau_{l{\rm max}}-\tau_{l{\rm min}}\vert\propto{\hat R}_P^{-1/2}$ for ${\hat R}_P\gg1$ when $\theta_P$ has the value $\theta_{PlS}\vert_{{\hat R}_P\to\infty}$ for any given $\alpha$, i.e., when $\theta_P$ is such that $\vert\tau_{l{\rm max}}-\tau_{l{\rm min}}\vert$ shrinks to zero as ${\hat R}_P$ tends to infinity.  We shall see in \S~\ref{subsec:inverse-square} that this property of $f_{lC}$ results in a decay of the radiation's intensity  with distance in the direction of $\theta_{PlS}$ that is slower than predicted by the inverse-square law. 

Note that $f_{1C}$ and $f_{2C}$ for a given value of $n$ are equal at $\tau=0$ and differ by $2\pi$ at $\tau=\pi$ (see (\ref{4E21}) and (\ref{4E14})).  In cases where the absolute value of $f_{lC}\vert_{\tau=\pi}-f_{lC}\vert_{\tau=0}$ for either $l=1$ or $l=2$ is greater than $2\pi$ (e.g., when $\alpha=15^\circ$ and $\theta_P=65^\circ$) or $f_{lC}\vert_{\tau_{l{\rm min}}}$ and $f_{lC}\vert_{\tau_{l{\rm max}}}$ differ by more than $2\pi$ (e.g., when $\alpha=85^\circ$ and $\theta_P=5^\circ$), ordinates of the points on $f_{1C}$ and $f_{2C}$ in figure~\ref{PF9} also span intervals whose lengths exceed $2\pi$.  In such cases, therefore, $f_{1C}$ and $f_{2C}$ for several values of $n$ (i.e., several cycles of retarded time) simultaneously contribute toward the intensity of the pulse that is observed during a single period. 

The above discussion applies also to the modified functions ${\bar f}_{1C}$ and ${\bar f}_{2C}$: their critical points differ from those of $f_{1C}$ and $f_{2C}$ only in their positions, not in their nature.  The generic forms assumed by ${\bar f}_{1C}$ and ${\bar f}_{2C}$ are the same as those illustrated in figure~\ref{PF9} except that the role of $\alpha=\pi/3$ in figure~\ref{PF8} is played by $\alpha=0.8707129958$.  

Depending on relative positions of the coordinate $\theta_P$ of the observation point and the inclination angle $\alpha$, therefore, the four phase functions $f_{1C}$, $f_{2C}$, ${\bar f}_{1C}$ and ${\bar f}_{2C}$ in the integrand of the $\tau$-integral in~(\ref{4E45}) can jointly have a set of isolated stationary points with one to eight members (where $\partial f_{lC}/\partial\tau$ and/or $\partial{\bar f}_{lC}/\partial\tau$ for $l=1$ and/or $l=2$ vanish) or can have one or two degenerate stationary points (where $\partial ^2f_{lC}/\partial\tau^2$ and/or $\partial ^2{\bar f}_{lC}/\partial\tau^2$ simultaneously vanish with $\partial f_{lC}/\partial\tau$ and/or $\partial {\bar f}_{2C}/\partial\tau$).   The number of contributing stationary points is higher in cases where the ordinates of the curves depicted in figure \ref{PF9} span intervals whose lengths exceed $2\pi$ and so the contributions from several cycles of retarded time are received during a single period of observation time.

\subsection{A uniform asymptotic approximation to the integral over the colatitude of the source elements for large $\vert k\vert$}
\label{subsec:tau}

Since this paper is concerned with determining the radiation field ${\bf E}^{\rm uc}$ also at observation points for which the turning points of the phase functions $f_{lC}$ and ${\bar f}_{lC}$ are close to one another or coalescent, we need to obtain an asymptotic approximation to the value of the $\tau$-integral in (\ref{4E45}) that is uniform with respect to the interval separating the nearby saddle points of these phase functions (see~\cite{ChesterC:Extstd,BleisteinN:Asei}).

In cases where $f_{lC}$ and ${\bar f}_{lC}$ each have two turning points, one at their maxima $\tau_{l{\rm max}}$ and ${\bar\tau}_{l{\rm max}}$ and one at their minima $\tau_{l{\rm min}}$ and ${\bar\tau}_{l{\rm min}}$, each of these functions can be tansformed into a cubic function via
\begin{equation}
f_{lC}=\textstyle{\frac{1}{3}}\lambda^3-\sigma_{l1}^2\lambda+\sigma_{l2},\qquad{\rm and}\qquad {\bar f}_{lC}=\textstyle{\frac{1}{3}}\lambda^3-{\bar\sigma}_{l1}^2\lambda+{\bar\sigma}_{l2},
\label{4E48}
\end{equation}
where
\begin{equation}
\sigma_{l1}=\left[\textstyle{\frac{3}{4}}\left(f_{lC}\left\vert_{\tau=\tau_{l{\rm max}}}-f_{lC}\left\vert_{\tau=\tau_{l{\rm min}}}\right.\right.\right)\right]^{1/3},
\label{4E49}
\end{equation}
\begin{equation}
{\bar\sigma}_{l1}=\left[\textstyle{\frac{3}{4}}\left({\bar f}_{lC}\left\vert_{\tau={\bar\tau}_{l{\rm max}}}-{\bar f}_{lC}\left\vert_{\tau={\bar\tau}_{l{\rm min}}}\right.\right.\right)\right]^{1/3},
\label{4E50}
\end{equation}
and
\begin{equation}
\sigma_{l2}=\textstyle{\frac{1}{2}}\left(f_{lC}\left\vert_{\tau=\tau_{l{\rm max}}}+f_{lC}\left\vert_{\tau=\tau_{l{\rm min}}}\right.\right.\right),
\label{4E51}
\end{equation}
\begin{equation}
{\bar\sigma}_{l2}=\textstyle{\frac{1}{2}}\left({\bar f}_{lC}\left\vert_{\tau={\bar\tau}_{l{\rm max}}}+{\bar f}_{lC}\left\vert_{\tau={\bar\tau}_{l{\rm min}}}\right.\right.\right).
\label{4E52}
\end{equation}
The transformation of the integration variable in (\ref{4E45}) from $\tau$ to $\lambda$ thus results in
\begin{eqnarray}
{\bf E}^{\rm uc}&\simeq&\frac{2j_0}{3\omega}\sum_{l=1}^2\int_0^\infty{\rm d}k\int_0^{\pi}{\rm d}\lambda\,b\,\bigg\{\bigg[\frac{{\rm i}}{3}{\rm H}(k_2-k){\bf P}_l+\sqrt{\frac{2}{\pi a k}}\exp\left(\frac{{\rm i}\pi}{4}\right)\nonumber\\*
&&\times\left({\rm H}(k-k_1){\bf Q}_l-\frac{1}{3a}{\rm H}(k-k_2){\bf P}_l\right)\bigg]\frac{{\rm d}\tau}{{\rm d}\lambda}\bigg\vert_{f_{lC}}\exp\left[{\rm i}k\left(\textstyle{\frac{1}{3}}\lambda^3-\sigma_{l1}^2\lambda+\sigma_{l2}\right)\right]\nonumber\\*
&&+\sqrt{\frac{2}{\pi a k}}\exp\left(-\frac{{\rm i}\pi}{4}\right)\left({\rm H}(k-k_1){\bf Q}_l-\frac{1}{a}{\rm H}(k-k_2){\bf P}_l\right)\nonumber\\*
&&\times\frac{{\rm d}\tau}{{\rm d}\lambda}\bigg\vert_{{\bar f}_{lC}}\exp\left[{\rm i}k\left(\textstyle{\frac{1}{3}}\lambda^3-{\bar\sigma}_{l1}^2\lambda+{\bar\sigma}_{l2}\right)\right]\bigg\},
\label{4E53}
\end{eqnarray}
whose asymptotic value for large $k$ can be written, as shown by~\cite{ChesterC:Extstd}, as
\begin{eqnarray}
{\bf E}^{\rm uc}&\simeq &\frac{2j_0}{3\omega}\sum_{l=1}^2\int_0^\infty{\rm d}k\bigg\{\exp({\rm i}k\sigma_{l2})\bigg[\mbox{{\boldmath${\cal P}$}}_l\int_{-\infty}^\infty{\rm d}\lambda\,\exp[{\rm i}k(\textstyle{\frac{1}{3}}\lambda^3-\sigma_{l1}^2\lambda)]\nonumber\\*
&&+\mbox{{\boldmath${\cal Q}$}}_l\int_{-\infty}^\infty{\rm d}\lambda\,\lambda\exp[{\rm i}k(\textstyle{\frac{1}{3}}\lambda^3-\sigma_{l1}^2\lambda)]\bigg]+\exp({\rm i}k{\bar\sigma}_{l2})\nonumber\\*
&&\times\bigg[\mbox{{\boldmath${\bar{\cal P}}$}}_l\int_{-\infty}^\infty{\rm d}\lambda\,\exp[{\rm i}k(\textstyle{\frac{1}{3}}\lambda^3-{\bar\sigma}_{l1}^2\lambda)]+\mbox{{\boldmath${\bar{\cal Q}}$}}_l\int_{-\infty}^\infty{\rm d}\lambda\,\lambda\exp[{\rm i}k(\textstyle{\frac{1}{3}}\lambda^3-{\bar\sigma}_{l1}^2\lambda)]\bigg]\bigg\},\nonumber\\*
\label{4E54}
\end{eqnarray}
where
\begin{equation}
\mbox{{\boldmath${\cal P}$}}_l=\frac{1}{2}\left(\mbox{{\boldmath${\cal K}$}}_l\left\vert_{\tau=\tau_{l{\rm min}}}+\mbox{{\boldmath${\cal K}$}}_l\left\vert_{\tau=\tau_{l{\rm max}}}\right.\right.\right),
\label{4E55}
\end{equation}
\begin{equation}
\mbox{{\boldmath${\cal Q}$}}_l=\frac{1}{2\sigma_{l1}}\left(\mbox{{\boldmath${\cal K}$}}_l\left\vert_{\tau=\tau_{l{\rm min}}}-\mbox{{\boldmath${\cal K}$}}_l\left\vert_{\tau=\tau_{l{\rm max}}}\right.\right.\right),
\label{4E56}
\end{equation}
\begin{equation}
\mbox{{\boldmath${\bar{\cal P}}$}}_l=\frac{1}{2}\left(\mbox{{\boldmath${\bar{\cal K}}$}}_l\left\vert_{\tau={\bar\tau}_{l{\rm min}}}+\mbox{{\boldmath${\bar{\cal K}}$}}_l\left\vert_{\tau={\bar\tau}_{l{\rm max}}}\right.\right.\right),
\label{4E57}
\end{equation}
\begin{equation}
\mbox{{\boldmath${\bar{\cal Q}}$}}_l=\frac{1}{2{\bar\sigma}_{l1}}\left(\mbox{{\boldmath${\bar{\cal K}}$}}_l\left\vert_{\tau={\bar\tau}_{l{\rm min}}}-\mbox{{\boldmath${\bar{\cal K}}$}}_l\left\vert_{\tau={\bar\tau}_{l{\rm max}}}\right.\right.\right),
\label{4E58}
\end{equation}
with
\begin{eqnarray}
\mbox{{\boldmath${\cal K}$}}_l&=&b\frac{{\rm d}\tau}{{\rm d}\lambda}\bigg\vert_{f_{lC}}\bigg[\frac{{\rm i}}{3}{\rm H}(k_2-k){\bf P}_l+\sqrt{\frac{2}{\pi a k}}\exp\left(\frac{{\rm i}\pi}{4}\right)\nonumber\\*
&&\times\left({\rm H}(k-k_1){\bf Q}_l-\frac{1}{3a}{\rm H}(k-k_2){\bf P}_l\right)\bigg],
\label{4E59}
\end{eqnarray}
\begin{equation}
\mbox{{\boldmath${\bar{\cal K}}$}}_l=b\sqrt{\frac{2}{\pi a k}}\exp\left(-\frac{{\rm i}\pi}{4}\right)\frac{{\rm d}\tau}{{\rm d}\lambda}\bigg\vert_{{\bar f}_{lC}}\left[{\rm H}(k-k_1){\bf Q}_l-\frac{1}{a}{\rm H}(k-k_2){\bf P}_l\right]
\label{4E60}
\end{equation}
(see also~\cite{BleisteinN:Asei}).  Evaluating the integrals over $\lambda$ (cf.~\cite{Olver}), we obtain
\begin{eqnarray}
{\bf E}^{\rm uc}&\simeq&\frac{4\pi j_0}{3\omega}\sum_{l=1}^2\int_0^\infty{\rm d}k\,k^{-1/3}\bigg\{\exp({\rm i}k\sigma_{l2})\nonumber\\*
&&\times\left[\mbox{{\boldmath${\cal P}$}}_l{\rm Ai}(-k^{2/3}\sigma_{l1}^2)-{\rm i}k^{-1/3}\mbox{{\boldmath${\cal Q}$}}_l{\rm Ai}^\prime(-k^{2/3}\sigma_{l1}^2)\right]+\exp({\rm i}k{\bar\sigma}_{l2})\nonumber\\*
&&\times\left[\mbox{{\boldmath${\bar{\cal P}}$}}_l{\rm Ai}\left(-k^{2/3}{\bar\sigma}_{l1}^2\right)-{\rm i}k^{-1/3}\mbox{{\boldmath${\bar{\cal Q}}$}}_l{\rm Ai}^\prime\left(-k^{2/3}{\bar\sigma}_{l1}^2\right)\right]\bigg\},
\label{4E61}
\end{eqnarray}
where ${\rm Ai}$ and ${\rm Ai}^\prime$ are the Airy function and the derivative of the Airy function with respect to its argument, respectively.  Note that $(\mbox{{\boldmath${\cal P}$}}_l,\mbox{{\boldmath${\cal Q}$}}_l)$ are different from the vectors $({\bf P}_l,{\bf Q}_l)$ defined in (\ref{4E26})--(\ref{4E27}).

The indeterminate quantities $({\rm d}\tau/{\rm d}\mu)_{f_{lC}}\vert_{\tau={\tau_{l{\rm min}}}}$ and $({\rm d}\tau/{\rm d}\mu)_{f_{lC}}\vert_{\tau={\tau_{l{\rm max}}}}$ that appear in (\ref{4E61}) have to be found by repeated differentiation of (\ref{4E48}) with respect to $\lambda$ and the evaluation of the resulting relations
\begin{equation}
\frac{\partial f_{lC}}{\partial\tau}\frac{{\rm d}\tau}{{\rm d}\lambda}=\lambda^2-\sigma_{l1}^2,
\label{4E62}
\end{equation} 
and
\begin{equation}
\frac{\partial^2f_{lC}}{\partial\tau^2}\left(\frac{{\rm d}\tau}{{\rm d}\lambda}\right)^2+\frac{\partial f_{lC}}{\partial\tau}\frac{{\rm d}^2\tau}{{\rm d}\lambda^2}=2\lambda,
\label{4E63}
\end{equation}
at $\tau={\tau_{l{\rm min}}}$ and $\tau={\tau_{l{\rm max}}}$.  This procedure results in
\begin{equation}
\left(\frac{{\rm d}\tau}{{\rm d}\lambda}\right)_{f_{lC}}\Bigg\vert_{\lambda=\pm\sigma_{l1}}=\left(\frac{2\lambda}{\partial^2f_{lC}/\partial\tau^2}\Big\vert_{\lambda=\pm\sigma_{l1}}\right)^{1/2},
\label{4E64}
\end{equation}
in which $\lambda=\pm\sigma_{l1}$ are the images of $\tau={\tau_{l{\rm min}}}$ and $\tau={\tau_{l{\rm max}}}$, respectively.  Likewise,
\begin{equation}
\left(\frac{{\rm d}\tau}{{\rm d}\lambda}\right)_{{\bar f}_{lC}}\Bigg\vert_{\lambda=\pm{\bar\sigma}_{l1}}=\left(\frac{2\lambda}{\partial^2{\bar f}_{lC}/\partial\tau^2}\Big\vert_{\lambda=\pm{\bar\sigma}_{l1}}\right)^{1/2},
\label{4E65}
\end{equation}
a result that can be obtained by applying the procedure described above to the function ${\bar f}_{lC}$.  In this expression, too, $\lambda=\pm{\bar\sigma}_{l1}$ are the images of $\tau={{\bar\tau}_{l{\rm min}}}$ and $\tau={{\bar\tau}_{l{\rm max}}}$, respectively. 

Note that every one of the terms appearing in (\ref{4E61}) would contribute toward the value of the radiation field only when the phase functions $f_{1C}$, $f_{2C}$, ${\bar f}_{1C}$ and ${\bar f}_{2C}$ each have two turning points (see figure~\ref{PF8}).  If any one of these functions varies monotonically with $\tau$, for the prescribed values of $\alpha$ and $\theta_P$, then the terms entailing the (non-existent) $\tau$ coordinates of its maximum and minimum should be omitted from (\ref{4E61}). 

\subsection{The remaining integration with respect to $k$}
\label{subsec:kIntegration}
 
To perform the $k$-integration in (\ref{4E61}) we first need to render the $k$-dependence of the coefficients $\mbox{{\boldmath${\cal P}$}}_l$, $\mbox{{\boldmath${\cal Q}$}}_l$, $\mbox{{\boldmath${\bar{\cal P}}$}}_l$ and $\mbox{{\boldmath${\bar{\cal Q}}$}}_l $ that appear in this equation explicit.  This can be done by rewriting (\ref{4E55})--(\ref{4E60}) as
\begin{equation}
\left[\matrix{\mbox{{\boldmath${\cal P}$}}_l\cr\mbox{{\boldmath${\cal Q}$}}_l\cr}\right]={\rm H}(k_2-k)\left[\matrix{\mbox{{\boldmath${\cal P}$}}^{(0)}_l\cr\mbox{{\boldmath${\cal Q}$}}^{(0)}_l\cr}\right]+\left({\rm H}(k-k_1)\left[\matrix{\mbox{{\boldmath${\cal P}$}}^{(1)}_l\cr\mbox{{\boldmath${\cal Q}$}}^{(1)}_l\cr}\right]+{\rm H}(k-k_2)\left[\matrix{\mbox{{\boldmath${\cal P}$}}^{(2)}_l\cr\mbox{{\boldmath${\cal Q}$}}^{(2)}_l\cr}\right]\right)\,k^{-1/2},
\label{4E66}
\end{equation}
\begin{equation}
\left[\matrix{\mbox{{\boldmath${\bar{\cal P}}$}}_l\cr\mbox{{\boldmath${\bar{\cal Q}}$}}_l\cr}\right]=\left({\rm H}(k-k_1)\left[\matrix{\mbox{{\boldmath${\bar{\cal P}}$}}^{(1)}_l\cr\mbox{{\boldmath${\bar{\cal Q}}$}}^{(1)}_l\cr}\right]+{\rm H}(k-k_2)\left[\matrix{\mbox{{\boldmath${\bar{\cal P}}$}}^{(2)}_l\cr\mbox{{\boldmath${\bar{\cal Q}}$}}^{(2)}_l\cr}\right]\right)\,k^{-1/2},
\label{4E67}
\end{equation}
where
\begin{equation}
\mbox{{\boldmath${\cal P}$}}^{(j)}_l=\frac{1}{2}\left(\mbox{{\boldmath${\cal K}$}}^{(j)}_l\left\vert_{\tau=\tau_{l{\rm min}}}+\mbox{{\boldmath${\cal K}$}}^{(j)}_l\left\vert_{\tau=\tau_{l{\rm max}}}\right.\right.\right),\qquad j=0,1,2,
\label{4E68}
\end{equation}
\begin{equation}
\mbox{{\boldmath${\cal Q}$}}^{(j)}_l=\frac{1}{2\sigma_{l1}}\left(\mbox{{\boldmath${\cal K}$}}^{(j)}_l\left\vert_{\tau=\tau_{l{\rm min}}}-\mbox{{\boldmath${\cal K}$}}^{(j)}_l\left\vert_{\tau=\tau_{l{\rm max}}}\right.\right.\right),\qquad j=0,1,2,
\label{4E69}
\end{equation}
\begin{equation}
\mbox{{\boldmath${\bar{\cal P}}$}}^{(j)}_l=\frac{1}{2}\left(\mbox{{\boldmath${\bar{\cal K}}$}}^{(j)}_l\left\vert_{\tau={\bar\tau}_{l{\rm min}}}+\mbox{{\boldmath${\bar{\cal K}}$}}^{(j)}_l\left\vert_{\tau={\bar\tau}_{l{\rm max}}}\right.\right.\right),\qquad j=1,2,
\label{4E70}
\end{equation} 
\begin{equation}
\mbox{{\boldmath${\bar{\cal Q}}$}}^{(j)}_l=\frac{1}{2{\bar\sigma}_{l1}}\left(\mbox{{\boldmath$\bar{{\cal K}}$}}^{(j)}_l\left\vert_{\tau={\bar\tau}_{l{\rm min}}}-\mbox{{\boldmath${\bar{\cal K}}$}}^{(j)}_l\left\vert_{\tau={\bar\tau}_{l{\rm max}}}\right.\right.\right),\qquad j=1,2,
\label{4E71}
\end{equation} 
with
\begin{equation}
\mbox{{\boldmath${\cal K}$}}^{(0)}_l=\frac{{\rm i}}{3}b\,\frac{{\rm d}\tau}{{\rm d}\lambda}\bigg\vert_{f_{lC}}{\bf P}_l ,
\label{4E72}
\end{equation}
\begin{equation}
\left[\matrix{\mbox{{\boldmath${\cal K}$}}^{(1)}_l\cr\mbox{{\boldmath${\cal K}$}}^{(2)}_l\cr}\right]=b\,\sqrt{\frac{2}{\pi a}}\frac{{\rm d}\tau}{{\rm d}\lambda}\bigg\vert_{f_{lC}}\exp\left(\frac{{\rm i}\pi}{4}\right)\left[\matrix{{\bf Q}_l \cr-\frac{1}{3a}{\bf P}_l\cr}\right],
\label{4E73}
\end{equation} 
\begin{equation}
\left[\matrix{\mbox{{\boldmath${\bar{\cal K}}$}}^{(1)}_l\cr\mbox{{\boldmath${\bar{\cal K}}$}}^{(2)}_l\cr}\right]=b\,\sqrt{\frac{2}{\pi a}}\frac{{\rm d}\tau}{{\rm d}\lambda}\bigg\vert_{{\bar f}_{lC}}\exp\left(-\frac{{\rm i}\pi}{4}\right)\left[\matrix{{\bf Q}_l \cr-\frac{1}{a}{\bf P}_l\cr}\right].
\label{4E74}
\end{equation} 
The quantities appearing in (\ref{4E68})--(\ref{4E74}) are all independent of $k$, so that the remaining integrals in (\ref{4E61}) are--according to (\ref{4E66})--of two types: those that extend over a finite interval $0\le k\le k_i$ and those that extend over a semi-infinite interval $k_i\le k<\infty$, where $i$ is either $1$ or $2$.  

The rapid oscillations of the Airy functions for large $k$ (cf.~\cite{Olver}) ensure that the integrals over $0\le k\le k_i$ in (\ref{4E61}) receive their main contributions from the vicinity of $k=0$.  The ranges of these integrals can therefore be extended to $0<k<\infty$ without introducing an appreciable error.  Moreover, each of the integrals over $k_i<k<\infty$ is accurately approximated (according to numerical integrations) if it is written as the difference between two integrals with the same integrands but with the ranges $0\le k<\infty$ and $0\le k\le k_i$ and the Airy function in the integrand of the integral over $0\le k\le k_i$ is replaced by its value at $k=0$, as in (\ref{4E80}) below.  

Once these approximations are applied to (\ref{4E61}) and the Airy functions that appear in the resulting equation are expressed in terms of Bessel functions, the $k$-integrals in question assume the form of tabulated Fourier transforms (cf.,~\cite{Olver, Erdelyi}) and can be evaluated analytically to arrive at
\begin{equation}
{\bf E}^{\rm uc}={\bf E}^{\rm uc}_{\cal P}+{\bf E}^{\rm uc}_{\cal Q}
\label{4E75}
\end{equation}
with
\begin{eqnarray}
{\bf E}^{\rm uc}_{\cal P}&\simeq&\frac{2^{4/3}\pi j_0}{3^{1/3}\omega}\sum_{l=1}^2\left\{\sigma_{l1}^{-2}{\cal G}_1(\eta_l) \mbox{{\boldmath${\cal P}$}}^{(0)}_l+\left(\frac{2}{3\sigma_{l1}}\right)^{1/2}\left[{\cal G}_3(\eta_l,\kappa_1) \mbox{{\boldmath${\cal P}$}}^{(1)}_l+{\cal G}_3(\eta_l,\kappa_2) \mbox{{\boldmath${\cal P}$}}^{(2)}_l\right]\right.\nonumber\\*
&&\left.+\left(\frac{2}{3{\bar\sigma}_{l1}}\right)^{1/2}\left[{\cal G}_3({\bar\eta}_l,{\bar\kappa}_1) \mbox{{\boldmath${\bar{\cal P}}$}}^{(1)}_l+{\cal G}_3({\bar\eta}_l,{\bar\kappa}_2) \mbox{{\boldmath${\bar{\cal P}}$}}^{(2)}_l\right]\right\}
\label{4E76}
\end{eqnarray}
and
\begin{eqnarray}
{\bf E}^{\rm uc}_{\cal Q}&\simeq&\frac{2^{5/3}\pi j_0}{3^{2/3}{\rm i}\,\omega}\sum_{l=1}^2\left\{\sigma_{l1}^{-1}{\cal G}_2(\eta_l) \mbox{{\boldmath${\cal Q}$}}^{(0)}_l+\left(\frac{2\sigma_{l1}}{3}\right)^{1/2}\left[{\cal G}_4(\eta_l,\kappa_1) \mbox{{\boldmath${\cal Q}$}}^{(1)}_l+{\cal G}_4(\eta_l,\kappa_2) \mbox{{\boldmath${\cal Q}$}}^{(2)}_l\right]\right.\nonumber\\*
&&\left.+\left(\frac{2{\bar\sigma}_{l1}}{3}\right)^{1/2}\left[{\cal G}_4({\bar\eta}_l,{\bar\kappa}_1) \mbox{{\boldmath${\bar{\cal Q}}$}}^{(1)}_l+{\cal G}_4({\bar\eta}_l,{\bar\kappa}_2) \mbox{{\boldmath${\bar{\cal Q}}$}}^{(2)}_l\right]\right\}
\label{4E77}
\end{eqnarray}
where
\begin{eqnarray}
{\cal G}_1(x)&=&\int_0^\infty{\rm d}\kappa\exp({\rm i}\kappa x)\kappa^{-1/3}{\rm Ai}\left[-\left({\textstyle\frac{3}{2}}\kappa\right)^{2/3}\right]\nonumber\\*
&=&\left({\textstyle\frac{2}{3}}\right)^{2/3}(1-x^2)^{-1/2}[{\rm H}(x+1)-{\rm H}(x-1)]\cos\left({\textstyle\frac{1}{3}}\arcsin x\right)+18^{-1/3}{\rm i}(x^2-1)^{-1/2}\nonumber\\*
&&\times\left\{{\rm H}(x-1)\left[\exp({\rm i}\pi/6)\left(\sqrt{x^2-1}+x\right)^{-1/3}+\exp(-{\rm i}\pi/6)\left(\sqrt{x^2-1}+x\right)^{1/3}\right]\right.\nonumber\\*
&&\left.-{\rm H}(-x-1)\left[\exp(-{\rm i}\pi/6)\left(\sqrt{x^2-1}-x\right)^{-1/3}+\exp({\rm i}\pi/6)\left(\sqrt{x^2-1}-x\right)^{1/3}\right]\right\},\nonumber\\*
\label{4E78}
\end{eqnarray} 
\begin{eqnarray}
{\cal G}_2(x)&=&\int_0^\infty{\rm d}\kappa\exp({\rm i}\kappa x)\kappa^{-2/3}{\rm Ai}^\prime\left[-\left({\textstyle\frac{3}{2}}\kappa\right)^{2/3}\right]\nonumber\\*
&=&\left({\textstyle\frac{2}{3}}\right)^{1/3}{\rm i}(1-x^2)^{-1/2}[{\rm H}(x+1)-{\rm H}(x-1)]\sin\left({\textstyle\frac{2}{3}}\arcsin x\right)+12^{-1/3}{\rm i}(x^2-1)^{-1/2}\nonumber\\*
&&\times\left\{{\rm H}(x-1)\left[\exp({\rm i}\pi/3)\left(\sqrt{x^2-1}+x\right)^{-2/3}-\exp(-{\rm i}\pi/3)\left(\sqrt{x^2-1}+x\right)^{2/3}\right]\right.\nonumber\\*
&&\left.-{\rm H}(-x-1)\left[\exp(-{\rm i}\pi/3)\left(\sqrt{x^2-1}-x\right)^{-2/3}-\exp({\rm i}\pi/3)\left(\sqrt{x^2-1}-x\right)^{2/3}\right]\right\},\nonumber\\*
\label{4E79}
\end{eqnarray} 
\begin{eqnarray}
{\cal G}_3(x,\kappa_0)&=&\int_0^\infty{\rm d}\kappa\,\exp({\rm i}\kappa x)\kappa^{-5/6}{\rm Ai}\left[-\left({\textstyle\frac{3}{2}}\kappa\right)^{2/3}\right]-{\rm Ai}(0)\int_0^{\kappa_0}{\rm d}\kappa\,\exp({\rm i}\kappa x)\kappa^{-5/6}\nonumber\\*
&=&{\rm H}(x-1)\left[\frac{\Gamma(5/6)}{6^{2/3}x^{5/6}\Gamma(4/3)}\exp\left(\frac{5{\rm i}\pi}{12}\right)\,{}_2F_1\left(\frac{11}{12},\frac{5}{12};\frac{4}{3};\frac{1}{x^2}\right)\right.\nonumber\\*
&&\left.+\frac{\Gamma(1/6)}{3^{2/3}x^{1/6}\Gamma(2/3)}\exp\left(\frac{{\rm i}\pi}{12}\right)\,{}_2F_1\left(\frac{7}{12},\frac{1}{12};\frac{2}{3};\frac{1}{x^2}\right)\right]\nonumber\\*
&&+\frac{2^{2/3}\pi}{3^{1/6}}[{\rm H}(x+1)-{\rm H}(x-1)]\left[\frac{1}{\Gamma(7/12)\Gamma(11/12)}\,{}_2F_1\left(\frac{5}{12},\frac{1}{12};\frac{1}{2};x^2\right)\right.\nonumber\\*
&&\left.+\frac{2{\rm i}x}{\Gamma(5/12)\Gamma(1/12)}\,{}_2F_1\left(\frac{11}{12},\frac{7}{12};\frac{3}{2};x^2\right)\right]\nonumber\\*
&&+{\rm H}(-x-1)\left[\frac{\Gamma(5/6)}{6^{2/3}\vert x\vert^{5/6}\Gamma(4/3)}\exp\left(-\frac{5{\rm i}\pi}{12}\right)\,{}_2F_1\left(\frac{11}{12},\frac{5}{12};\frac{4}{3};\frac{1}{x^2}\right)\right.\nonumber\\*
&&\left.+\frac{\Gamma(1/6)}{3^{2/3}\vert x\vert^{1/6}\Gamma(2/3)}\exp\left(-\frac{{\rm i}\pi}{12}\right)\,{}_2F_1\left(\frac{7}{12},\frac{1}{12};\frac{2}{3};\frac{1}{x^2}\right)\right]\nonumber\\*
&&-{\rm Ai}(0)\exp\left[\frac{{\rm i}\pi}{12}{\rm sgn}(x)\right]\vert x\vert^{-1/6}\left[\Gamma\left(\frac{1}{6}\right)-\Gamma\left(\frac{1}{6},-{\rm i}\kappa_0 x\right)\right],
\label{4E80}
\end{eqnarray} 
\begin{eqnarray}
{\cal G}_4(x,\kappa_0)&=&\left\{\int_0^\infty{\rm d}\kappa\,\exp({\rm i}\kappa x)\kappa^{\nu-1}{\rm Ai}^\prime\left[-\left({\textstyle\frac{3}{2}}\kappa\right)^{2/3}\right]-{\rm Ai}^\prime(0)\int_0^{\kappa_0}{\rm d}\kappa\,\exp({\rm i}\kappa x)\kappa^{\nu-1}\right\}_{\nu=-1/6}\nonumber\\*
&=&{\rm H}(x-1)\left[\frac{\Gamma(7/6)}{48^{1/3}x^{7/6}\Gamma(5/3)}\exp\left(\frac{7{\rm i}\pi}{12}\right)\,{}_2F_1\left(\frac{13}{12},\frac{7}{12};\frac{5}{3};\frac{1}{x^2}\right)\right.\nonumber\\*
&&\left.-\frac{x^{1/6}\Gamma(-1/6)}{3^{1/3}\Gamma(1/3)}\exp\left(-\frac{{\rm i}\pi}{12}\right)\,{}_2F_1\left(\frac{5}{12},-\frac{1}{12};\frac{1}{3};\frac{1}{x^2}\right)\right]\nonumber\\*
&&+12^{1/6}\pi[{\rm H}(x+1)-{\rm H}(x-1)]\left[\frac{1}{\Gamma(13/12)\Gamma(5/12)}\,{}_2F_1\left(\frac{7}{12},-\frac{1}{12};\frac{1}{2};x^2\right)\right.\nonumber\\*
&&\left.-\frac{2{\rm i}x}{\Gamma(7/12)\Gamma(-1/12)}\,{}_2F_1\left(\frac{13}{12},\frac{5}{12};\frac{3}{2};x^2\right)\right]\nonumber\\*
&&+{\rm H}(-x-1)\left[\frac{\Gamma(7/6)}{48^{1/3}\vert x\vert^{7/6}\Gamma(5/3)}\exp\left(-\frac{7{\rm i}\pi}{12}\right)\,{}_2F_1\left(\frac{13}{12},\frac{7}{12};\frac{5}{3};\frac{1}{x^2}\right)\right.\nonumber\\*
&&\left.-\frac{\vert x\vert^{1/6}\Gamma(-1/6)}{3^{1/3}\Gamma(1/3)}\exp\left(\frac{{\rm i}\pi}{12}\right)\,{}_2F_1\left(\frac{5}{12},-\frac{1}{12};\frac{1}{3};\frac{1}{x^2}\right)\right]\nonumber\\*
&&-{\rm Ai}^\prime(0)\exp\left[-\frac{{\rm i}\pi}{12}{\rm sgn}(x)\right]\vert x\vert^{1/6}\left[\Gamma\left(-\frac{1}{6}\right)-\Gamma\left(-\frac{1}{6},-{\rm i}\kappa_0 x\right)\right],
\label{4E81}
\end{eqnarray} 
and
\begin{equation}
\eta_l=\frac{3\sigma_{l2}}{2\sigma_{l1}^3}=\frac{f_{lC}\vert_{\tau=\tau_{l{\rm max}}}+f_{lC}\vert_{\tau=\tau_{l{\rm min}}}}{f_{lC}\vert_{\tau=\tau_{l{\rm max}}}-f_{lC}\vert_{\tau=\tau_{l{\rm min}}}},
\label{4E82}
\end{equation}
\begin{equation}
{\bar\eta}_l=\frac{3{\bar\sigma}_{l2}}{2{\bar\sigma}_{l1}^3}=\frac{{\bar f}_{lC}\vert_{\tau={\bar\tau}_{l{\rm max}}}+{\bar f}_{lC}\vert_{\tau={\bar\tau}_{l{\rm min}}}}{{\bar f}_{lC}\vert_{\tau={\bar \tau}_{l{\rm max}}}-{\bar f}_{lC}\vert_{\tau={\bar\tau}_{l{\rm min}}}}.
\label{4E83}
\end{equation}
In the above expressions, ${}_2F_1$ is the hypergeometric function, $\Gamma(\nu,x)$ is the incomplete gamma function and the integration variable $\kappa$ is related to $k$ via $\kappa={\textstyle\frac{2}{3}}\sigma_{l1}^3k$.  Note that ${\cal G}_4$ is here found by performing the integrations in the first line of (\ref{4E81}) for $\nu>0$ and evaluating the analytic continuation of the resulting expression at $\nu=-1/6$.  This yields a value for ${\cal G}_4$ that exactly agrees with the outcome of the numerical evaluation of the integrals defining this function. 

According to (\ref{4E82}), the variable $\eta_l$ that appears in the arguments of the functions ${\cal G}_1,\cdots,{\cal G}_4$ equals 1 when $f_{lC}\vert_{\tau=\tau_{l{\rm min}}}=0$ and equals $-1$ when $f_{lC}\vert_{\tau=\tau_{l{\rm max}}}=0$.  This holds true, as indicated by (\ref{4E83}), also for ${\bar\eta}_l$ when ${\bar f}_{lC}$ vanishes at its maximum or minimum.  Moreover, $\eta_l$ assumes an infinitely large value at the point where the maximum and minimum of $f_{lC}$ coalesce and so $\sigma_{l1}$ vanishes, an unbounded upper limit that is also approached by ${\bar\eta}_l$ as the turning points of ${\bar f}_{lC}$ coalesce to form an inflection point. 

\begin{figure}
\centerline{\includegraphics[width=14cm]{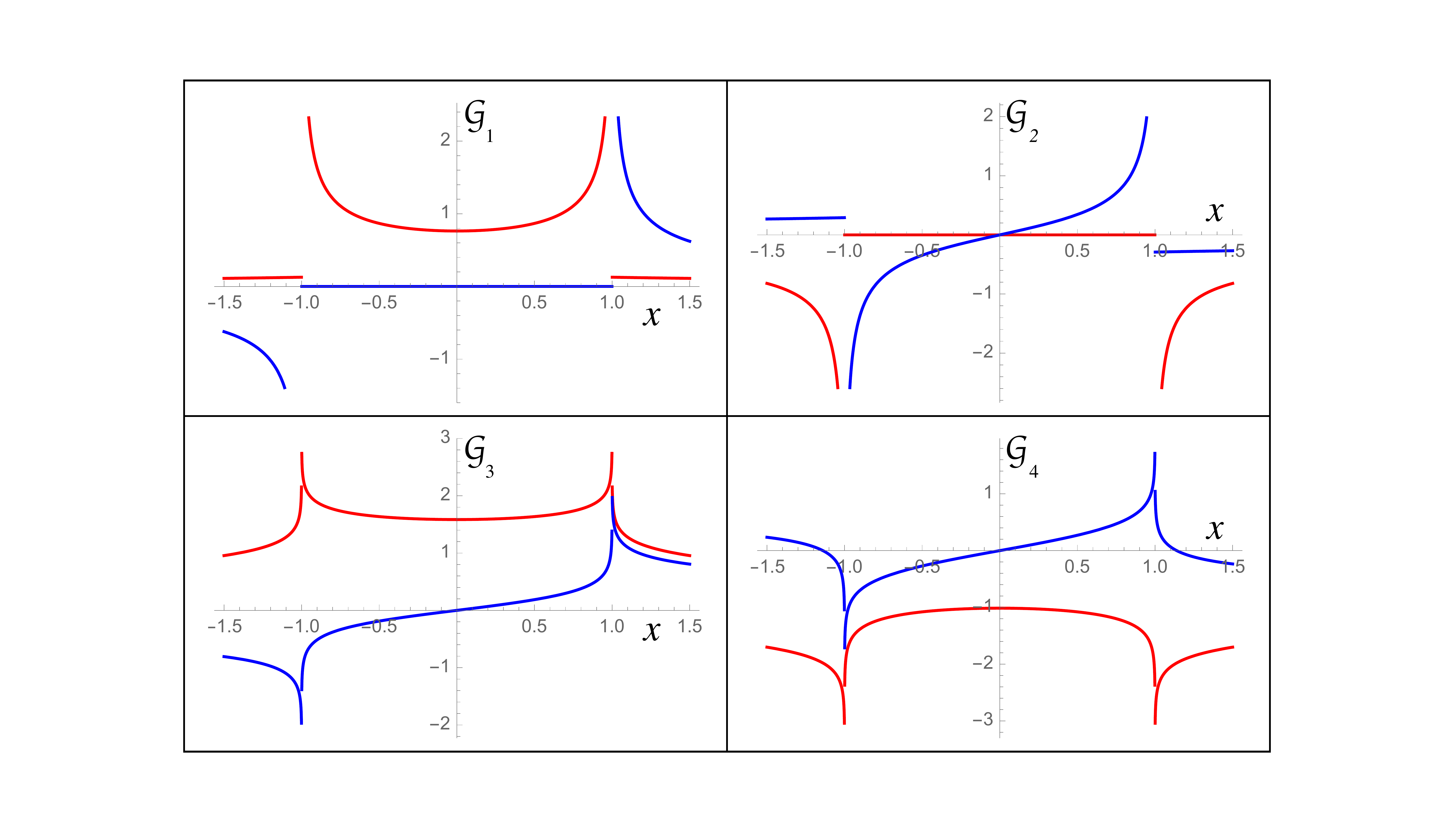}}
\caption{Real part (in red) and imaginary part (in blue) of the functions ${\cal G}_1,\cdots,{\cal G}_4$ given by (\ref{4E78})--(\ref{4E82}) for $\kappa_1=10^{-2}$ and $\kappa_2=4\times10^{-2}$.  The limiting values of these functions across their discontinuities at $x=\pm1$ are given by (\ref{4E84})--(\ref{4E87}).}
\label{PF10}
\end{figure}

\subsection{The divergence that arises from the vanishing width of the current sheet and its regularization}
\label{subsec:divergencence}

The functions ${\cal G}_1,\cdots,{\cal G}_4$ that appear in (\ref{4E76}) and (\ref{4E77}) approach the following divergent values as their argument $x$ tends to $\pm1$ from inside and outside the interval $-1<x<1$ (see figure~\ref{PF10}):  
\begin{equation}
\lim_{x\to\pm1}{\cal G}_1=(2^2\times3)^{-1/6}\left[\pm{\rm i}\,{\rm H}(\pm x-1)(x^2-1)^{-1/2}+{\rm H}(1\mp x)(1-x^2)^{-1/2}\right],
\label{4E84}
\end{equation} 
\begin{equation}
\lim_{x\to\pm1}{\cal G}_2=(2^{-4}\times3)^{1/6}\left[-{\rm H}(\pm x-1)(x^2-1)^{1/2}\pm{\rm i}\,{\rm H}(1\mp x)(1-x^2)^{-1/2}\right],
\label{4E85}
\end{equation} 
\begin{equation}
\lim_{x\to\pm1}{\cal G}_3=(2^5\times3)^{-1/6}\pi^{-1/2}\exp\left(\mp\frac{3{\rm i}\pi}{4}\right)\left[{\rm H}(\pm x-1)\ln(x^2-1)+{\rm H}(1\mp x)\ln(1-x^2)\right],
\label{4E86}
\end{equation} 
\begin{equation}
\lim_{x\to\pm1}{\cal G}_4=\frac{3^{1/6}}{2^{7/6}\pi^{1/2}}\exp\left(\mp\frac{{\rm i}\pi}{4}\right)\left[{\rm H}(\pm x-1)\ln(x^2-1)+{\rm H}(1\mp x)\ln(1-x^2)\right]
\label{4E87}
\end{equation}
(cf., \S 15.4 of~\cite{Olver}).  The radiation field ${\bf E}^{\rm uc}$ correspondingly diverges when $\eta_l=\pm1$ (or ${\bar\eta}_l=\pm1$), i.e., when $f_{lC}$ (or ${\bar f}_{lC}$) vanishes at one of its turning points.

The above singularities in the expression for the radiation field stem from assigning a zero width to the current sheet.  Because its charge and current densities are proportional to a Dirac delta function, the current sheet described by (\ref{2E18}) and (\ref{2E19}) has a vanishing thickness.  The vanishing thickness of the current sheet in turn results in an infinitely wide range of values for the variable $k$ that appears in its Fourier representation (see~(\ref{4E11})). But, given that it is created by the coordinated motion of aggregates of subluminally moving particles, a superluminally moving source is necessarily volume-distributed: it can neither be point-like nor be distributed over a line or a surface~\cite{BolotovskiiBM:Radbcm}.  In a physically more realistic model of the magnetosphere, where the processes that occur on plasma scales within the current sheet are taken into account, this sheet would have a non-zero thickness and the singularities in question would not occur.  

To circumvent the divergence that arises from overlooking the finite width of the current sheet, we will here replace the integration domain $0\le k <\infty$ in (\ref{4E61}) with $0\le k\le k_{\rm u}$ and treat the upper limit $k_{\rm u}$ $(\gg1)$ on the range of values of $\vert k\vert$ as a free parameter.  The thickness of the current sheet is dictated by microphysical processes that are not well understood: the standard Harris solution of the Vlasov-Maxwell equations that is commonly used in analysing a current sheet~\cite{Harris1962} is not applicable in the present case because the current sheet in the magnetosphere of a neutron star moves faster than light and so has no rest frame.  Introducing the upper limit $k_{\rm u}$ is tantamount to assuming that the (unknown) thickness of the current sheet is of the order of $1/k_{\rm u}$ in units of the light-cylinder radius $c/\omega$.

\begin{figure}
\centerline{\includegraphics[width=12cm]{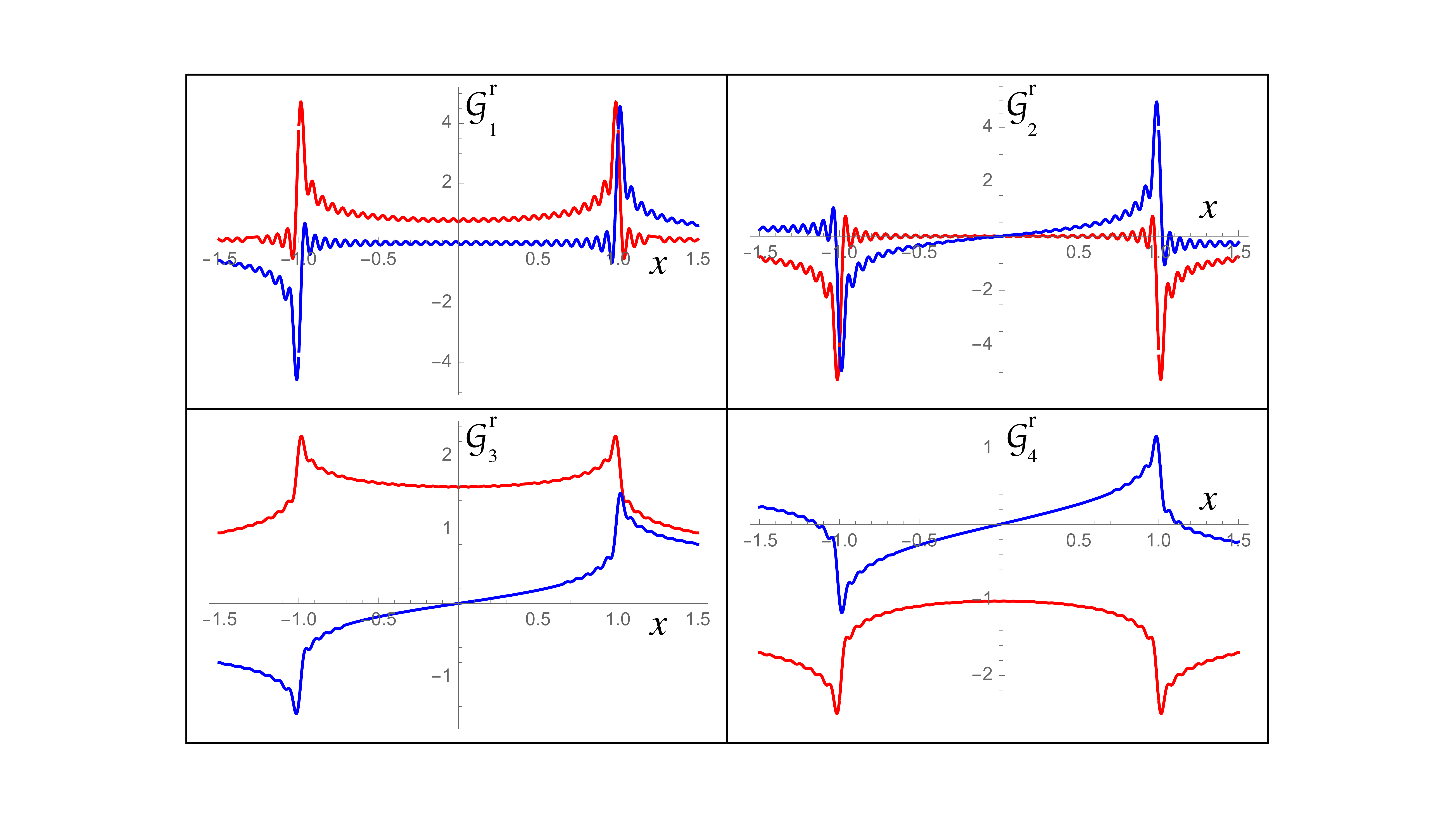}}
\caption{Real part (in red) and imaginary part (in blue) of the regularized versions (${\cal G}^{\rm r}_1,\cdots,{\cal G}^{\rm r}_4$) of the functions ${\cal G}_1,\cdots,{\cal G}_4$ for $\kappa_1=10^{-2}$, $\kappa_2=4\times10^{-2}$ and $\kappa_{\rm u}=10^2$.  The higher the value of $\kappa_{\rm u}$, the larger are the absolute values of the maxima and minima of these functions and the shorter are the wavelengths of their modulations (microstructure).  In the limit $\kappa_{\rm u}\to\infty$, these functions approach those depicted in figure~\ref{PF10}.}
\label{PF11}
\end{figure} 

The singularities that arise from $k=\infty$ can thus be regularized by (i) changing the ranges of those integrals in (\ref{4E78})--(\ref{4E82}) that extend over $0\le k<\infty$ to $0\le k\le k_{\rm u}$, (ii) writing the resulting integrals over $0\le k\le k_{\rm u}$ as
\begin{equation}
\int_0^{k_{\rm u}}{\rm d}k\cdots=\left(\int_0^\infty-\int_{k_{\rm u}}^\infty\right){\rm d}k\,\cdots,
\label{4E88}
\end{equation} 
(iii) equating the integrals over $(0,\infty)$ on the right-hand side of (\ref{4E88}) to the expressions found in (\ref{4E78})--(\ref{4E82}) and (iv) replacing the Airy functions in the integrands of the integrals over $k_{\rm u}\le k<\infty$ by the leading terms
in their asymptotic expansions for large $k$ (cf.,~\cite{Olver}).  The integrals over $k_{\rm u}\le k<\infty$ can then be performed analytically to arrive at the following regularized versions, ${\cal G}^{\rm r}_1,\cdots,{\cal G}^{\rm r}_4$, of ${\cal G}_1,\cdots,{\cal G}_4$:
\begin{equation}
{\cal G}^{\rm r}_i={\cal G}_i-{\cal G}^{\rm u}_i,\qquad i=1,\cdots,4,
\label{4E89}
\end{equation}
in which the subtracted contributions arising from $k_{\rm u}\le k<\infty$ are given by
\begin{eqnarray}
{\cal G}_1^{\rm u}&=&(2^5\times3)^{-1/6}\pi^{-1/2}\left[\exp\left(\frac{{\rm i}\pi}{4}\left[{\rm sgn}(x+1)-1\right]\right)\Gamma\left(\frac{1}{2},-{\rm i}\kappa_{\rm u}(x+1)\right)\vert x+1\vert^{-1/2}\right.\nonumber\\*
&&\left.+\exp\left(\frac{{\rm i}\pi}{4}\left[{\rm sgn}(x-1)+1\right]\right)\Gamma\left(\frac{1}{2},-{\rm i}\kappa_{\rm u}(x-1)\right)\vert x-1\vert^{-1/2}\right],
\label{4E90}
\end{eqnarray}
\begin{eqnarray}
{\cal G}_2^{\rm u}&=&-\frac{3^{1/6}{\rm i}}{2^{7/6}\pi^{1/2}}\left[\exp\left(\frac{{\rm i}\pi}{4}\left[{\rm sgn}(x+1)-1\right]\right)\Gamma\left(\frac{1}{2},-{\rm i}\kappa_{\rm u}(x+1)\right)\vert x+1\vert^{-1/2}\right.\nonumber\\*
&&\left.-\exp\left(\frac{{\rm i}\pi}{4}\left[{\rm sgn}(x-1)+1\right]\right)\Gamma\left(\frac{1}{2},-{\rm i}\kappa_{\rm u}(x-1)\right)\vert x-1\vert^{-1/2}\right],
\label{4E91}
\end{eqnarray}
\begin{eqnarray}
{\cal G}_3^{\rm u}&=&(2^5\times3)^{-1/6}\pi^{-1/2}\exp\left(\frac{3{\rm i}\pi}{4}\right)\left\{{\rm Ci}\left(\kappa_{\rm u}\vert x+1\vert\right)-{\rm i}\,{\rm sgn}(x+1)\left[\frac{\pi}{2}-{\rm Si}\left(\kappa_{\rm u}\vert x+1\vert\right)\right]\right.\nonumber\\*
&&\left.+{\rm i}\,{\rm Ci}\left(\kappa_{\rm u}\vert x-1\vert\right)+{\rm sgn}(x-1)\left[\frac{\pi}{2}-{\rm Si}\left(\kappa_{\rm u}\vert x-1\vert\right)\right]\right\},
\label{4E92}
\end{eqnarray}
\begin{eqnarray}
{\cal G}_4^{\rm u}&=&\frac{3^{1/6}}{2^{7/6}\pi^{1/2}}\exp\left(-\frac{3{\rm i}\pi}{4}\right)\left\{-{\rm Ci}\left(\kappa_{\rm u}\vert x+1\vert\right)+{\rm i}\,{\rm sgn}(x+1)\left[\frac{\pi}{2}-{\rm Si}\left(\kappa_{\rm u}\vert x+1\vert\right)\right]\right.\nonumber\\*
&&\left.+{\rm i}\,{\rm Ci}\left(\kappa_{\rm u}\vert x-1\vert\right)+{\rm sgn}(x-1)\left[\frac{\pi}{2}-{\rm Si}\left(\kappa_{\rm u}\vert x-1\vert\right)\right]\right\}.\label{4E93}
\end{eqnarray}
Here, $\kappa_{\rm u}={\textstyle\frac{2}{3}}\sigma_{l1}^3k_{\rm u}$ and ${\rm Ci}$ and ${\rm Si}$ are the cosine and sine integrals, respectively.  It can be easily verified that the singularities of the functions ${\cal G}^{\rm u}_i$ at $x=\pm1$ are the same as those of the functions ${\cal G}_i$ that were derived in (\ref{4E78})--(\ref{4E82}) and so cancel out in ${\cal G}_i^{\rm r}$ when ${\cal G}_i^{\rm u}$ are subtracted from ${\cal G}_i$ (see figure~\ref{PF11}).  

The following singularity-free expressions for the two parts ${\bf E}^{\rm uc}_{\cal P}$ and ${\bf E}^{\rm uc}_{\cal P}$ of the radiation field ${\bf E}^{\rm uc}$ are thus obtained by replacing the functions ${\cal G}_i$ in (\ref{4E76}) and (\ref{4E77}) by their regularized versions ${\cal G}_i^{\rm r}$:\begin{eqnarray}
{\bf E}^{\rm uc}_{\cal P}&\simeq&\frac{2^{4/3}\pi j_0}{3^{1/3}\omega}\sum_{l=1}^2\left\{\sigma_{l1}^{-2}{\cal G}^{\rm r} _1(\eta_l,\kappa_{\rm u}) \mbox{{\boldmath${\cal P}$}}^{(0)}_l+\left(\frac{2}{3\sigma_{l1}}\right)^{1/2}\left[{\cal G}^{\rm r}_3(\eta_l,\kappa_1,\kappa_{\rm u}) \mbox{{\boldmath${\cal P}$}}^{(1)}_l+{\cal G}^{\rm r}_3(\eta_l,\kappa_2,\kappa_{\rm u}) \mbox{{\boldmath${\cal P}$}}^{(2)}_l\right]\right.\nonumber\\*
&&\left.+\left(\frac{2}{3{\bar\sigma}_{l1}}\right)^{1/2}\left[{\cal G}^{\rm r}_3({\bar\eta}_l,{\bar\kappa}_1,{\bar\kappa}_{\rm u}) \mbox{{\boldmath${\bar{\cal P}}$}}^{(1)}_l+{\cal G}^{\rm r}_3({\bar\eta}_l,{\bar\kappa}_2,{\bar\kappa}_{\rm u}) \mbox{{\boldmath${\bar{\cal P}}$}}^{(2)}_l\right]\right\}
\label{4E94}
\end{eqnarray}
and
\begin{eqnarray}
{\bf E}^{\rm uc}_{\cal Q}&\simeq&\frac{2^{5/3}\pi j_0}{3^{2/3}{\rm i}\,\omega}\sum_{l=1}^2\left\{\sigma_{l1}^{-1}{\cal G}^{\rm r}_2(\eta_l,\kappa_{\rm u}) \mbox{{\boldmath${\cal Q}$}}^{(0)}_l+\left(\frac{2\sigma_{l1}}{3}\right)^{1/2}\left[{\cal G}^{\rm r}_4(\eta_l,\kappa_1,\kappa_{\rm u}) \mbox{{\boldmath${\cal Q}$}}^{(1)}_l+{\cal G}^{\rm r}_4(\eta_l,\kappa_2,\kappa_{\rm u}) \mbox{{\boldmath${\cal Q}$}}^{(2)}_l\right]\right.\nonumber\\*
&&\left.+\left(\frac{2{\bar\sigma}_{l1}}{3}\right)^{1/2}\left[{\cal G}^{\rm r}_4({\bar\eta}_l,{\bar\kappa}_1,{\bar\kappa}_{\rm u}) \mbox{{\boldmath${\bar{\cal Q}}$}}^{(1)}_l+{\cal G}^{\rm r}_4({\bar\eta}_l,{\bar\kappa}_2,{\bar\kappa}_{\rm u}) \mbox{{\boldmath${\bar{\cal Q}}$}}^{(2)}_l\right]\right\},
\label{4E95}
\end{eqnarray}
where ${\bar\kappa}_{\rm u}={\textstyle\frac{2}{3}}{\bar\sigma}_{l1}^3k_{\rm u}$.  The right-hand sides of (\ref{4E94}) and (\ref{4E95}) are indeterminate at points where the maximum and minimum of $f_{lC}$ (or ${\bar f}_{lC}$) coalesce and $\sigma_{l1}$ and $\eta_l$ (or ${\bar\sigma}_{l1}$ and ${\bar\eta}_l$) respectively assume the values zero and infinity simultaneously.  The limiting values of ${\bf E}^{\rm uc}_{\cal P}$ and ${\bf E}^{\rm uc}_{\cal P}$ at such points, given by l'H\^opital's rule, are however finite and regular.  Note that in the limit $k_{\rm u}\to\infty$, the functions ${\cal G}^{\rm r}_1$ and ${\cal G}^{\rm r}_2$ diverge as $k_{\rm u}^{1/2}$ while  the functions ${\cal G}^{\rm r}_3$ and ${\cal G}^{\rm r}_4$ diverge as $\ln(k_{\rm u})$.

\section{Characteristics of the resulting radiation}
\label{sec:characteristics}
\setcounter{equation}{0}
\setcounter{figure}{0}

\subsection{Pulse profiles and polarization position angles}
\label{subsec:profiles}

As a result of receiving contributions from multiple stationary points of the phase functions $f_{lC}$ and ${\bar f}_{lC}$, the unconventional component ${\bf E}^{\rm uc}$ of the radiation field ${\bf E}$ has an amplitude that exceeds that of the conventional component ${\bf E}^{\rm v}$ of this field by many orders of magnitude (see (\ref{3E40})--(\ref{3E42}) and (\ref{4E28}) and the figures below).  We can therefore calculate the Stokes parameters of the present radiation from the expressions
\begin{equation}
I=\vert E_{\varphi_P}^{\rm uc}\vert^2+ \vert E_{\theta_P}^{\rm uc}\vert^2,\quad Q=\vert E_{\varphi_P}^{\rm uc}\vert^2- \vert E_{\theta_P}^{\rm uc}\vert^2, 
\label{5E1}
\end{equation}
\begin{equation}
U=2\Re\left(E_{\varphi_P}^{\rm uc} {E_{\theta_P}^{\rm uc}}^*\right),\quad V=-2\Im\left(E_{\varphi_P}^{\rm uc}{E_{\theta_P}^{\rm uc}}^*\right),
\label{5E2}
\end{equation}
\begin{equation}
L=(Q^2+U^2)^{1/2},\quad \psi=\frac{1}{2}\arctan\frac{U}{Q},
\label{5E3}
\end{equation}
based on the spherical components $E_{\varphi_P}^{\rm uc}$ and $E_{\theta_P}^{\rm uc}$ of ${\bf E}^{\rm uc}$ alone.  (The superscript $*$ in the above expressions denotes complex conjugation.) 

It can be seen from (\ref{4E75}) and (\ref{4E94})--(\ref{4E95}) that the unconventional radiation field ${\bf E}^{\rm uc}$ consists of the sum of two distinct parts:  one part, ${\bf E}^{\rm uc}_{\cal P}$, depending on the vectors $\mbox{{\boldmath${\cal P}$}}^{(j)}_l$ and $\mbox{{\boldmath${\bar{\cal P}}$}}^{(j)}_l$ and another part, ${\bf E}^{\rm uc}_{\cal Q}$, depending on the vectors $\mbox{{\boldmath${\cal Q}$}}^{(j)}_l$ and $\mbox{{\boldmath${\bar{\cal Q}}$}}^{(j)}_l$.  As suggested by the occurrence of the factor ${\rm i}$ in (\ref{4E95}), these two parts turn out to be out of phase with one another by approximately $\pi/2$.  By calculating the Stokes parameters for ${\bf E}^{\rm uc}_{\cal P}$ and ${\bf E}^{\rm uc}_{\cal Q}$ separately, we will show here that the polarization position angles associated with these two fields are approximately orthogonal to one another in general.  Accordingly, the two distinct parts of the radiation field defined by ${\bf E}^{\rm uc}_{\cal P}$ and ${\bf E}^{\rm uc}_{\cal Q}$ are respectively referred to in this paper as the $P$ and $Q$ polarization modes.  

In this section we begin with evaluating the Stokes parameters $(I,V,L)$ of the radiation analysed in \S 4 in units of 
\begin{equation}
I_0=\left(\frac{B_0{\hat r}_{s0}^2}{{\hat R}_P}\right)^2
\label{5E4}
\end{equation}
(see (\ref{2E22}) and (\ref{4E94})--(\ref{4E95})) as functions of the longitude $\varphi_P$ at ${\hat R}_P\gg1$ for various values of the colatitude $\theta_P$, the inclination angle $\alpha$ and the lower bound $k_{\rm u}^{-1}$ on the width (in units of the light-cylinder radius) of the current sheet.  We will evaluate the exact expression for the radiated field at a suitably large value of ${\hat R}_P$, rather than proceeding to the far-field limit $R_P\to\infty$, because as we have already pointed out in \S~\ref{subsec:critical} the relative positions of the stationary points of the phase functions $f_{lC}$ and ${\bar f}_{lC}$ depend on ${\hat R}_P$ at colatitudes for which the maxima and minima of these functions are close to one another.  However, away from such colatitudes the shapes of the pulse profiles and position-angle distributions we will be discussing do not change perceptibly with distance once the value of ${\hat R}_P$ exceeds $10^6$.  We will therefore use ${\hat R}_P=10^6$ for plotting most of the figures in this section.  

The number of components of a pulse profile is determined by the total number of stationary points of the four phase functions $f_{1C}$, $f_{2C}$, ${\bar f}_{1C}$ and ${\bar f}_{2C}$ described in \S~\ref{subsec:critical} and the number of values of $n$ (i.e., the number of cycles of retarded time) that contribute to the radiation received during a single period of observation time.  The longitudinal interval occupied by various components of a pulse profile is determined by the separation between the $\tau$ coordinates of the stationary points of these phase functions.  To be able to depict the pulse profiles over the entire longitudinal intervals occupied by their various components while displaying the finite width of each component, we will plot the profiles of high-intensity pulses for $k_{\rm u}=10^2$.  It should be borne in mind, however, that intensity at the peaks of the main pulses in these figures is a linearly increasing function of $k_{\rm u}$ (see the last paragraph of \S~\ref{subsec:divergencence}).  The higher the value of $k_{\rm u}$ is, the narrower are the rapid low-amplitude modulations (microstructure) of the pulse profile.  For $k_{\rm u}=10^4$ and higher, these modulations are too sharp and dense to show up in most of the figures plotted here.  We will also adopt those branches of  the multi-valued function arctan appearing in (\ref{5E3}) that yield continuous polarization position-angle distributions across various components of a given pulse. 

In most of the examples given below, the inclination angle of the magnetic axis and the colatitude of the observation point are set in the upper hemisphere $0 <\theta_P< 90^\circ$.  At any given value of the inclination angle $\alpha$, the pulse observed at $180^\circ-\theta_P$ differs from that observed at $\theta_P$ only in that the intensity $V$ of its circularly polarized part is replaced by $-V$ and the longitude $\varphi_P$ is replaced by $\varphi_P+180^\circ$.  Moreover, the results for $\alpha > 90^\circ$ follow from those for $\alpha < 90^\circ$ by replacing $\theta_P$, $\varphi_P$ and $V$ by  $180^\circ-\theta_P$, $\varphi_P+180^\circ$ and $-V$, respectively (see \S~\ref{subsec:critical}).  Note also that in these examples the range of values of the azimuthal coordinate $\varphi_P$ that is spanned by the observation point across the pulse window differs from that given by (\ref{4E94})--(\ref{4E95}) for any $n$ (see also (\ref{4E21})): the origin of this coordinate is shifted in each case to place the starting point of the plotted pulse profile at $\varphi_P=0$. 

\begin{figure}
\centerline{\includegraphics[width=16cm]{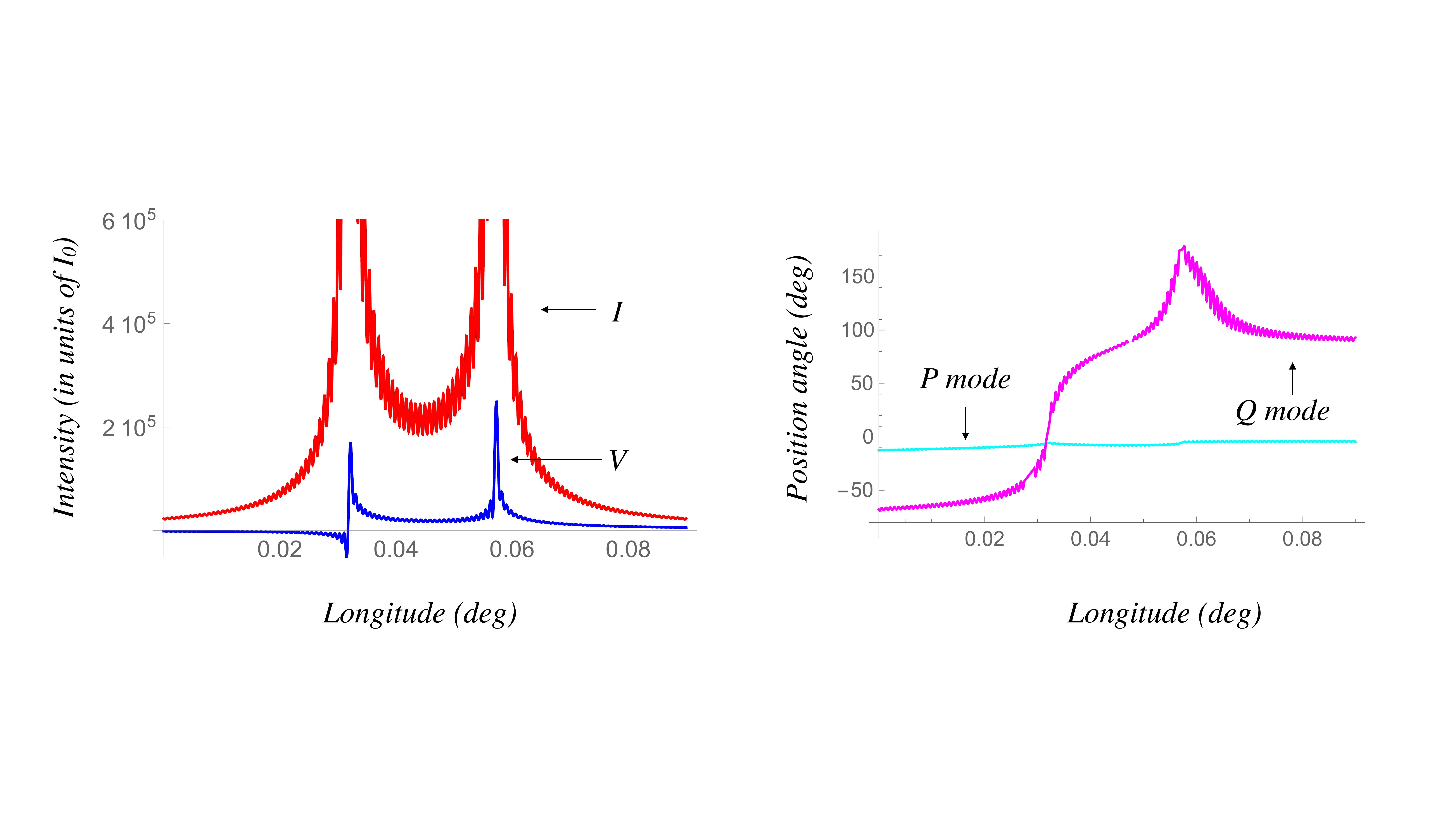}}
\caption{The Stokes parameters $I$, $V$ and the position angles $\psi$ of the polarization modes $P$ and $Q$ at an observation point with the colatitude $\theta_P=5^\circ$ for the inclination angle $\alpha=5^\circ$ and $k_{\rm u}=10^2$.  In this case, only the stationary points of $f_{C2}$ and ${\bar f}_{C2}$ contribute toward the field.  The large value of the intensity and the short duration of this pulse stems from the proximity of the maximum and minimum of $f_{2C}$.  At its peak, the right-hand component of this pulse has the intensity $I=1.05\times10^{12}I_0$ and the longitudinal width $6.76\times{10^{-9}}$ second when $k_{\rm u}$ is $10^7$.  This is the only pulse occurring in the entire pulse window.}
\label{PF12}
\end{figure} 

\begin{figure}
\centerline{\includegraphics[width=16cm]{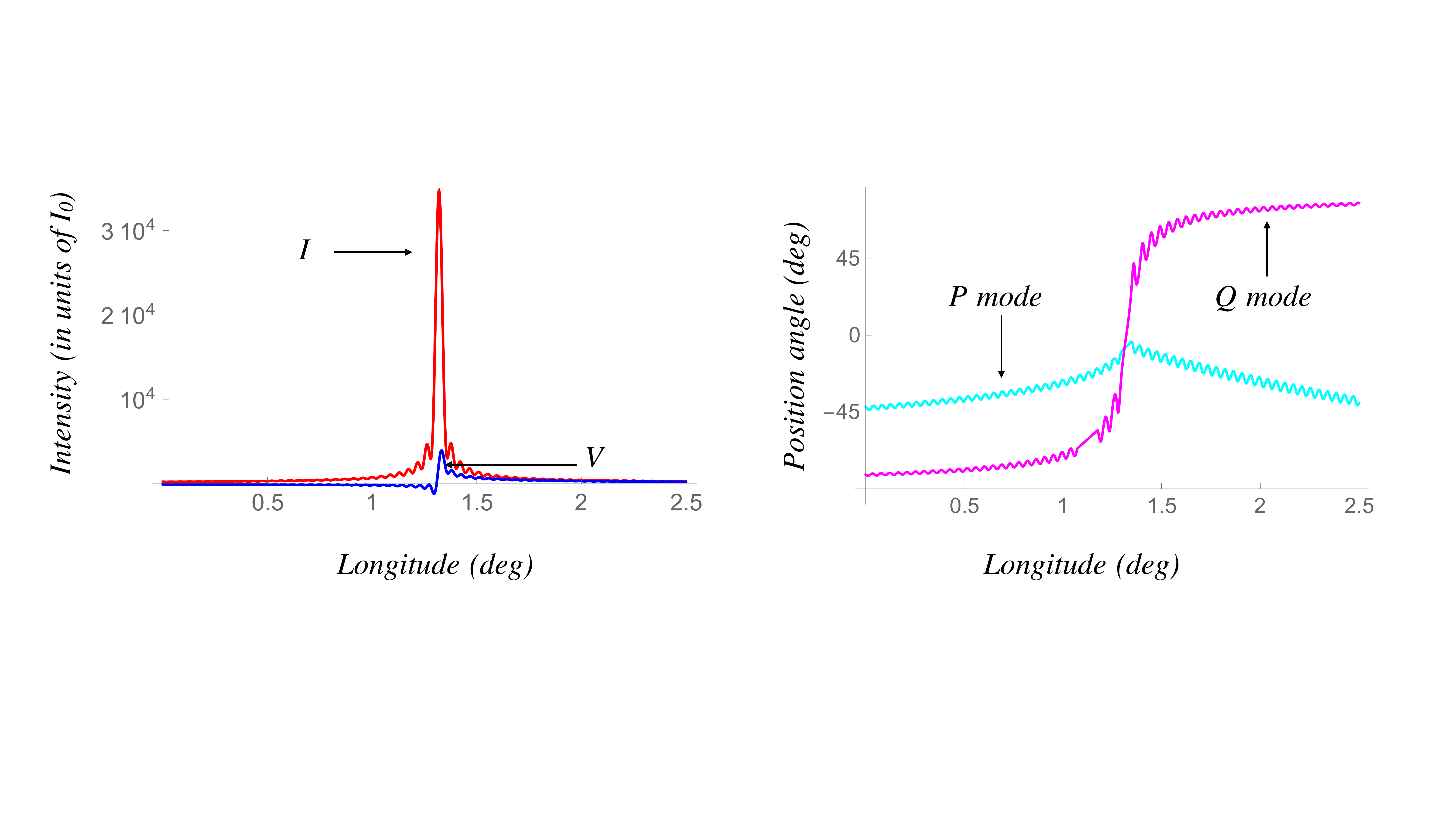}}
\caption{The Stokes parameters $I$, $V$ and the position angles $\psi$ of the polarization modes $P$ and $Q$ at an observation point with the colatitude $\theta_P=2.5^\circ$ for the inclination angle $\alpha=5^\circ$ and $k_{\rm u}=10^4$.  In this case, only the stationary points of $f_{C2}$ and ${\bar f}_{C2}$ contribute toward the field.  The pulse window encompasses another similar pulse at a longitudinal distance of about $140^\circ$ from this one.  Note that not only does $V$ change sign across the pulse profile but also the position angle of the $Q$ mode swings through $180^\circ$ across the depicted interval and differs from that of the $P$ mode by approximately $90^\circ$ on the right and $45^\circ$ on the left of the pulse.}
\label{PF13}
\end{figure} 

\begin{figure}
\centerline{\includegraphics[width=16cm]{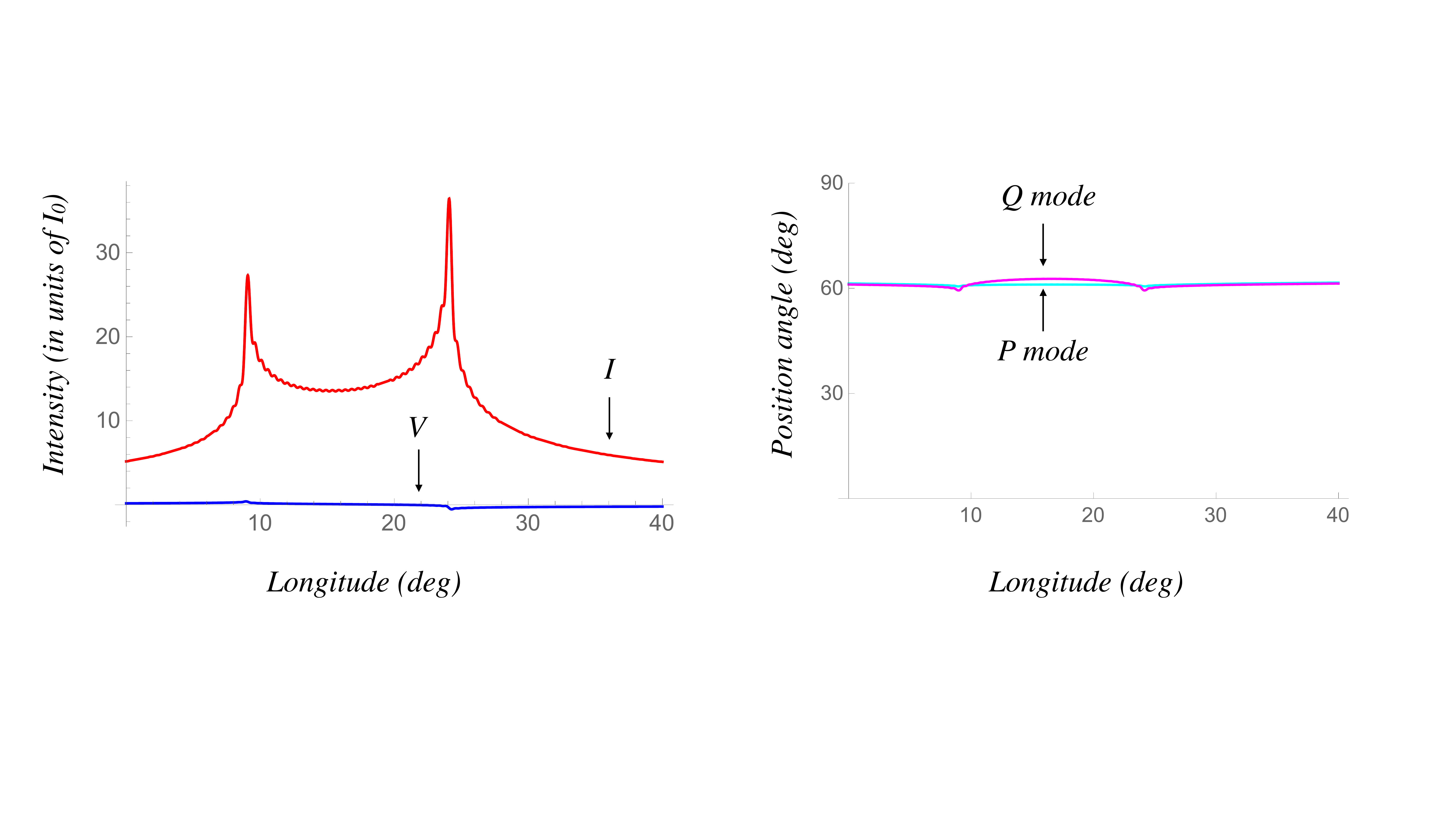}}
\caption{The Stokes parameters $I$, $V$ and the position angles $\psi$ of the polarization modes $P$ and $Q$ at an observation point with the colatitude $\theta_P=20^\circ$ for the inclination angle $\alpha=5^\circ$ and $k_{\rm u}=10^2$.  In this case, the stationary points of ${\bar f}_{C2}$ alone contribute toward the field.  Although not easily discernible because of the low value of its intensity $V$, circular polarization reverses sense across the right-hand component of this pulse.  In contrast to those shown in the previous figures, on the other hand, the position angles of the two polarization modes are essentially coincident in this case.  This is the only pulse occurring in the entire pulse window.}
\label{PF14}
\end{figure} 

\begin{figure}
\centerline{\includegraphics[width=16cm]{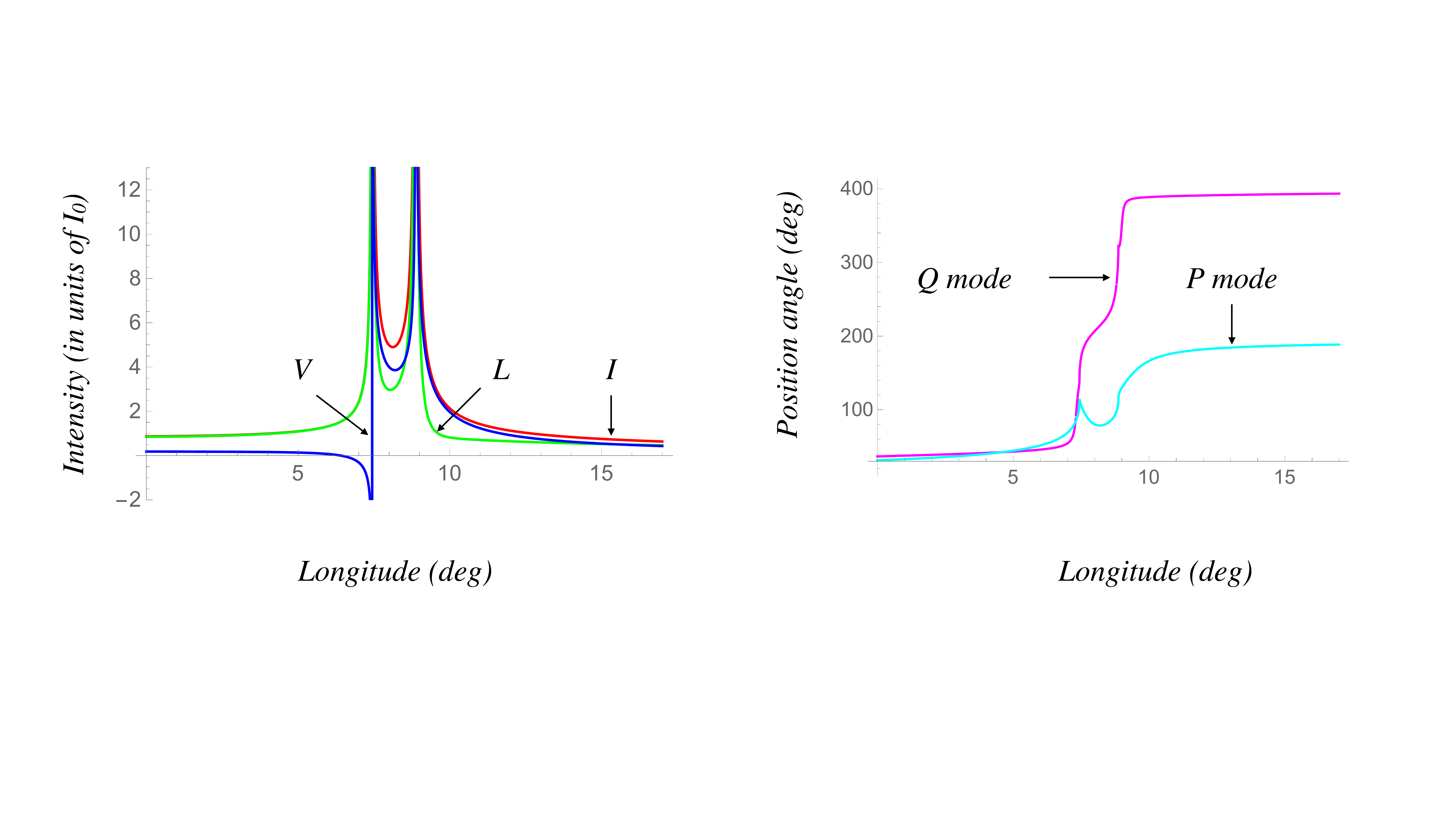}}
\caption{The Stokes parameters $(I, V, L)$ and the position angles $\psi$ of the polarization modes $P$ and $Q$ at an observation point with the colatitude $\theta_P=55^\circ$ for the inclination angle $\alpha=45^\circ$ and $k_{\rm u}=10^2$.  In this case, only the stationary points of $f_{C2}$ and ${\bar f}_{C2}$ contribute toward the field.  Note that the polarization of this pulse changes from linear to circular across it and the position angle of the Q mode swings through $180^\circ$ across each one of its two components.  The pulse window encompasses another pulse at a longitudinal distance of about $115^\circ$ from this one}
\label{PF15}
\end{figure} 

\begin{figure}
\centerline{\includegraphics[width=16cm]{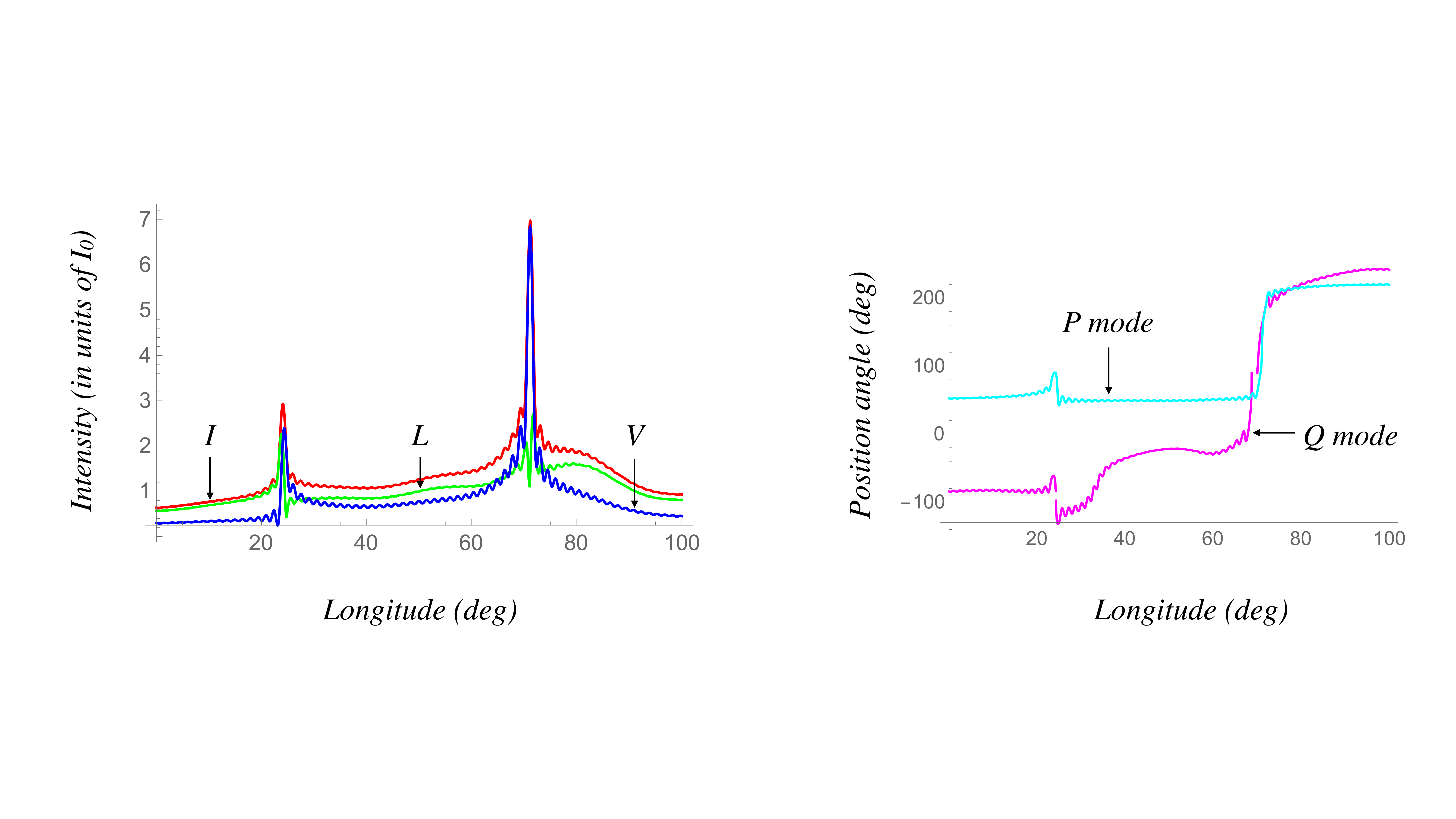}}
\caption{The Stokes parameters $(I, V, L)$ and the position angles $\psi$ of the polarization modes $P$ and $Q$ at an observation point with the colatitude $\theta_P=35^\circ$ for the inclination angle $\alpha=45^\circ$ and $k_{\rm u}=10^2$.  In this case, only the stationary points of $f_{C2}$ and ${\bar f}_{C2}$ contribute toward the field.  Position angle of the $Q$ mode turns through $360^\circ$ across the narrow gap at longitude $70^\circ$ in the curve depicting the distribution of the position angle of this polarization mode.  These are the only pulses occurring in the entire pulse window.}
\label{PF16}
\end{figure} 

\begin{figure}
\centerline{\includegraphics[width=16cm]{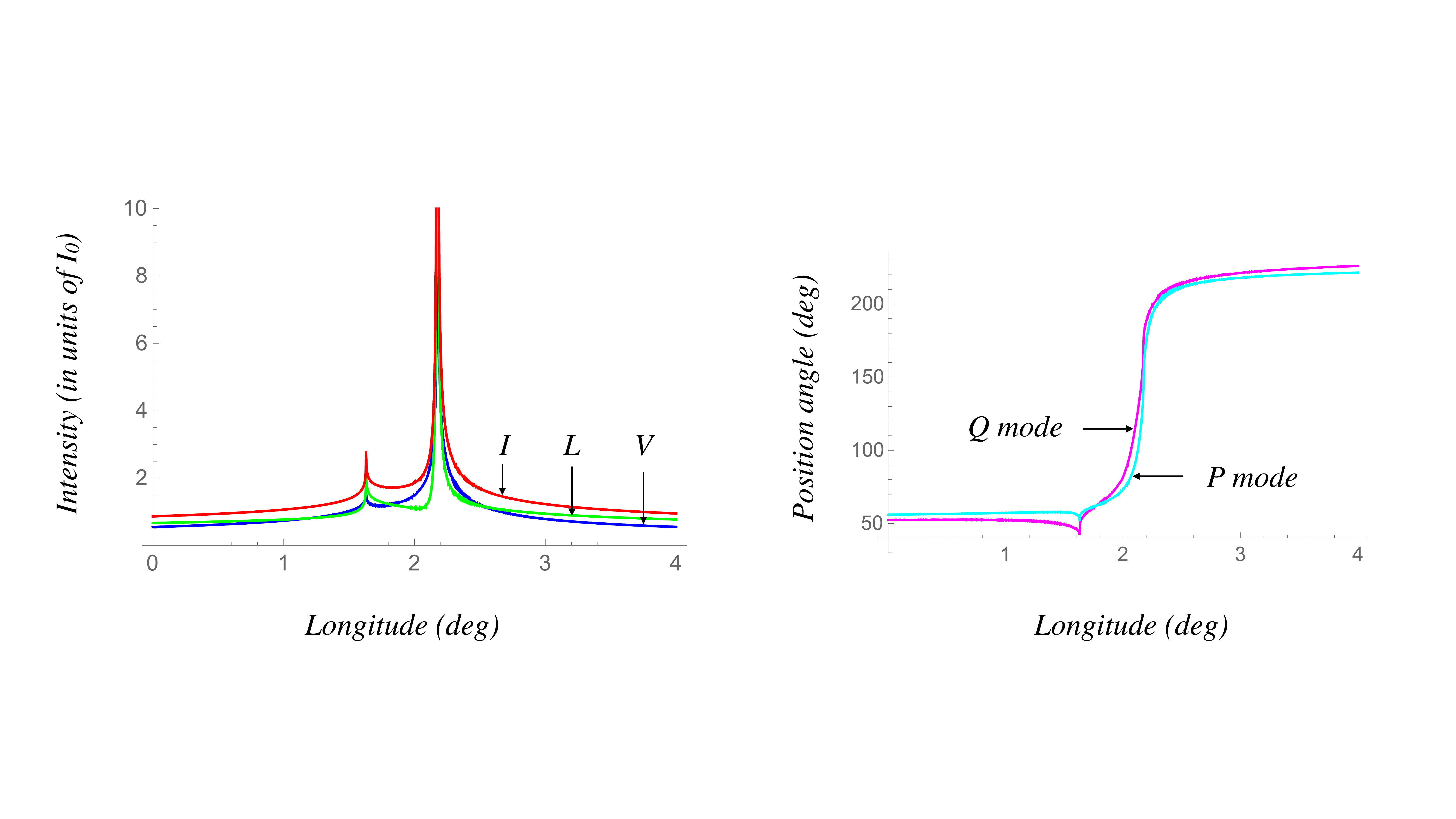}}
\caption{The Stokes parameters $(I, V, L)$ and the position angles $\psi$ of the polarization modes $P$ and $Q$ at an observation point with the colatitude $\theta_P=30^\circ$ for the inclination angle $\alpha=65^\circ$ and $k_{\rm u}=10^2$.  In this case, only the stationary points of $f_{C2}$ and ${\bar f}_{C2}$ contribute toward the field.  Note that the position angles of both modes swing through $180^\circ$ across the right-hand component of the above pulse.  Note also the high degree of circular polarization of the pulse throughout the depicted longitudes.  The pulse window encompasses in addition a weaker pulse at a longitudinal distance of about $20^\circ$ from these ones.}
\label{PF17}
\end{figure} 

\begin{figure}
\centerline{\includegraphics[width=16cm]{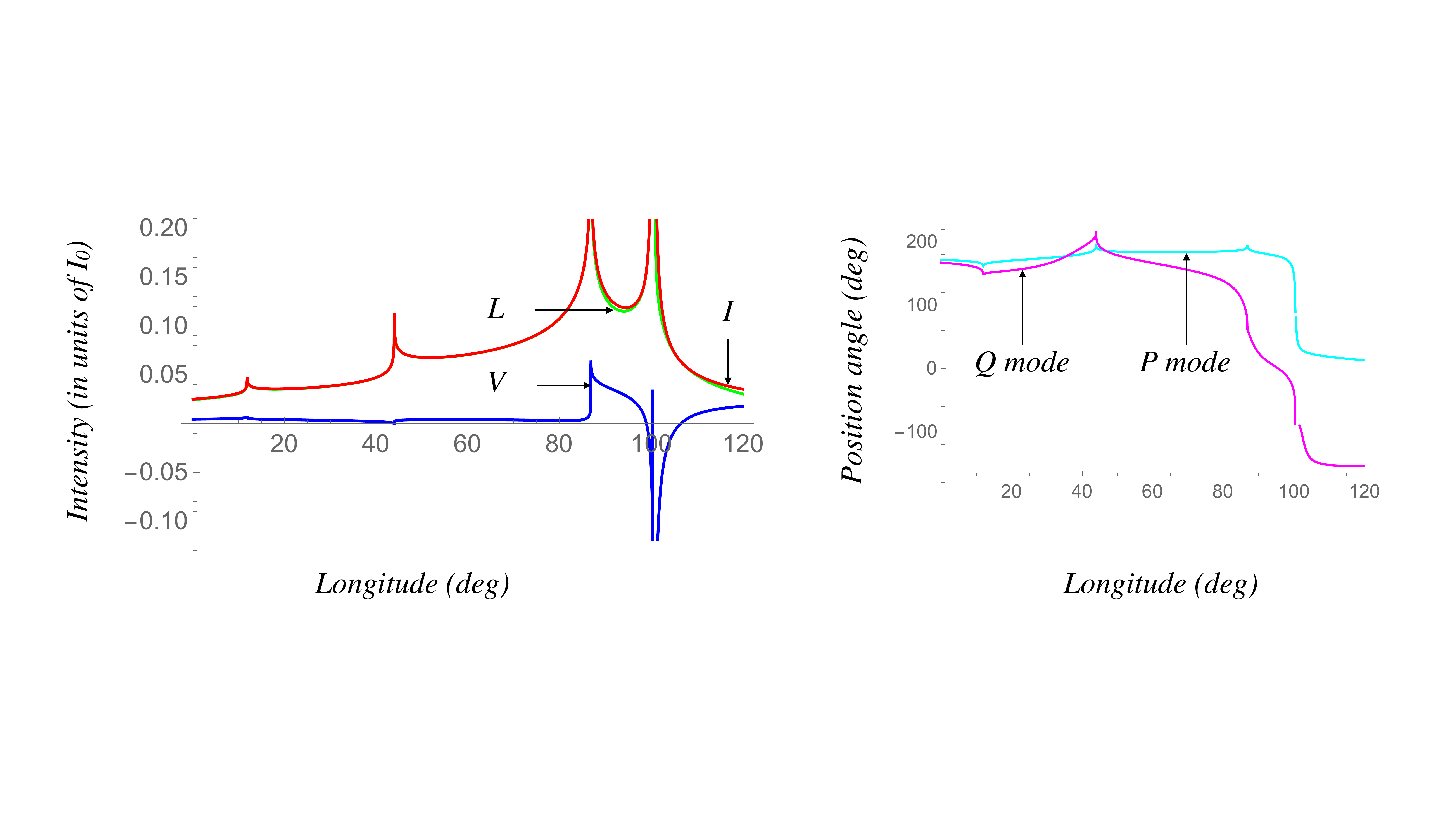}}
\caption{The Stokes parameters $(I, V, L)$ and the position angles $\psi$ of the polarization modes $P$ and $Q$ at an observation point with the colatitude $\theta_P=77.5^\circ$ for the inclination angle $\alpha=65^\circ$ and $k_{\rm u}=10^7$.  In this case, the stationary points of all of the phase functions ($f_{C1}$, ${\bar f}_{C2}$, ${\bar f}_{C1}$, and ${\bar f}_{C2}$) contribute toward the field.  Position angle of the $Q$ mode turns through $360^\circ$ across the narrow gap at longitude $100^\circ$ in the curve depicting the distribution of the position angle of this polarization mode.  Note the approximate orthogonality of the position angles of the two modes across the right-hand components of the pulse and their approximate coincidence across the left-hand components.}
\label{PF18}
\end{figure} 

\begin{figure}
\centerline{\includegraphics[width=16cm]{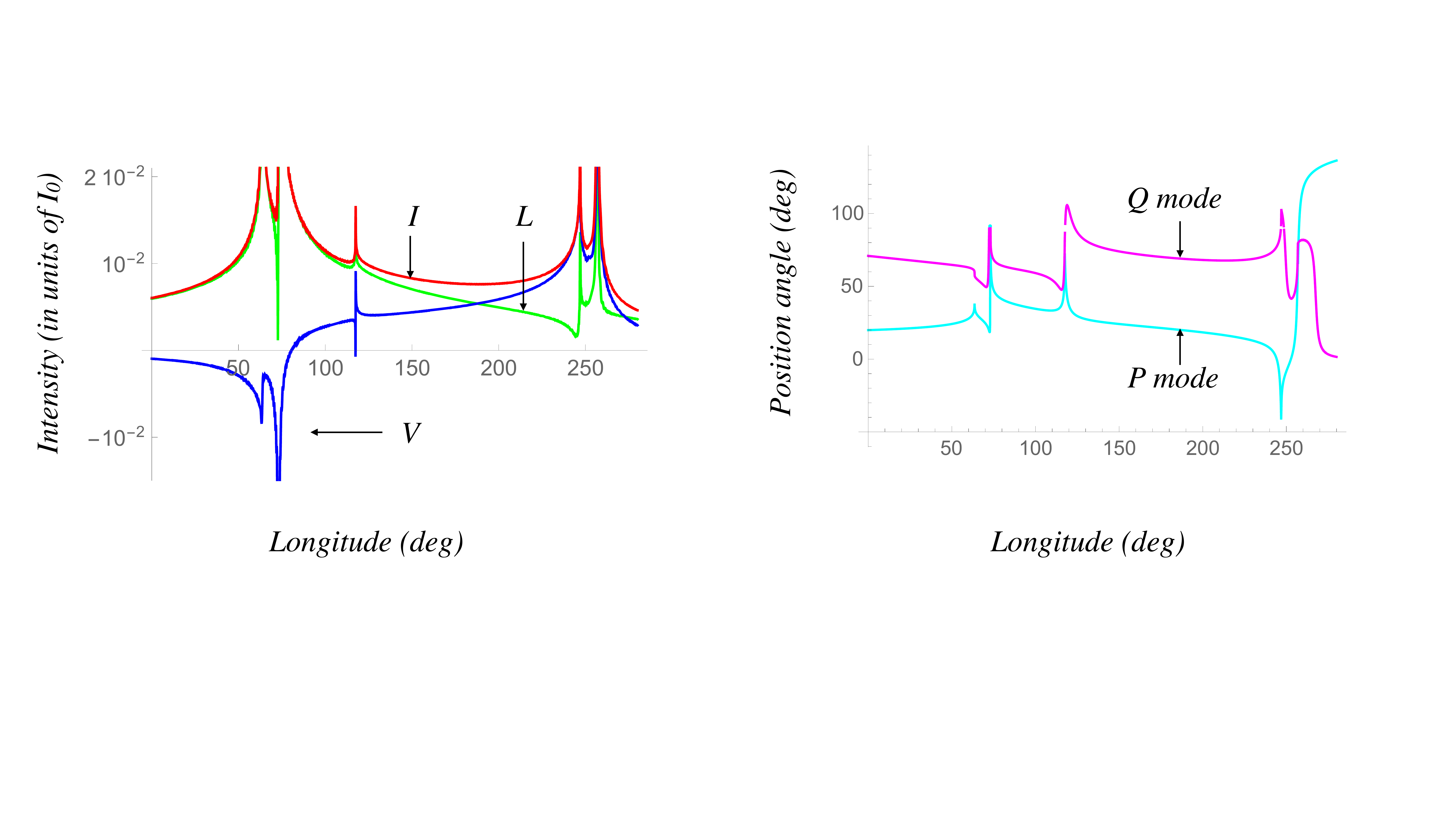}}
\caption{The Stokes parameters $(I, V, L)$ and the position angles $\psi$ of the polarization modes $P$ and $Q$ at an observation point with the colatitude $\theta_P=110^\circ$ for the inclination angle $\alpha=80^\circ$, $k_{\rm u}=10^4$ (pulse profile) and $k_{\rm u}=10^7$ (position angle).  In this case, the stationary points of all of the phase functions ($f_{C1}$, ${\bar f}_{C2}$, ${\bar f}_{C1}$, and ${\bar f}_{C2}$) contribute toward the field.  This is an example of a case in which the Stokes parameters are comparable in magnitude over some longitudinal intervals and the pulse profile entails several widely separated components.  It is an example also of a multi-component pulse the position angles of whose different components have differing longitudinal variations.}
\label{PF19}
\end{figure} 

\begin{figure}
\centerline{\includegraphics[width=16cm]{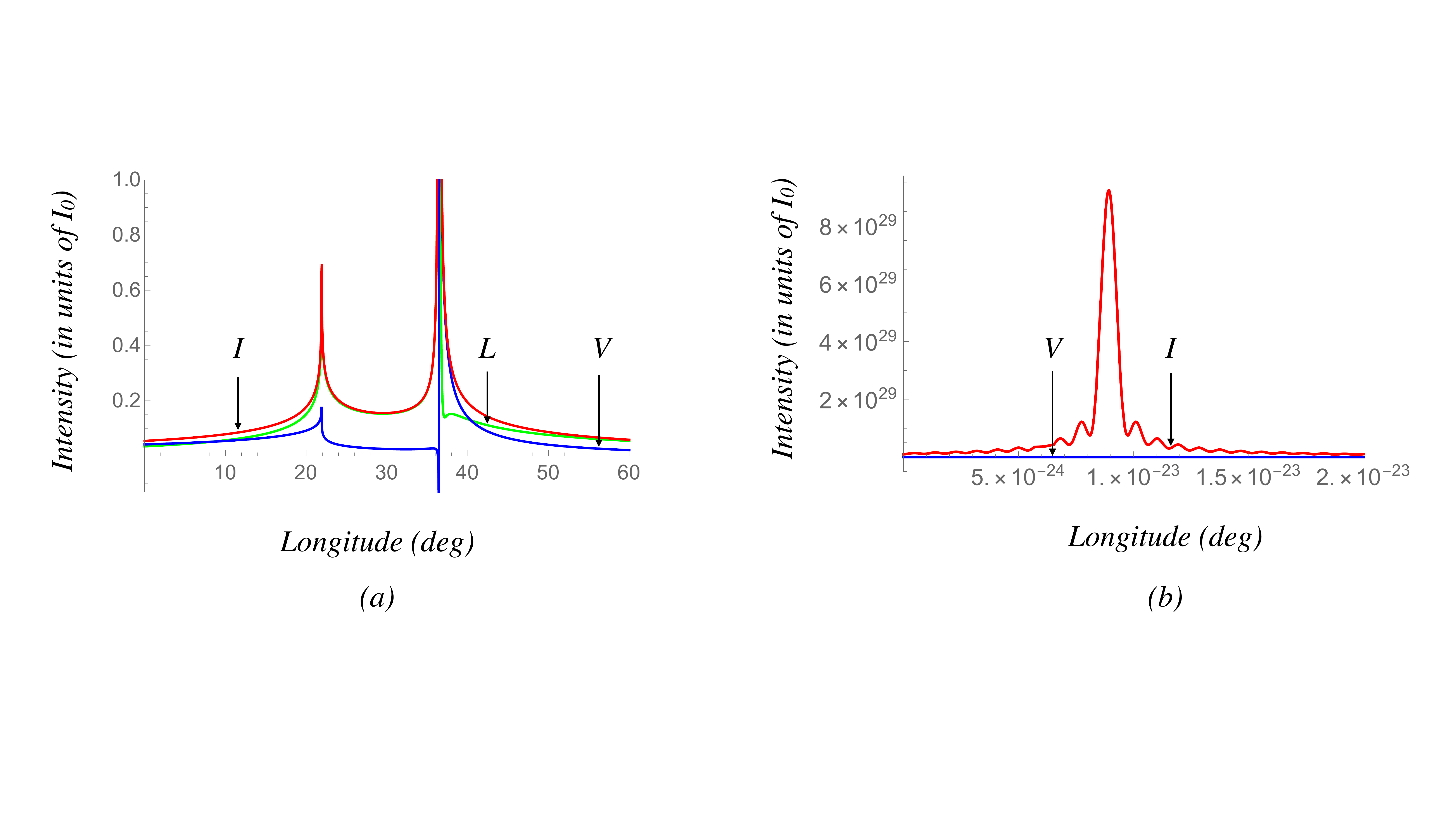}}
\caption{({\it a}) The Stokes parameters $(I, V, L)$ at an observation point with the coordinates ${\hat R}_P=10^{13}$ and $\theta_P=90^\circ$ for the inclination angle $\alpha=60^\circ$ and $k_{\rm u}=10^7$.  In this case, the stationary points of all of the phase functions ($f_{C1}$, ${\bar f}_{C2}$, ${\bar f}_{C1}$, and ${\bar f}_{C2}$) contribute toward the field.  ({\it b}) The right-hand component of the pulse depicted in {\it a} is here plotted over a sufficiently short longitudinal interval to resolve its peak and width.  The values of $\alpha$, $\theta_P$, $k_{\rm u}$ and distance in {\it b} are the same as in {\it a} but the origin of longitude is shifted in {\it b} for clarity.  The shape of the pulse depicted in {\it b} is the same in all cases of this type.  The extraordinary values of the amplitudes and widths of such pulses, illustrated by the above example, are what underpin the high brightness temperatures and broad frequency spectra of the radiation generated by the current sheet.  (For the value of the azimuthal coordinate $\varphi_P$ of the peak, see the text.)}
\label{PF20}
\end{figure} 

Some examples of the linearly and circularly polarized intensity and polarization position angle distributions of the pulses described by (\ref{4E75}), (\ref{4E94}) and (\ref{4E95}) are shown in figures~\ref{PF12}--\ref{PF19}.

As suggested by figure~\ref{PF20}, a radically different type of pulse is detected when the colatitude $\theta_P$ of the observation point has a value close to (or equal to) $\theta_{PlS}({\hat R}_P,\alpha)$ or ${\bar\theta}_{PlS}({\hat R}_P,\alpha)$ for which the extrema of one (or more) of the phase functions $f_{lC}$ and ${\bar f}_{lC}$ coalesce into an inflection point (see \S~\ref{subsec:critical}).  In the example plotted in figure~\ref{PF20}, the critical angle $\theta_{P2S}$ happens to lie within a distance of the order of $1/{\hat R}_P$ radians from the longitude $\pi/2$ of the observation point.  When sampled over a wide longitudinal interval, the profile that is shown in part $a$ of figure~\ref{PF20} does not radically differ from those of the pulses shown in other figures.  The exceptionally high intensity and narrow width of the right-hand component of this pulse at its peak, i.e., what is depicted in part $b$ of figure~\ref{PF20}, shows up only when the exact position of this peak is resolved.  Here, we have set
${\hat R}_P= 10^{13}$ and have determined the value $\varphi_P=134.80887902374020168766219^\circ$ (where $n=2$ in (\ref{4E21})) of the azimuthal coordinate of this peak graphically: by plotting the distribution of the Stokes parameter $I$ over successively shorter longitudinal intervals centred on the peak of the distribution until the maximum value of $I$ stops growing.  This zooming in procedure reveals not only the pulse shown in figure~\ref{PF20}({\it b}) but also another as narrow and as intense pulse at a longitudinal distance of $4\times10^{-18}$ deg from it.  In addition, the coincidence of the limiting values of $\theta_{P1S}$ and ${\bar\theta}_{P2S}$ for $R_P\to\infty$ results, in this case, in two identical pulses whose longitudes are separated by $180^\circ$.

The features exhibited by the pulse in figure~\ref{PF20} can be inferred from (\ref{4E75}), (\ref{4E94}) and (\ref{4E95}) also analytically.  The only variables in the expression for ${\bf E}^{\rm uc}$ that depend on the observer's longitude are $\sigma_{l2}$ and ${\bar\sigma}_{l2}$ which vary linearly with $\varphi_P$ (see~(\ref{4E51}) and (\ref{4E52}) and note that, according to (\ref{4E21}), $\varphi_P$ drops out of the expressions for $\sigma_{l1}$ and ${\bar\sigma}_{l1}$).  The variables $\sigma_{l2}$ and ${\bar\sigma}_{l2}$, on the other hand, appear in (\ref{4E94})--(\ref{4E95}) only in the combinations $\sigma_{l2}/\sigma_{l1}^3$ and ${\bar\sigma}_{l2}/{{\bar\sigma}_{l1}}^3$.  Hence, in cases where the turning points of the phase functions are sufficiently close to one another for $\sigma_{l1}$ or ${\bar\sigma}_{11}$ to be appreciably smaller than $1$, as in figure~\ref{PF20}, the arguments of the functions ${\cal G}^{\rm r}_i$ that appear in the expression for ${\bf E}^{\rm uc}$ are highly sensitive functions of $\varphi_P$.  Not only the widths but, as indicated by (\ref{4E78})--(\ref{4E81}) and (\ref{4E89})--(\ref{4E93}), also the amplitudes of ${\cal G}^{\rm r}_i$ vary sharply with $\varphi_P$ when $\sigma_{l1}$ or ${\bar\sigma}_{11}$ assume values that are close to zero.     

\subsection{Brightness temperature}
\label{subsec:brightness}

The brightness temperature $T_{\rm b}$ of the present radiation can be calculated by equating the magnitude of the Poynting flux of this radiation ($c\vert{\bf E}^{\rm uc}\vert^2/4\pi$) to the Rayleigh-Jeans law ($2k_{\rm B}T_{\rm b}\nu^2\Delta\nu/c^2$) for the energy that a black body of the same temperature would emit per unit time per unit area into the frequency band $\Delta\nu$ centred on the frequency $\nu$, where $k_{\rm B}$ is the Boltzmann constant.  The resulting equation can then be solved for $T_{\rm b}$ to obtain 
\begin{equation}
T_{\rm b}=\frac{c^3}{8\pi k_{\rm B}\nu^2\Delta\nu}\vert{\bf E}^{\rm uc}\vert^2.
\label{5E5}
\end{equation}
This in conjunction with (\ref{5E1}) shows that $T_{\rm b}$ is related to the dimensionless Stokes parameter ${\hat I}=I/I_0$ by
\begin{equation}
T_{\rm b}=\frac{5.37\times10^{-4}\,c\,\omega^2 r_{s0}^4B_0^2}{k_{\rm B}{R}_P^2\nu^2\Delta \nu}{\hat I},
\label{5E6}
\end{equation}
where ${r}_{s0}$ and $B_0$ are the radius of the star and the magnitude of the star's dipolar field at its magnetic pole, respectively (see the first paragraph of \S~\ref{sec:magnetosphere}).  

Replacing the parameters that appear in (\ref{5E6}) by
\begin{equation}
B_0=10^{12}{\hat B}_0\,\,{\rm Gauss},\qquad  R_P=D\,\,{\rm kpc}=3.085\times10^{21}D\,\,{\rm cm},
\label{5E7}
\end{equation}
\begin{equation}
\omega=10^2{\hat P}^{-1}\,\,{\rm rad}/{\rm sec},\qquad r_{s0}=10^6 d\,\,{\rm cm},\qquad \nu=10^8 {\hat\nu}\,\,{\rm Hz},\qquad \Delta\nu=10^6\Delta{\hat\nu}\,\,{\rm Hz},
\label{5E8}
\end{equation}
we obtain 
\begin{equation}
T_{\rm b}=1.23\times10^{10}\,{\hat I}\,{\hat T}_{\rm b}\,\,\,\,{}^{\circ}{\rm K}\qquad{\rm with}\qquad{\hat T}_{\rm b}=\frac{{\hat B}_0^2d^4}{{\hat P}^2D^2{\hat \nu}^2\Delta{\hat\nu}}
\label{5E9}
\end{equation}
in which the value of ${\hat I}$ is specified, as in the case of figures~\ref{PF12}--\ref{PF20}, by the numerical evaluation of the Stokes parameter $I$ in units of $I_0$ at the highest peak of the pulse detected at ${\hat R}_P=10^{13}$ (i.e., at ${\hat R}_P=1.028\times10^{13}D{\hat P}^{-1}$ when the factor $1.028D{\hat P}^{-1}$ equals unity). 

The brightness temperature implied by (\ref{5E9}) and $k_{\rm u}=10^7$ is listed in Table~\ref{T1} for the pulses depicted in figures~\ref{PF12}, \ref{PF15} and \ref{PF20} and for a pair of examples of the pulses that are detected at the critical colatitudes $\lim_{R_P\to\infty}\theta_{PlS}$.  Table~\ref{T1} also shows the full width at half maximum $\delta\varphi_P$ of the listed pulses (see figure~\ref{PF20}({\it b})).  Once resolved, the longitudinal distributions of the narrow pulses that stem from the focusing of the radiation when $\sigma_{l1}$ and ${\bar\sigma}_{l1}$ are small all have the same shape as that of the pulse shown in figure~\ref{PF20}({\it b}).  Note that, as indicated by the last column of Table~\ref{T1}, the pulse profiles depicted in figures~\ref{PF12}--\ref{PF19} have to be plotted on considerably shorter longitudinal scales before their peaks assume the shape shown in figure~\ref{PF20}({\it b}) and the maximum values of their dimensionless intensity ${\hat I}$ can be discerned graphically (see \S~\ref{subsec:profiles}).  In general, as one reduces the longitudinal interval over which $I$ is plotted, the peak of the pulse splits in two before the finite widths of either of the partitioned pulses are visible.

Values of $T_{\rm b}$ higher than those listed in Table~\ref{T1} are predicted by (\ref{5E9}) when the colatitude of the observation point lies closer to one of the critical angles $\theta_{PlS}$ or  ${\bar\theta}_{PlS}$.  In the case of $\alpha=60^\circ$, for example, the listed value ($1.17\times10^{40}\,{}^\circ$K) of $T_{\rm b}$ corresponds to $\theta_P=\lim_{R_P\to\infty}\theta_{P1S}=\lim_{R_P\to\infty}\theta_{P2S}=\pi/2$.  For an observation point whose colatitude is closer to the critical angles in question than $\pi/2$ is, e.g., for $\theta_P=\theta_{P1S}+10^{-20}$ rad, $T_{\rm b}$ and $\delta\varphi_P$ have the values $1.17\times10^{54}\,{}^\circ$K and $2.26\times10^{-34}$ deg, respectively.

\begin{table}
\centering
\begin{tabular}{ c | c | c | c }     
$\alpha\,\, ({\rm deg})$ & $\theta_P\,\, ({\rm deg})$ & $T_{\rm b}/{\hat T}_{\rm b}\,\, ({}^{\circ}{\rm K})$ & $\delta\varphi_P\,\, ({\rm deg})$ \\       \hline
 $5$ & $5$ & $1.29\times10^{22}$ & $ 6.76\times10^{-9} $\\  \hline
 $30$ & $\lim_{R_P\to\infty}\theta_{P2S}$ & $6.89\times10^{40}$ & $ 2.39\times10^{-24} $\\  \hline
 $45$ & $55$ & $2.34\times10^{17}$ & $ 3.81\times10^{-7} $ \\  \hline
 $60$ & $90$ & $1.17\times10^{40}$ & $ 6.94\times10^{-25} $ \\  \hline
 $75$ & $\lim_{R_P\to\infty}\theta_{P1S}$ & $4.06\times10^{38}$ & $ 5.73\times10^{-25} $ \\  \hline
\end{tabular}
\caption{Brightness temperature $T_{\rm b}$ and full width at half maximum $\delta\varphi_P$ of the pulse detected at colatitude $\theta_P$ for the inclination angle $\alpha$ and ${\hat R}_P=10^{13}$, $k_{\rm u}=10^7$.  The dimensionless factor ${\hat T}_{\rm b}$, defined by (\ref{5E7})--(\ref{5E9}), is of the order of unity in the case of most radio pulsars.  The limiting values of $\theta_{P1S}$ and $\theta_{P2S}$ in the second column are $48.533945294618400228^\circ$ and $33.932818533330613261^\circ$, respectively.}
\label{T1}
\end{table}

\subsection{Frequency spectrum}
\label{sec:frequency}

Given that the radiation field ${\bf E}^{\rm uc}$ depends on the observation time $t_P$ only in the combination $\varphi_P-\omega t_P$, the frequency spectrum of the present radiation is equally well described by the Fourier decomposition of the field ${\bf E}^{\rm uc}$ with respect to the azimuthal angle $\varphi_P$ of the observation point.  In the present case, the content of this spectrum stems from two factors.  One factor is the thickness of the current sheet ($\simeq c/(k_{\rm u}\omega$) which is manifested in the sharp small-amplitude modulations (microstructure) of the pulse profile (see figures~\ref{PF12} and \ref{PF16}):  the wavelengths of such modulations are proportional to $k_{\rm u}^{-1}$.  The other factor is the full width at half maximum ($\delta\varphi_P$) of the pulse with the highest peak in the pulse profile (see figure~\ref{PF20}({\it b})): the fraction  $\delta\varphi_P/2\pi$ of a period during which such narrow pulses propagate past a detector is by many orders of magnitude smaller than the fraction $k_{\rm u}^{-1}$ of the light-cylinder-radius in cases where the colatitude of the observation point is close to or coincident with one of the critical angles described in \S~\ref{subsec:critical} (see Table~\ref{T1}).  While the Fourier decomposition of the fluctuations associated with the first factor yields a frequency spectrum centred on radio waves when $k_{\rm u}\gtrsim10^5$, that of the fluctuations associated with the second factor yields a wide spectral distribution extending to gamma rays: the value $\delta\varphi_P=1.21\times10^{-26}$ radian appearing in the last column of Table~\ref{T1} corresponds to a frequency spectrum that extends as far as $\nu\simeq\omega/(2\pi\delta\varphi_P)\simeq1.31\times10^{27} {\hat P}^{-1}$ Hz.

Our replacing the Dirac delta function in (\ref{4E5}) by its Fourier representation (\ref{4E11}) is tantamount to Fourier analysing the fluctuations of the radiation field that arise from the short thickness of the current sheet with respect to $\varphi_P$ since the argument of that delta function depends on $\varphi_P$ linearly.  The spectral distribution of the part of the radiation that stems from the thickness of the current sheet is therefore given by the $k$ dependence of 
\begin{eqnarray}
{\tilde I}&\simeq&\left(\frac{4\pi j_0}{3\omega}\right)^2\sum_{l=1}^2 k^{-2/3}\bigg\vert\exp({\rm i}k\sigma_{l2})\nonumber\\*
&&\times\left[\mbox{{\boldmath${\cal P}$}}_l{\rm Ai}(-k^{2/3}\sigma_{l1}^2)-{\rm i}k^{-1/3}\mbox{{\boldmath${\cal Q}$}}_l{\rm Ai}^\prime(-k^{2/3}\sigma_{l1}^2)\right]+\exp({\rm i}k{\bar\sigma}_{l2})\nonumber\\*
&&\times\left[\mbox{{\boldmath${\bar{\cal P}}$}}_l{\rm Ai}\left(-k^{2/3}{\bar\sigma}_{l1}^2\right)-{\rm i}k^{-1/3}\mbox{{\boldmath${\bar{\cal Q}}$}}_l{\rm Ai}^\prime\left(-k^{2/3}{\bar\sigma}_{l1}^2\right)\right]\bigg\vert^2,
\label{5E10}
\end{eqnarray}
i.e., the $k$ dependence of the square of the modulus of the integrand that appears in (\ref{4E61}).  The frequency $\nu$ of the radiation is related to $k$ through $\nu=(2\pi k)/\omega$.  

At  harmonic numbers $k$ for which the arguments of the Airy functions in (\ref{5E10}) are smaller or of the order of unity, these functions assume values that are independent of frequency.  When their arguments are large, on the other hand, they reduce to  
\begin{equation}
{\rm Ai}(-k^{2/3}\sigma_{l1}^2)\simeq\pi^{-1/2}\sigma_{l1}^{-1/2}k^{-1/6}\cos\left({\textstyle\frac{2}{3}}k\sigma_{l1}^3-{\textstyle\frac{1}{4}}\pi\right),
\label{5E11}
\end{equation}
and
\begin{equation}
{\rm Ai}^\prime(-k^{2/3}\sigma_{l1}^2)\simeq\pi^{-1/2}\sigma_{l1}^{1/2}k^{1/6}\sin\left({\textstyle\frac{2}{3}}k\sigma_{l1}^3-{\textstyle\frac{1}{4}}\pi\right).
\label{5E12}
\end{equation}
Equation (\ref{5E10}) and these limiting values of the Airy functions jointly yield the dependence
\begin{equation}
{\tilde I}\propto k^{-\beta}
\label{5E13}
\end{equation}
of the radiation intensity ${\tilde I}$ on frequency and the values that the spectral index $\beta$ can assume in various regimes. 

The variables $\sigma_{l1}$ and $\sigma_{l2}$ and the vector functions ${\bf P}_l$ and ${\bf Q}_l$ that appear in (\ref{5E10}) are independent of $k$.  When $\vert{\bf P}_l\vert\gg k^{-1/2}\vert{\bf Q}_l\vert$, the vector $\mbox{{\boldmath${\cal K}$}}_l$ and hence $\mbox{{\boldmath${\cal P}$}}_l$ and $\mbox{{\boldmath${\cal Q}$}}_l$ are also independent of $k$ and are by a factor of the order of $k^{1/2}$ larger than $\mbox{{\boldmath${\bar{\cal P}}$}}_l$ and $\mbox{{\boldmath${\bar{\cal Q}}$}}_l$ (see (\ref{4E55})--(\ref{4E60})).  In this case, the possible values of the spectral index $\beta$ are determined by the relative magnitudes of $\vert\mbox{{\boldmath${\cal P}$}}_l\vert$ and $\vert\mbox{{\boldmath${\cal Q}$}}_l\vert$ only.  If $\vert\mbox{{\boldmath${\cal P}$}}_l\vert\gg k^{-1/3}\vert \mbox{{\boldmath${\cal Q}$}}_l\vert$, then $\beta=2/3$ when the Airy functions in the first square bracket in (\ref{5E10}) are of the order of unity and $\beta=1$ when these Airy functions have the limiting values given by (\ref{5E11}) and (\ref{5E12}) and so the first square bracket in (\ref{5E10}) decays as $k^{-1/6}$.   If $\vert\mbox{{\boldmath${\cal P}$}}_l\vert\ll k^{-1/3}\vert \mbox{{\boldmath${\cal Q}$}}_l\vert$, then $\beta=4/3$ when the arguments of the Airy functions in question are of the order of unity and $\beta=1$ when the first square bracket in (\ref{5E10}) decays as $k^{-1/6}$.  When $\sigma_{l1}$ is small and the second term of the first square bracket in (\ref{5E10}) dominates, there is a short frequency interval in which the spectral intensity ${\tilde I}$ increases with increasing $k$. 

In the opposite regime $\vert{\bf P}_l\vert\ll k^{-1/2}\vert{\bf Q}_l\vert$, the factor $k^{-1/2}$ multiplying ${\bf Q}_l$ in (\ref{4E59}) and (\ref{4E60}) reduces the value of the spectral index $\beta$ by $1$ everywhere (see Table~\ref{T2}).  

\begin{table}
\centering
\begin{tabular}{ l | l | l | l | r | r}     
{} & $k\ll\sigma_{l1}^{-3}$ & $k\gg\sigma_{l1}^{-3}$ \\       \hline
 $\vert{\bf P}_l\vert\gg k^{-1/2}\vert{\bf Q}_l\vert$ and $\vert\mbox{{\boldmath${\cal P}$}}_l\vert\gg k^{-1/3}\vert \mbox{{\boldmath${\cal Q}$}}_l\vert$&$\quad2/3$ & $\quad\,\,\,1$\\  \hline
$\vert{\bf P}_l\vert\gg k^{-1/2}\vert{\bf Q}_l\vert$ and $\vert\mbox{{\boldmath${\cal P}$}}_l\vert\ll k^{-1/3}\vert \mbox{{\boldmath${\cal Q}$}}_l\vert$ & $\quad4/3$ & $\quad\,\,\,1$\\ \hline 
$\vert{\bf P}_l\vert\ll k^{-1/2}\vert{\bf Q}_l\vert$ and $\vert\mbox{{\boldmath${\cal P}$}}_l\vert\gg k^{-1/3}\vert \mbox{{\boldmath${\cal Q}$}}_l\vert$& $\quad5/3$&$\quad\,\,\,2$\\  \hline 
$\vert{\bf P}_l\vert\ll k^{-1/2}\vert{\bf Q}_l\vert$ and $\vert\mbox{{\boldmath${\cal P}$}}_l\vert\ll k^{-1/3}\vert \mbox{{\boldmath${\cal Q}$}}_l\vert$&$\quad7/3$&$\quad\,\,\,2$\\  \hline
\end{tabular}
\caption{Values, in various regimes, of the spectral index $\beta$ (defined in (\ref{5E13})) for the part of the radiation associated with the sharp small-amplitude modulations (microstructure) of the pulse profile.}
\label{T2}
\end{table}

\subsection{Flux density and its rate of decay with distance}
\label{subsec:inverse-square}

Flux density of a radiation is given, in general, by the magnitude of the Poynting vector, $c\vert{\bf E}^{\rm uc}\vert^2/4\pi$, which has the dimensions of erg/(cm${}^2\times$sec) in cgs units.  In the present case, however, the linear extents in the azimuthal direction, $R_P\delta\varphi_P$, of the focused radiation beams that embody the high-frequency (optical to gamma-ray) radiation are invariably smaller than $1$ cm at $R_P=1$ kpc (see Table~\ref{T1}).  The amount of energy that crosses a unit area per unit time is therefore given by 
\begin{eqnarray}
{\cal S}&=&\frac{c}{4\pi}\vert{\bf E}^{\rm uc}\vert^2\delta\varphi_P\nonumber\\*
&=&2.79\times10^{-3}{\hat I}\delta\varphi_P\left(\frac{{\hat B}_0d^2}{{\hat P}D}\right)^2\quad \frac{{\rm erg}}{{\rm cm}^2\times{\rm sec}} ,
\label{5E14}
\end{eqnarray}
in which $\delta\varphi_P$ is in radians (see (\ref{5E1}), (\ref{5E4}), (\ref{5E7}) and (\ref{5E8})).  Note that the linear extents in the latitudinal direction, $R_P\delta\theta_P$, of the focused radiation beams that embody the high-frequency radiation are of the order of the light-cylinder radius, $c/\omega$, in general: these beams remain fully in focus at all distances ${\hat R}_P$ over the latitudinal interval $\delta\theta_P\simeq\vert\theta_{PlS}-\lim_{R_P\to\infty}\theta_{PlS}\vert$ which turns out to be of the order of ${\hat R}_P^{-1}$ independently of the values of the other parameters. 

In the case of $\alpha=60^\circ$, $\theta_P=90^\circ$, $k_{\rm u}=10^7$ and $D=1$ kpc depicted in figure~\ref{PF20}, for example, the flux density ${\cal S}$ has the value $32.1\, {\cal S}_0$ erg/(sec$\times$cm${}^2$), where
\begin{equation}
{\cal S}_0=\left(\frac{{\hat B}_0d^2}{{\hat P}}\right)^2.
\label{5E15}
\end{equation}
At latitudes closer to or further away from the critical angle for this example ($\theta_P=90^\circ$), the degree of focusing of the radiation beam and so the value of the flux density ${\cal S}$ is, respectively, higher or lower.

As pointed out in \S~\ref{subsec:critical}, the length of the interval $\vert\tau_{l{\rm max}}-\tau_{l{\rm min}}\vert$ separating the $\tau$ coordinates of the maximum and minimum of the phase function $f_{lC}$ decreases as ${\hat R}_P^{-1/2}$ with increasing ${\hat R}_P$ in a case where this interval is small, i.e., when the colatitude $\theta_P$ of the observation point has the critical value $\theta_{PlS}(L,\alpha)$, with $L > {\hat R}_P$.  In particular, if the observation point has the colatitude $\lim_{R_P\to\infty}\theta_{PlS}$ (or $\pi-\lim_{R_P\to\infty}\theta_{PlS}$), then the maximum and minimum of $f_{lC}$ coalesce into an inflection point only at ${\hat R}_P\to\infty$, rather than at a finite distance $L$.  (These statements hold true also when $f_{lC}$ and $\theta_{PlS}$ are replaced by ${\bar f}_{lC}$ and ${\bar\theta}_{PlS}$, respectively.)  In the case illustrated in figure~\ref{PF20}, for example, the colatitude of the observation point equals $\lim_{R_P\to\infty}\theta_{P1S}=90^\circ$ so that at the finite distance ${\hat R}_P=10^{13}$, the $\tau$ coordinates of  maximum and minimum of $f_{1C}$ are separated by the short interval $3.05\times10^{-5}$ degrees. It follows from the expression for $\partial f_{1C}/\partial\tau$ in (\ref{A1}) that this separation has the value $3.05\times10^{-5}({\hat R}_P/10^{13})^{-1/2}$ for all ${\hat R}_P$.

The enhanced focusing of the radiation with distance that is caused by this shortening of the separation between the turning points of the phase functions results in a slower decay rate of the flux density with distance than that predicted by the inverse-sqaure law.  Along colatitudes close to $\theta_{PlS}$ or ${\bar\theta}_{PlS}$, the flux density ${\cal S}$ of the radiation diminishes with increasing distance from its source as ${\hat R}_P^{-3/2}$ instead of ${\hat R}_P^{-2}$.  This dependence of ${\cal S}$ on ${\hat R}_P$, or equivalently $D$, is illustrated in figure~\ref{PF21} in the case where $\alpha=60^\circ$, $\theta_P=90^\circ$, $k_{\rm u}=10^7$, and $D$ ranges from $0.1$ to $10^5$ kpc, i.e., from a galactic to a cosmological distance.

\begin{figure}
\centerline{\includegraphics[width=14cm]{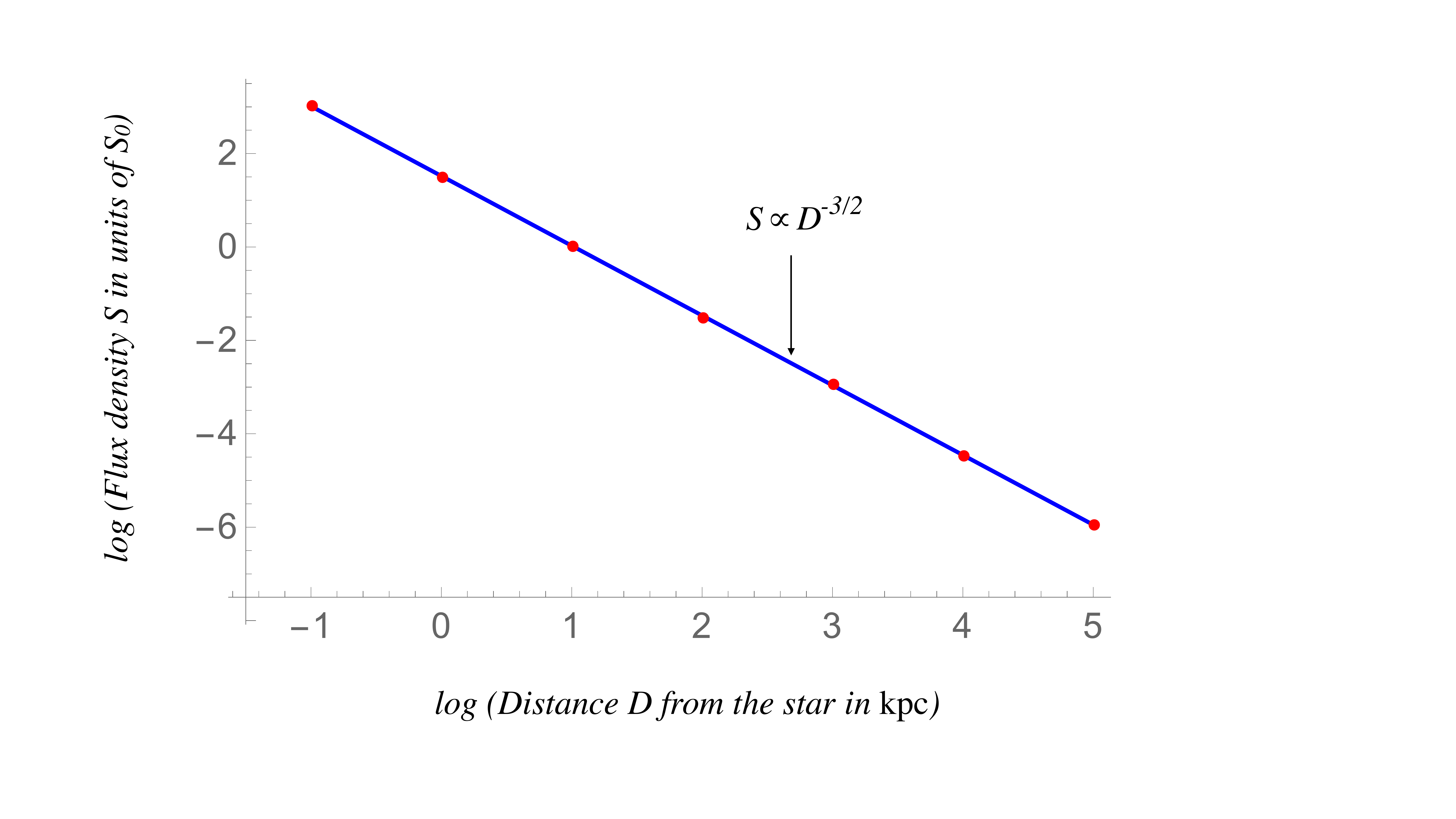}}
\caption{$log({\cal S}/{\cal S}_0)$ versus $log(D)$ for $\alpha=60^\circ$, $\theta_P=90^\circ$ and $k_{\rm u}=10^7$.  The blue line with the slope $-3/2$ is the best fit to the red dots whose coordinates are determined by evaluating (\ref{5E14}).  The violation of the inverse-square law illustrated in this figure remains in force all the way to infinity whenever the colatitude of the observation point coincides with or is close to one of the eight angles given by $\lim_{R_P\to\infty}\theta_{PlS}$,  $\lim_{R_P\to\infty}{\bar\theta}_{PlS}$, $\pi-\lim_{R_P\to\infty}\theta_{PlS}$ and $\pi-\lim_{R_P\to\infty}{\bar\theta}_{PlS}$.}
\label{PF21}
\end{figure} 

The violation of the inverse-square law encountered here (i.e., the fact that ${\cal S}$ is proportional to $D^{-3/2}$ rather than being proportional to $D^{-2}$) is not incompatible with the requirements of the conservation of energy because the radiation process discussed in this paper is intrinsically transitive.  Temporal rate of change of the energy density of the radiation generated by this process has a time-averaged value that is negative (instead of being zero as in a conventional radiation) at points where the envelopes of of the wave fronts emanating from the constituent volume elements of the source distribution are cusped.  The difference in the fluxes of power across any two spheres centred on the star is in this case balanced by the change with time of the energy contained inside the shell bounded by those spheres (see appendix C of~\cite{Ardavan_JPP} for a detailed discussion of this point).

\clearpage

\appendix
\section{Derivatives of the phase functions $f_{lC}$ and ${\bar f}_{lC}$}
\label{AppA}
\setcounter{equation}{0}

The first two derivatives with respect to $\tau$ of the function $f_{lC}$ that appears in the phase of the first exponential in (\ref{4E45}) are given by 
\begin{equation}
\frac{\partial f_{lC}}{\partial\tau}=\frac{\sin\alpha\sin\tau}{\sin\theta}\frac{\partial f_{lC}}{\partial\theta},
\label{A1}
\end{equation}
\begin{equation}
\frac{\partial ^2f_{lC}}{\partial\tau^2}=\frac{\sin\alpha\cos^2\alpha\cos\tau}{\sin^3\theta}\frac{\partial f_{lC}}{\partial\theta}+\frac{\sin^2\alpha\sin^2\tau}{\sin^2\theta}\frac{\partial^2 f_{lC}}{\partial\theta^2},
\label{A2}
\end{equation}
with
\begin{eqnarray}
\frac{\partial f_{lC}}{\partial\theta}&=&({\hat r}_P^2-1)^{-1/2}({\hat R}_P^2\sin^2\theta-1)^{-1/2}\left[\csc\theta({\hat z}_P-{\hat r}_{sC}\cos\theta)({\hat r}_P^2{\hat r}_{sC}^2\sin^2\theta-1)^{1/2}\right.\nonumber\\*
&&\left.-{\hat r}_{sC}{\hat r}_P\sin\theta({\hat r}_P{\hat r}_{sC}\cos\theta-\cot\theta_P)\right]\nonumber\\*
&&-(-1)^l\cot\alpha\csc^2\theta(1-\cot^2\alpha\cot^2\theta)^{-1/2},
\label{A3}
\end{eqnarray}
and
\begin{eqnarray}
\frac{\partial ^2f_{lC}}{\partial\theta^2}&=&({\hat r}_P^2-1)^{-1/2}({\hat R}_P^2\sin^2\theta-1)^{-1/2}\Big\{({\hat r}_P^2-1)^{-1/2}({\hat R}_P^2\sin^2\theta-1)^{-1/2}\nonumber\\*
&&\times({\hat z}_P-{\hat r}_{sC}\cos\theta)\Big[{\hat r}_P^2{\hat r}_{sC}^2({\hat r}_P^2{\hat r}_{sC}^2\sin^2\theta-1)^{-1/2}({\hat z}_P-{\hat r}_{sC}\cos\theta)+{\hat z}_P{\hat r}_{sC}\Big]\nonumber\\*
&&-\csc\theta({\hat r}_P^2{\hat r}_{sC}^2\sin^2\theta-1)^{1/2}\Big[\cos\theta\frac{\partial{\hat r}_{sC}}{\partial\theta}+\csc\theta({\hat z}_P\cos\theta-{\hat r}_{sC})\Big]\nonumber\\*
&&-{\hat R}_P^2\cos\theta({\hat R}_P^2\sin^2\theta-1)^{-1}\Big[({\hat z}_P-{\hat r}_{sC}\cos\theta)({\hat r}_P^2{\hat r}_{sC}^2\sin^2\theta-1)^{1/2}\nonumber\\*
&&-{\hat r}_P{\hat r}_{sC}\sin^2\theta({\hat r}_P{\hat r}_{sC}\cos\theta-\cot\theta_P)\Big]-{\hat r}_P^2{\hat r}_{sC}\Big(\sin2\theta\frac{\partial{\hat r}_{sC}}{\partial\theta}+{\hat r}_{sC}\cos2\theta\Big)\Big\}\nonumber\\*
&&+(-1)^l\cos\theta\csc^5\theta\cot\alpha(1-\cot^2\alpha\cot^2\theta)^{-3/2}(1-\csc^2\alpha\cos2\theta),\nonumber\\*
\label{A4}
\end{eqnarray}
where
\begin{equation}
\frac{\partial{\hat r}_{sC}}{\partial\theta}=-\frac{{\hat r}_P{\hat r}_{sC}\sin\theta({\hat r}_P{\hat r}_{sC}\cos\theta-\cot\theta_P)}{({\hat r}_P^2-1)^{1/2}({\hat R}_P^2\sin^2\theta-1)^{1/2}}
\label{A5}
\end{equation}
(see~(\ref{4E16}), (\ref{4E21}) and (\ref{4E23})).

The corresponding derivatives of ${\bar f}_{lC}$ follow from (\ref{A3}), (\ref{A4}) and the following derivatives of the function $a$ defined in (\ref{4E22}):
\begin{eqnarray}
\frac{\partial a}{\partial\theta}&=&\frac{\sin\theta}{({\hat r}_P^2-1)^{1/2}({\hat R}_P^2\sin^2\theta-1)^{1/2}}\Bigg\{\frac{{\hat r}_P^2{\hat r}_{sC}^2({\hat z}_P-{\hat r}_{sC}\cos\theta)}{({\hat r}_P^2{\hat r}_{sC}^2\sin^2\theta-1)^{1/2}}\left[\frac{2a}{({\hat r}_P^2{\hat r}_{sC}^2\sin^2\theta-1)^{1/2}}+1\right]\nonumber\\*
&&-\frac{({\hat r}_P^2\cos\theta-{\hat z}_P/{\hat r}_{sC})({\hat r}_P^2{\hat r}_{sC}^2\sin^2\theta-1)^{1/2}}{({\hat r}_P^2-1)^{1/2}({\hat R}_P^2\sin^2\theta-1)^{1/2}}+\frac{{\hat R}_P^2\cos\theta({\hat r}_P^2{\hat r}_{sC}^2\sin^2\theta-1)^{1/2}}{{\hat R}_P^2\sin^2\theta-1}\nonumber\\*
&&\times\left[\left(\frac{1}{{\hat r}_{sC}}-{\hat r}_{sC}\sin^2\theta_P\right)-({\hat r}_P^2{\hat r}_{sC}^2\sin^2\theta-1)^{1/2}\right]\Bigg\}
\label{A6}
\end{eqnarray}
and
\begin{eqnarray}
\frac{\partial^2 a}{\partial\theta^2}&=&-\frac{\cot\theta}{{\hat R}_P^2\sin^2\theta-1}\frac{\partial a}{\partial\theta}+\frac{\sin^2\theta}{({\hat r}_P^2-1)({\hat R}_P^2\sin^2\theta-1)}\Bigg[{\hat r}_P^2{\hat r}_{sC}^2({\hat z}_P-{\hat r}_{zC}\cos\theta)\nonumber\\*
&&\times\bigg\{-\frac{{\hat r}_P^2{\hat r}_{sC}^2({\hat z}_P-{\hat r}_{zC}\cos\theta)}{({\hat r}_P^2{\hat r}_{sC}^2\sin^2\theta-1)^2}\left[4a+({\hat r}_P^2{\hat r}_{sC}^2\sin^2\theta-1)^{1/2}\right]\nonumber\\*
&&-\frac{{\hat r}_P^2\cos\theta-{\hat z}_P/{\hat r}_{sC}}{({\hat r}_P^2-1)^{1/2}({\hat R}_P^2\sin^2\theta-1)^{1/2}({\hat r}_P^2{\hat r}_{sC}^2\sin^2\theta-1)^{1/2}}+\frac{{\hat R}_P^2\cos\theta}{{\hat R}_P^2\sin^2\theta-1}\nonumber\\*
&&\times\left[\frac{1-{\hat r}_{sC}^2\sin^2\theta_P}{{\hat r}_{sC}({\hat r}_P^2{\hat r}_{sC}^2\sin^2\theta-1)^{1/2}}-2\right]\bigg\}-{\hat r}_{sC}({\hat r}_P^2{\hat r}_{sC}\cos\theta-{\hat z}_P)\nonumber\\*
&&\times\Bigg\{\frac{{\hat r}_P^2{\hat r}_{sC}(2{\hat z}_P-3{\hat r}_{sC}\cos\theta)}{{\hat r}_P^2{\hat r}_{sC}^2\sin^2\theta-1}\left[2a+({\hat r}_P^2{\hat r}_{sC}^2\sin^2\theta-1)^{1/2}\right]-\frac{({\hat r}_P^2{\hat r}_{sC}^2\sin^2\theta-1)^{1/2}}{{\hat r}_{sC}^2({\hat R}_P^2\sin^2\theta-1)}\nonumber\\*
&&\times\left[\frac{{\hat z}_P({\hat R}_P^2\sin^2\theta-1)^{1/2}}{({\hat r}_P^2-1)^{1/2}}+{\hat R}_P^2\cos\theta(1+{\hat r}_{sC}^2\sin^2\theta_P)\right]\Bigg\}\nonumber\\*
&&+\frac{{\hat r}_P^2{\hat r}_{sC}^2({\hat r}_P^2-1)^{1/2}({\hat R}_P^2\sin^2\theta-1)^{1/2}}{{\hat r}_P^2{\hat r}_{sC}^2\sin^2\theta-1}\left[2\csc\theta({\hat z}_P-{\hat r}_{sC}\cos\theta)\frac{\partial a}{\partial\theta}+2{\hat r}_{sC}a\right.\nonumber\\*
&&\left.+{\hat r}_{sC}({\hat r}_P^2{\hat r}_{sC}^2\sin^2\theta-1)^{1/2}\right]+\frac{{\hat R}_P^2({\hat r}_P^2{\hat r}_{sC}^2\sin^2\theta-1)^{1/2}}{{\hat R}_P^2\sin^2\theta-1}\left({\hat r}_P^2-\sin^2\theta_P-\frac{{\hat z}_P}{{\hat r}_{sC}}\cos\theta\right)\nonumber\\*
&&-\frac{{\hat R}_P^2({\hat r}_P^2-1)^{1/2}({\hat r}_P^2{\hat r}_{sC}^2\sin^2\theta-1)^{1/2}}{({\hat R}_P^2\sin^2\theta-1)^{3/2}}[{\hat R}_P^2(1+\cos^2\theta)-1]\nonumber\\*
&&\times\left[\frac{1}{{\hat r}_{sC}}-{\hat r}_{sC}\sin^2\theta_P-({\hat r}_P^2{\hat r}_{sC}^2\sin^2\theta-1)^{1/2}\right]\Bigg].
\label{A7}
\end{eqnarray}
These derivatives can be converted into $\partial a/\partial\tau$ and $\partial^2 a/\partial\tau^2$ in the same way as was done for the derivatives of $f_{lC}$ in (\ref{A1}) and (\ref{A2}).